%% file: main.tex
\documentclass[reprint,aip,amsiath,amssymb,onecolumn]{revtex4-1}
\usepackage{graphicx}
\usepackage{dcolumn}
\usepackage{bm}
\usepackage[table]{xcolor}
\usepackage{tikz}
\usepackage[utf8]{inputenc}
\usepackage[T1]{fontenc}
\usepackage{mathptmx}
\usepackage{amsmath}
\usepackage{chngcntr}
\usepackage{pifont}
\newcommand{\cmark}{\ding{51}}%
\newcommand{\xmark}{\ding{55}}%

\newcommand{\fl}{}
\usepackage[UKenglish]{babel}
\usepackage{bbold}
\usepackage{appendix}
\usepackage{tabularx}
\usepackage[table]{xcolor}
\newcommand{\ii}{\text{i}}
\newcommand{\p}{\nu}
\newcommand{\g}{g}
\newcommand{\bmrz}{\bm{r}_0}
\newcommand{\T}{\rm{T}}
\newcommand{\cinf}{\bm{\chi}_{\infty}}
\newcommand{\eps}{\epsilon}
\newcommand{\be}{\begin{equation}}
\newcommand{\ee}{\end{equation}}
\newcommand{\ba}{\begin{eqnarray}}
\newcommand{\ea}{\end{eqnarray}}

\newcommand{\order}{\mathcal{O}}

\newcommand{\bdx}{\delta\bm{x}}

\newcommand{\dx}{\delta x}

\newcommand{\ebb}{\bm{E}_{\rm bb}}

\newcommand*\xbar[1]{%
  \hbox{%
    \vbox{%
      \hrule height 0.5pt 
      \kern0.2ex
      \hbox{%
        \kern-0.2em
        \ensuremath{#1}%
        \kern-0.1em
      }%
    }%
  }%
}

\newcommand{\uhat}{\hat{\bm{U}}}

\newcommand{\uinf}{\bm{U}_\infty}

\newcommand{\bu}{\bm{U}}

\newcommand{\rinf}{\bm{r}_{\infty}}
\newcommand{\rinfz}{\bm{r}_{0\,\infty}}
\newcommand{\rfhat}{\hat{\bm{r}}}
\newcommand{\rfhatz}{\hat{\bm{r}}_0}

\newcommand{\beginsupplement}{%
        \setcounter{table}{0}
        \renewcommand{\thetable}{S\arabic{table}}%
        \setcounter{figure}{0}
        \renewcommand{\thefigure}{S\arabic{figure}}%
     }
\newcommand{\red}{\textcolor{black}}

\definecolor{darkpastelgreen}{rgb}{0.01, 0.75, 0.24}

\begin{document}
\title[]{Systematic model reduction captures the dynamics of extrinsic noise in biochemical subnetworks}
\author{Barbara Bravi}
\email{barbarabravi@ymail.com}
\affiliation{Institute of Theoretical Physics, Ecole Polytechnique F\'ed\'erale de Lausanne (EPFL), CH-1015 Lausanne, Switzerland\\
Current affiliation: Laboratoire de physique de l’Ecole Normale Sup\'erieure, PSL University, CNRS,
Sorbonne Universit\'e, Universit\'e de Paris,
24 rue Lhomond, 75005 Paris, France}
\author{Katy J. Rubin}
\affiliation{Department of Mathematics, King's College London, Strand, London WC2R 2LS, UK}
\author{Peter Sollich}
\email{peter.sollich@uni-goettingen.de}
\affiliation{Department of Mathematics, King's College London, Strand, London WC2R 2LS, UK}
\affiliation{Institute for Theoretical Physics, Georg-August-University G\"ottingen, Friedrich-Hund-Platz 1, 37077 G\"ottingen, Germany}

\begin{abstract}
We consider the general problem of describing the dynamics of \emph{subnetworks} of larger biochemical reaction networks, e.g.\ protein interaction networks involving complex formation and dissociation reactions. We propose the use of model reduction strategies to understand the ``extrinsic'' sources of stochasticity arising from the rest of the network. Our approaches are based on subnetwork dynamical equations derived by projection methods and by path integrals. The results provide a principled derivation of the different components of the extrinsic noise that is observed experimentally in cellular biochemical reactions, over and above the intrinsic noise from the stochasticity of biochemical events in the subnetwork. We explore several intermediate approximations to assess systematically the relative importance of different extrinsic noise components, including initial transients, long-time plateaus, temporal correlations, multiplicative noise terms and nonlinear noise propagation. The best approximations achieve excellent accuracy in quantitative tests on a simple protein network and on the epidermal growth factor receptor signalling network. 
\end{abstract}

\maketitle
\section{Introduction}
Networks of biochemical reactions can comprise thousands of species, making mathematical modeling of their full structure a difficult task. The many parameters often cannot be estimated with the precision that would be required for reliable quantitative predictions. Focusing on a few, well characterized species, i.e.\ ``reducing'' the model to a subnetwork, is then necessary for the sake of analysis and qualitative understanding \cite{apri, ackermann, radulescu}. In this paper we address the question of how noise should be incorporated into such reduced models. The complex kinetics of e.g.\ intracellular biochemical networks is crucially determined by fluctuations, their interplay with nonlinearities giving rise to a rich set of dynamical behaviors such as signal amplification and damping, focusing of oscillations \cite{paulsson_fluct}, and bimodal profiles of gene expression \cite{del_giudice} and cell responses \cite{birtwistle}. These effects can make noise either detrimental or functional to signal processing within cells \cite{braviUCNC}. Accounting properly for fluctuations while retaining the benefits of reduced models thus becomes a key challenge; to address it one has to keep track of the effects of the embedding environment -- which we will call the ``bulk'' -- of a chosen subnetwork. The resulting dynamical information will be crucial in identifying and ultimately measuring the different drivers of cellular heterogeneity. This would be of great benefit in synthetic biology applications that aim to design circuits where the effects of the biomolecular context are fully characterized, and could even be exploited for specific functions.

Among possible approaches to model reduction \cite{ackermann, radulescu, apri, okino, sunnaker}, projection methods from statistical mechanics \cite{zwanzig, ritort} are attractive as they allow one to project a given dynamics systematically onto an arbitrary subspace of subnetwork observables. In protein-protein interaction networks this has been shown to improve the accuracy of quantitative predictions for the subnetwork dynamics~\cite{katy}, but the method could track only averages over the intrinsic subnetwork noise, with fluctuations arising solely from the potentially random bulk initial conditions. An alternative field-theoretic approach for reduction to subnetwork dynamics avoids this restriction~\cite{gaussianvar}. This approximate method relies on perturbations to a Gaussian theory and so will be referred to as ``3rd order corrected Gaussian Approximation'' (3GA) below. Its key advantage for our present purposes is that it can in principle account for all stochastic fluctuations including those from intrinsic subnetwork noise. In the limit of vanishing intrinsic noise the two schemes become essentially identical~\cite{katy,gaussianvar}: while they produce subnetwork dynamical equations with a different structure, one can prove~\cite{gaussianvar} that the resulting predictions agree up to quadratic order in deviations from a dynamical fixed point. 

The presence of the surrounding bulk system generates both memory terms in the subnetwork dynamics as well as \emph{extrinsic} noise \cite{swain,Paulsson2004}, which is the focus of our interest here. Hence, as our first new contribution, we extend the formalism of previous studies \cite{katy,gaussianvar} to extrinsic noise stemming from two sources: the fact that the initial deviation from steady state in the bulk is unknown and therefore can be taken as randomly distributed, and the intrinsic stochastic fluctuations of the dynamics, captured only by the 3GA framework. Importantly, our results do not rely on any timescale separation, while projection methods and more recently path integrals have so far been applied primarily to coarse-graining by elimination of fast bulk variables \cite{thomasproj,bo,england,sinitsyn}. As a second novel contribution, we propose and analyze approximations of the extrinsic noise terms that help us evaluate them more expediently and elucidate the relative contribution of different noise properties such as color, non-linearity and multiplicativity of the source.

The structure of this paper is as follows. We first recall key features of model reduction approaches for subnetwork dynamics in Sec.\ \ref{sec:modred}, starting from the case of linear reaction equations to build insight (Sec.\ \ref{sec:lin}). We then derive the extrinsic noise terms for the case of nonlinear dynamics, separately by the projection method (Sec.\ \ref{sec:proj}) and by 3GA (Sec.\ \ref{sec:3GA}). In Sec.\ \ref{sec:accuracyrf} we assess the accuracy of subnetwork equations with stochastic terms arising only from uncertain bulk initial state, for which a comparison between 3GA and projection methods can be made. As in general it will be complicated to estimate the full time-dependence of these stochastic terms from measurements on the bulk, we propose an approximation aimed at capturing the long and short time subnetwork behavior (Sec.\ \ref{sec:approx_rf_main}). The restriction to negligible intrinsic noise can be easily removed in the 3GA approach, allowing us to characterize all the intrinsic and extrinsic terms that contribute to the subnetwork noise. We compare the 3GA performance to the Steady State Linear Noise Approximation \cite{thomas,thomasproj} and to intermediate approximations that neglect either nonlinearity or time correlations in the extrinsic noise (Sec.\ \ref{sec:approx_chi}). In addition we estimate how the accuracy of all these approximations scales with the strength of extrinsic noise, in Secs.~\ref{sec:var_eps0} and \ref{sec:var_eps}. For this detailed analysis we use a simple network model; in the final Sec.\ \ref{sec:egfr} we extend this and explore the application of different model reduction approaches to a paradigmatic signalling pathway, the reaction network around the Epidermal Growth Factor Receptor.

\section{Model reduction methods for subnetwork dynamics}
\label{sec:modred}

\subsection{Subnetworks in protein-protein interaction networks}
\label{sec:setup}
We consider a reaction network of $N$ molecular species such as proteins and protein complexes, with concentrations $\bm{x}=(x_1,\ldots,x_N)$, evolving in time according to a set of Chemical Langevin Equations (CLE) \cite{vankampen, gardiner}
\begin{equation}
\label{eq:steq}
\frac{dx_i(t)}{dt}=[\bm{S}\bm{f}(\bm{x}(t))]_i +\xi_i(t) \qquad \quad i=1,\ldots,N
\end{equation}
 In the deterministic part $\bm{S}\bm{f}$, $\bm{S}$ denotes the stoichiometric matrix of size $N \times R$ ($R$ being the number of chemical reactions) and $\bm{f}$ is the $R-$dimensional vector of reaction fluxes. The elements of $\bm{S}\bm{f}$ read
\begin{equation}
\label{eq:sf}
\begin{split}
&[\bm{S}\bm{f}(\bm{x}(t))]_i= \sum_{
j \neq l}\left(k^{-}_{l,ij}x_i(t)- k^{+}_{ij,l}x_i(t) x_j(t)\right) + 
\frac{1}{2}\sum_{
j\neq l}\left(k^{+}_{jl,i}x_j(t)x_l(t)-k^{-}_{i,jl}x_i(t)\right)\\
&{}+\sum_{j}\big(\lambda_{ji}x_j(t)-\lambda_{ij}x_i(t)\big)+
\sum_{l}\left(2k^{-}_{l,ii}x_l(t)-k^{+}_{ii,l}x_i(t) x_i(t)\right)+\sum_{j}\left(\frac{1}{2}k^{+}_{jj,i}x_j(t)x_j(t)-k^{-}_{i,jj}x_i(t)\right)
\end{split}
\end{equation}
and captures binary reactions, such as complex formation (with rate constants $k^{+}_{ij,l}$, $k^{+}_{jl,i}$, $k^{+}_{ii,l}$, $k^{+}_{jj,i}$) and dissociation ($k^{-}_{l,ij}$, $k^{-}_{i,jl}$, $k^{-}_{l,ii}$, $k^{-}_{i,jj}$), as well as the unary reactions ($\lambda_{ij}$ and $\lambda_{ji}$), e.g.\ phosphorylation or conformational change. The final term $\xi_i(t)$ in (\ref{eq:steq}) is an intrinsic noise term from the stochasticity of reaction events, which is Gaussian distributed with zero mean and covariance
\begin{equation}
\label{eq:fluct}
\langle \xi_i(t)\xi_j(t') \rangle=\bm{\Sigma}_{ij}(\bm{x}(t))\delta(t-t') \qquad  \bm{\Sigma}(\bm{x}(t))= \epsilon\,\bm{S}\,\text{diag}(\bm{f}(\bm{x}(t)))\,\bm{S}^{\T}
\end{equation}
The $\bm{x}$-dependent correlations make this noise multiplicative, to be interpreted here in the It\^o convention \cite{vankampenito}, while the temporal correlations remain white. The scale of the noise variance is set by the inverse reaction volume~\cite{gardiner} $\epsilon=1/V$, implying that fluctuations become less important for large reaction volumes $V$ but have a non-negligible effect on reactions occurring in cellular subcompartments that involve small to moderate copy numbers.

We assume that only a small subnetwork of species is well-characterized (e.g.\ because their biochemical interactions are known or their time evolution can be measured experimentally with high precision) while the rest of the network (the bulk) is not. In this paper we address the problem of deriving an accurate description of the dynamics of the chosen subnetwork that incorporates also the dynamical effects of extrinsic noise due to the embedding into the bulk environment.

An important aspect of our treatment is that the choice of subnetwork is not driven by the model reduction procedure itself, but in principle arbitrary. 
To make explicit expressions shorter, we simplify the subnetwork equations by making specific plausible assumptions about the split into subnetwork and bulk: given that bulk species are assumed uninteresting or poorly characterized, complexes involving at least one bulk species are themselves assigned to the bulk, while complexes made up of two subnetwork species are assigned to the subnetwork. 

\subsection{Subnetwork modeling with a linearized Langevin dynamics}
\label{sec:lin}
We first look at the case of linearized dynamics, which can be treated by direct elimination of bulk variables. The subnetwork reduced dynamics found in this case is \emph{exact} (with respect to the already linearized dynamics) and gives useful insights into the structure that emerges from integrating out the bulk.
Let us rewrite concentrations as $\bm{x}(t) = \bm{y} + \delta\bm{x}(t)$, where the $\delta\bm{x}(t) = \bm{x}(t) - \bm{y}$
are deviations from the steady state $\bm{y}$, which is found from the deterministic part of the dynamics by solving $\bm{S}\bm{f}(\bm{y})=0$. We rewrite \eqref{eq:steq} in terms of $\delta\bm{x}(t)$ and by retaining only terms linear in $\delta\bm{x}(t)$ one obtains the dynamics linearized about the steady state
\begin{equation}
\label{eq:lin}
\frac{d}{dt}\delta\bm{x}^{\T}(t) = \delta\bm{x}^{\T}(t)\bm{L} + \bm{\xi}_0^{\T}(t)
\end{equation}
where ${\T}$ denotes the tranpose of a column vector. The effective rate matrix $\bm{L}$ is defined as
\begin{equation}
\label{eq:effdrift}
 \bm{L}_{ji}=\frac{\partial}{\partial x_j}[\bm{S}\bm{f}(\bm{x})]_i\Bigr|_{\substack{\bm{x}=\bm{y}}}
\end{equation}
The white Gaussian noise $\bm{\xi}_0(t)$ appearing in the linearized dynamics is \emph{additive}; we will use the subscript $0$ to indicate this property. Its time correlations are
\be
\label{eq:effdiffusiony}
\langle \bm{\xi}_0(t)\bm{\xi}_0^{\T}(t')\rangle= \bm{\Sigma}_0 \delta(t-t')
\ee
with a covariance matrix $\bm{\Sigma}_0 =\bm{\Sigma}(\bm{x}(t)=\bm{y})$ that is constant in time and equal to the one obtained by performing a Linear Noise Approximation \cite{LNA} around the steady state $\bm{y}$. Our linearization is in exactly the same spirit as the latter approximation, i.e.\ we keep all terms in an expansion around the steady state that can be included while retaining purely Gaussian fluctuations.

As we are interested in modeling a subnetwork of species, we separate the $N$-dimensional vector of concentration deviations $\delta\bm{x}(t)$ into a subnetwork part, $\delta \bm{x}^{\rm s}(t)$, and a bulk part, $\delta \bm{x}^{\rm b}(t)$. It is convenient to choose a numbering where $i=1,\ldots,N^{\rm s}$ labels subnetwork species while the remaining $N^{\rm b}=N-N^{\rm s}$ species $i=N^{\rm s}+1,\ldots,N$ belong to the bulk, giving for the rate matrix \eqref{eq:effdrift} the block decomposition
\[\bm{L}=\begin{pmatrix}
\label{eq:blocks}
\bm{L}^{\rm ss}   & \bm{L}^{\rm bs} \\
\bm{L}^{\rm sb}   & \bm{L}^{\rm bb} \\
\end{pmatrix}\]
The equations \eqref{eq:lin} can then be split into a set of equations for the subnetwork and one for the bulk
\begin{eqnarray}
  \label{eq:linreddy}
   \frac{d}{dt}\delta\bm{x}^{\rm s^{\T}}&= \delta\bm{x}^{\rm s^{\T}}\bm{L}^{\rm ss} + 
   \delta\bm{x}^{\rm b^{\T}}\bm{L}^{\rm bs} + \bm{\xi}_0^{\rm s^{\T}}\label{eqdxs}\\
   \frac{d}{dt}\delta\bm{x}^{\rm b^{\T}}&= \delta\bm{x}^{\rm b^{\T}}\bm{L}^{\rm bb} + \delta\bm{x}^{\rm s^{\T}}\bm{L}^{\rm sb} 
   + \bm{\xi}_0^{\rm b^{\T}} 
  \label{eqdxb}
 \end{eqnarray}
where the noise correlations are given by the block form of the diffusion matrix
\[\bm{\Sigma}_0=\begin{pmatrix}
\label{eq:sigblock}
\bm{\Sigma}^{\rm ss}_0   & \bm{\Sigma}^{\rm sb}_0 \\
\bm{\Sigma}^{\rm bs}_0   & \bm{\Sigma}^{\rm bb}_0 \\
\end{pmatrix} \qquad \bm{\Sigma}^{\rm bs}_0 = (\bm{\Sigma}^{\rm sb}_0)^{\T}\]\\
The reduction to subnetwork dynamics can now be achieved by direct elimination of the bulk degrees of freedom (d.o.f.): we first solve explicitly for $\delta\bm{x}^{\rm b}$ 
\be
\label{eq:solbulk}
\delta\bm{x}^{\rm b^{\T}}(t)= \delta\bm{x}^{\rm b^{\T}}(0)e^{\bm{L}^{\rm bb}t}+ \int_0^t dt'\,\left(\delta\bm{x}^{\rm s^{\T}}(t')\bm{L}^{\rm sb} 
+ \bm{\xi}^{\rm b^{\T}}_0(t')\right)e^{\bm{L}^{\rm bb}(t-t')}
\ee
and then substitute into \eqref{eqdxs} to obtain the reduced subnetwork dynamics
\begin{equation}
\label{eq:linearreduced}
\frac{d}{dt}\delta\bm{x}^{\rm s^{\T}}(t)=\delta\bm{x}^{\rm s^{\T}}(t)\bm{L}^{\rm ss} + 
\int_0^t dt' \delta \bm{x}^{\rm s^{\T}}(t')\bm{M}^{\rm ss}(t-t') + \bm{\chi}_0^{\T}(t)
\end{equation}
with
\be
\label{eq:mem0}
\bm{M}^{\rm ss}(t-t')=\bm{L}^{\rm sb} e^{\bm{L}^{\rm bb}(t-t')}\bm{L}^{\rm bs}
\ee
\be
\label{eq:chi0}
\bm{\chi}_0^{\T}(t)= \bm{\xi}_0^{\rm s^{\T}}(t)+ \int_{0}^t dt'\,\bm{\xi}_0^{\rm b^{\T}}(t')e^{\bm{L}^{\rm bb}(t-t')}\bm{L}^{\rm bs}
+ \delta\bm{x}^{\rm b^{\T}}(0)e^{\bm{L}^{\rm bb}t}\bm{L}^{\rm bs}
\ee
In \eqref{eq:linearreduced}, reactions occurring within the subnetwork appear in their original form via the subnetwork matrix $\bm{L}^{\rm{ss}}$ while the embedding environment (the bulk) introduces non-Markovian terms. A memory function \eqref{eq:mem0} appears because bulk d.o.f are influenced by the past behavior of the subnetwork and feed this influence back to the subnetwork at a later time. On the other hand, a colored (i.e.\ time-correlated) noise \eqref{eq:chi0} accounts for the intrinsic subnetwork noise $\bm{\xi}^{\rm s}_0(t)$, for the effect of bulk intrinsic noise $\bm{\xi}^{\rm b}_0(t')$ at a previous time $t'$ on the subnetwork at time $t$, and for the (potentially unknown) initial deviations from steady state of the bulk, $\delta\bm{x}^{\rm b}(0)$. The two model reduction strategies introduced in \cite{katy, gaussianvar}, the first based on projection methods \cite{zwanzig,ritort} and the second on a path integral representation of the stochastic dynamics \cite{PathMethods}, both give the exact memory function (\ref{eq:mem0}). The projection approach averages over stochastic noise so only retrieves the last term of (\ref{eq:chi0}). Importantly, both methods allow us to go beyond the linearized dynamics while still accounting explicitly for memory and extrinsic noise: we will refer to the latter as \emph{random force} in the projection methods and \emph{colored noise} in the path integral approach 3GA.

\subsection{Random Force in Projection Methods}
\label{sec:proj}
The application of projection methods to networks of unary and binary protein reactions as in \eqref{eq:steq} was studied in \cite{katy} with a focus on memory terms. We next summarize those results and give the corresponding ones for the random force that interests us here.

The basic idea of projection methods is to define two operators projecting onto the subspace of subnetwork d.o.f.\ and onto the orthogonal one, respectively. For nonlinear projected equations, even though a subnetwork has been fixed, a choice remains as to which set of subnetwork observables to project onto. We argued in \cite{katy} that the most convenient set consists of linear and quadratic observables, i.e.\ concentration deviations and their products. With this set fixed, one can find a matrix representation of the projection operators involved that is accurate up to corrections of order $\delta x^3$ (cubic in deviations from steady state) and $\mathcal{O}(\epsilon)$ (linear in the noise variance). Therefore the projected nonlinear equations will be exact up to $O(\delta x^2)$ and in the limit $\epsilon \to 0$. We use the equations as derived for $\epsilon\to 0$ but will see later that at least for the mean behavior they can also give reasonable predictions when $\epsilon>0$. Referring to \cite{katy} for details and to Sec.\ I of SM for a summary, the resulting projected equations read
\begin{eqnarray}
\label{eq:nonlinprojeq}
\frac{d}{dt}\delta \bm{x}^{\rm{s}^{\T}}(t)=& \,\bdx^{\rm{s}^{\T}}(t)\bm{L}^{\rm{ss}} + (\delta\bm{x}^{\rm s}(t) \circ \delta\bm{x}^{\rm s}(t))^{\T}\bm{L}^{\rm ss,s} +  
\int_0^tdt'\bdx^{\rm{s}^{\T}}(t')\bm{M}^{\rm ss}(t-t') \notag \\ &+ \int_0^t dt'\big(\delta\bm{x}^{\rm s}(t')\circ\delta\bm{x}^{\rm s}(t')\big)^{\T} \bm{M}_{\rm P}^{\rm ss,s}(t-t')+ \bm{r}^{\T}(t)
\end{eqnarray}
and generalize the linear reduced dynamics \eqref{eq:linearreduced} to products of concentrations (and thus to nonlinear reactions such as formation of a complex from two other species). Here $\bm{M}_{\rm P}^{\rm ss,s}$ denotes the nonlinear memory from the projection approach, see equation (4) of SM. Formally, the projected equations describe \emph{averages} over the noise of the subnetwork concentrations conditional on specified initial values. We do not use separate nomenclature for these averages because they coincide with the concentration themselves in the noiseless limit considered here. We have introduced the notation $\bm{a}\circ\bm{b}$, not to be confused with the Hadamard (elementwise) product or the more similar Kronecker (outer) product, to denote the outer product $\bm{a}\bm{b}^{\T}$ rearranged into a single (column) vector. The entries of this vector are all possible componentwise products $a_i b_j$ so that e.g.\ $\delta\bm{x}^{\rm s}\circ\delta\bm{x}^{\rm b}$ has dimension $N^{\rm s}N^{\rm b}$. Where the two vectors belong to the same subset (both ``s'' or both ``b'') we mean the ordered products only: $\delta\bm{x}^{\rm s}\circ\delta\bm{x}^{\rm s}$ is the vector containing the $N^{\rm s}(N^{\rm s}+1)/2$ products $\delta x_i \delta x_j$ with $1\leq i\leq j\leq N^{\rm s}$. The matrix $\bm{L}^{\rm ss,s}$ is a block of a larger $\bm{L}$ matrix representing the full dynamics of all linear and quadratic observables. $\bm{L}^{\rm ss,s}$ contains the reaction rates in such a way as to give
\be
\big[\big(\delta\bm{x}^{\rm s}\circ
\delta\bm{x}^{\rm s}\big)^{\T}\bm{L}^{\rm ss,s}\big]_i=\sum_{j,l=1, j\neq l}^{N^{\rm s}}\bigg(\frac{1}{2}k^{+}_{jl,i}\delta x_j\delta x_l-k^{+}_{ij,l}\delta x_i\delta x_j\bigg)+\frac{1}{2}\sum_{j =1}^{N^{\rm s}}k^{+}_{jj,i}\delta x_j\delta x_j -\sum_{l =1}^{N^{\rm s}}k^{+}_{ii,l}\delta x_i \delta x_i
\ee
Because of the ordering of the $\circ$ product, this requires the elements of $\bm{L}^{\rm ss,s}$ to be defined as
$\bm{L}_{jl,i} =  k^{+}_{jl,i} - (\delta_{ij}+\delta_{il})\sum_{m=1}^{N^{\rm s}} k^{+}_{jl,m}$ for $j<l$ and 
$\bm{L}_{jj,i} = \frac{1}{2} k^+_{jj,i} - \delta_{ij}\sum_{m=1}^{N^{\rm s}} k^+_{ii,m}$. These coefficients encode the drift terms \eqref{eq:sf} for single subnetwork variables that depend on products of subnetwork variables.

The main structural features of \eqref{eq:nonlinprojeq} resemble the ones of the linear reduced dynamics. For instance, all subnetwork reactions are retained directly in the local-in-time terms, with rates appropriately arranged in $\bm{L}^{\rm ss}$ and $\bm{L}^{\rm ss,s}$, while the bulk evolution appears via the additional nonlocal-in-time terms, divided by the projection approach into a memory contribution and a random force. The memory describes how the present state of a subnetwork species is affected by all past states of the subnetwork: the effect of single subnetwork species is modulated by a linear memory function $\bm{M}^{\rm ss}(t-t')$ given by \eqref{eq:mem0} whereas the effects of subnetwork products enters via a nonlinear memory function (see Eq.~(4) in Sec.\ I of SM); these memory functions have been analyzed in detail in \cite{katy}. 

Here our focus will be on the structure of the random force $\bm{r}(t)$. Similarly to the memory, we can write $\bm{r}(t)=\bm{r}_0(t)+\bm{r}_1(t)$ as the sum of a contribution from the linear dynamics $\bm{r}_0(t)$ and a nonlinear correction $\bm{r}_1(t)$. 
The former is given by
\be
\label{eq:rf0}
\bm{r}_{0}^{\T}(t)=\delta \bm{x}^{\rm b^{\T}}(0)\ebb(t)\bm{L}^{\rm bs}
\ee
Here we have introduced the shorthand $\bm{E}(t)=e^{\bm{QLQ}t}$, where $\bm{Q}$ is the matrix form of the projector onto the bulk, so $\ebb(t)=e^{\bm{L}^{\rm bb}(t)}$ (see Sec.\ I in SM). $\bm{r}_{0}(t)$ is simply the random force featuring in the reduced linearized dynamics so it corresponds to \eqref{eq:chi0} in the limit $\epsilon \to 0$. This limit represents the fact that the projection approach describes observables \emph{conditionally averaged} over the intrinsic noise, leaving for the random force only the stochastic effects due to fluctuations in the bulk initial conditions. 

The nonlinear random force $\bm{r}_1(t)$ cannot be calculated in closed form but like the matrix representation of the relevant operators can be calculated in an expansion in $\delta x$. This gives
 \begin{equation} 
 \label{eq:rf1}
 \begin{split}  
 \bm{r}_1^{\T}(t)= &\,(\delta\bm{x}^{\rm s}(0)\circ \delta\bm{x}^{\rm b}(0))^{\T}
 \left[\bm{E}_{\rm sb,b}(t)\bm{L}^{\rm bs} + \bm{E}_{\rm sb,\rm sb}(t)\bm{L}^{\rm sb,s}\right]+
 (\delta\bm{x}^{\rm b}(0)\circ \delta\bm{x}^{\rm b}(0))^{\T}\left[\bm{E}_{\rm bb,b}(t)\bm{L}^{\rm bs}+
 \bm{E}_{\rm bb,\rm sb}(t)\bm{L}^{\rm sb,s}\right]
 \end{split}  
 \end{equation}
if we include terms up to $O(\delta\bm{x}^{\rm b}(0)^2)$; higher order terms may reasonably be expected to be negligible in many (but not all) applications to reaction networks. $\bm{L}^{\rm sb,s}$ is defined similarly to $\bm{L}^{\rm ss,s}$, i.e.\ contains the coefficients of products such as $\delta\bm{x}^{\rm s}\circ \delta\bm{x}^{\rm b}$, which are considered part of the bulk subspace, in the time evolution equations for linear subnetwork observables. Eqs.~(\ref{eq:rf0},\ref{eq:rf1}) will be the basis for all analysis of extrinsic noise properties below, as predicted by the projection approach. From them, one can also see that random force terms are present only for nonzero entries of the matrix blocks $\bm{L}^{\rm bs}$ and $\bm{L}^{\rm sb,s}$, which contain the rates of the subnetwork biochemical reactions with bulk species. In other words, the only subnetwork concentrations whose evolution is affected by a random force (and a memory term) as in \eqref{eq:nonlinprojeq} are the ones directly reacting with bulk species: we will call such subnetwork species  ``boundary species''.

\subsection{Colored Noise in 3GA}
\label{sec:3GA}
We next set out how extrinsic noise is obtained within our 3GA approach \cite{gaussianvar}. The underlying path integral method follows the classical Martin-Siggia-Rose-Jansen-DeDominicis \cite{martin,janssen,dedominicis,kamenev,PathMethods} formulation for stochastic processes. It allows one to express stochastic dynamics, in our case of the biochemical concentrations in \eqref{eq:steq}, as an integral over all possible time evolutions (``paths'' or ``trajectories'') of the system. Reducing the dynamical description to the subnetwork in this representation becomes equivalent to \emph{marginalizing} over the bulk trajectories. This task would be extremely difficult in general but becomes feasible, both conceptually and technically, when the joint probability distribution over subnetwork and bulk trajectories are Gaussian (as the marginal of a joint Gaussian is again Gaussian). The main idea of the 3GA \cite{gaussianvar} is thus to start from a Gaussian approximation for this probability distribution, which leads to a linearization of the equations \eqref{eq:steq} around the (in principle) time-dependent means. This Gaussian approximation is then used as a tractable reference distribution for obtaining corrections from nonlinear terms by perturbation theory, leading to a nonlinear subnetwork-reduced dynamics analogous to \eqref{eq:nonlinprojeq}. (The method was referred to as Gaussian Variational Approximation in \cite{gaussianvar} where we used a variational approach to derive the baseline approximation.) To simplify the final expressions, we choose to linearize about the deterministic steady states, as we did in Sec.~\ref{sec:lin}. 

Here we extend the analysis in \cite{gaussianvar} to include the intrinsic, multiplicative white noise $\bm{\xi}(t)$ with covariance \eqref{eq:fluct}; see Sec.\ II in SM for the full derivation. We thus retain \emph{all} stochastic terms, including the $\bm{x}$-dependencies of $\bm{\Sigma}(\bm{x})$, in the perturbative expansion. Substituting $\bm{x}(t)$ by $\bm{y} + \delta\bm{x}(t)$, we decompose \eqref{eq:fluct} as
\be
\label{eq:fluctdx}
\bm{\Sigma}(\bm{x}(t)) =
\epsilon\,\bm{S}\,\text{diag}(\bm{f}(\bm{x}))\,\bm{S}^{\T} = \epsilon\,\bm{S}\,[\text{diag}(\bm{f}_0) + \text{diag}(\bm{f}_1(\delta\bm{x})) + \text{diag}(\bm{f}_2(\delta\bm{x}^2))]\,
\bm{S}^{\T} =  \bm{\Sigma}_0+\bm{\Sigma}_1+\bm{\Sigma}_2
\ee
where $\bm{f}_0$, $\bm{f}_1(\delta\bm{x})$ and $\bm{f}_2(\delta\bm{x}^2)$ denote respectively the reaction fluxes for $\delta\bm{x}=0$, the $\delta\bm{x}$-dependent and the $\delta\bm{x}^2$-dependent pieces. The additive noise of the linearized dynamics corresponds to the first term, i.e.\ $\epsilon\,\bm{S}\,\text{diag}(\bm{f}_0)\,\bm{S}^{\T} = \bm{\Sigma}_0$; similarly we identify $\epsilon\,\bm{S}\,\text{diag}(\bm{f}_1(\delta\bm{x}))\,\bm{S}^{\T} \equiv \bm{\Sigma}_1$ and $\epsilon\,\bm{S}\,\text{diag}(\bm{f}_2(\delta\bm{x}^2))\,\bm{S}^{\T} \equiv \bm{\Sigma}_2$. When we develop the perturbative expansion, we focus on first order corrections to the linear dynamics and so discard the last term $\bm{\Sigma}_2$ (with $O(\delta\bm{x}^2)$ terms) that would contribute only at higher order. The resulting nonlinear 3GA-reduced dynamical equations are
\begin{eqnarray}
\label{eq:nonlin3GA}
\frac{d}{dt}\delta \bm{x}^{\rm{s}^{\T}}(t)=& \bdx^{\rm{s}^{\T}}(t)\bm{L}^{\rm{ss}} + 
(\delta\bm{x}^{\rm s}(t) \circ \delta\bm{x}^{\rm s}(t))^{\T}\bm{L}^{\rm ss,s} +  
\int_0^tdt'\bdx^{\rm{s}^{\T}}(t')\bm{M}^{\rm ss}(t-t') \notag \\ &+ \int_0^t dt'\int_0^{t'}dt''\big(\delta\bm{x}^{\rm s}(t')\circ\delta\bm{x}^{\rm s}(t'')\big)^{\T} \bm{M}_{\rm 3GA}^{\rm ss,s}(t,t',t'')+
\bm{\chi}^{\T}(t)
\end{eqnarray}
where the times in $\bm{M}_{\rm 3GA}^{\rm ss,s}(t,t',t'')$ are ordered as $t>t'>t''$. We have used a subscript to distinguish $\bm{M}_{\rm 3GA}^{\rm ss,s}(t,t',t'')$ explicitly from the nonlinear memory function of the projection approach $\bm{M}_{\rm P}^{\rm ss,s}(t-t')$, which is rather different as clear already from its dependence on only a single time difference.

Note that the reduced dynamics (\ref{eq:nonlin3GA}) is valid for any noise level $\epsilon$, i.e.\ all the intrinsic noise terms are still present in contrast to the projected equations (\ref{eq:nonlinprojeq}) that relate to conditional averages over the intrinsic fluctuations. The overall effective noise in the 3GA on the subnetwork is given by $\bm{\chi}(t)$ in \eqref{eq:nonlin3GA}. Similarly to the random force, we can split this as $\bm{\chi}(t) = \bm{\chi}_0(t)+\bm{\chi}_1(t)$, where $\bm{\chi}_0(t)$ is the noise of the linearized dynamics \eqref{eq:chi0} while $\bm{\chi}_1(t)$ stems from the nonlinear corrections, hence it contains $\delta\bm{x}^{\rm{s}}$-dependent terms as follows
\begin{eqnarray}    
 \label{eq:chi1}   
  \fl &&\bm{\chi}_1^{\T}(t)=\bm{\xi}_1^{\rm s^{\T}}(t) + \int_{0}^t dt''\,\bm{\xi}_1^{\rm b^{\T}}(t'')\bm{E}_{\rm bb}(t-t'')\bm{L}^{\rm bs}+
     \big(\delta\bm{x}^{\rm{s}}(t) \circ \bm{\lambda}(t) \big)^{\T}\bm{L}^{\rm sb,s}+\\
  \fl && \qquad \qquad \int_0^t dt'' \bigg[\big(\delta\bm{x}^{\rm{s}}(t'')\circ \bm{\lambda}(t'')\big)^{\T}\bm{L}^{\rm sb,b}
  + \big(\bm{\lambda}(t'')\circ\,\bm{\lambda}(t'')\big)^{\T}\bm{L}^{\rm bb,b}\bigg]\bm{E}_{\rm bb}(t-t'')
     \bm{L}^{\rm bs}+\notag \\ 
 \fl &&\int_0^t dt'\int_0^{t'}dt''\bigg[\big(\bm{\lambda}(t')\circ \bm{E}^{\T}_{\rm bb}(t'-t'')(\bm{L}^{\rm sb})^{\T}\delta\bm{x}^{\rm{s}}(t'')
 +\bm{E}^{\T}_{\rm bb}(t'-t'')(\bm{L}^{\rm sb})^{\T}\delta\bm{x}^{\rm{s}}(t'')\circ \bm{\lambda}(t')\big)^{\T}\bm{L}^{\rm bb,b}\bigg]\bm{E}_{\rm bb}(t-t')\bm{L}^{\rm bs}  \notag
\end{eqnarray}
There is some notation to explain here. Like in Sec.~\ref{sec:proj}, the matrices $\bm{L}^{\rm sb,s}$, $\bm{L}^{\rm sb,b}$, $\bm{L}^{\rm bb,b}$ contain the sets of rate constants involved, respectively, in the reaction of a bulk and subnetwork species to give a complex in the subnetwork ($\bm{L}^{\rm sb,s}$) or in the bulk ($\bm{L}^{\rm sb,b}$), or in two bulk species ($\bm{L}^{\rm bb,b}$) forming another bulk species. For brevity, and in line with our assumptions on subnetwork modeling set out in Sec.~\ref{sec:setup}, we have already set $\bm{L}^{\rm bb,s}=\bm{L}^{\rm ss,b}\equiv 0$, i.e.\ we assume there are no subnetwork complexes formed from two bulk species or conversely bulk complexes from two subnetwork species; the same convention was used tacitly already in \eqref{eq:rf1} above. The relevant matrix blocks can easily be re-instated if desired (see for example \cite{gaussianvar} and Sec.\ II in SM). A subscript $0$ on $\bm{\xi}$ labels a noise component whose covariance is independent of $\delta\bm{x}$, while the $\bm{\xi}_1$ carry the remaining $\delta\bm{x}$-dependent parts
\be
\label{eq:sigmasub}
\langle \bm{\xi}_0^{\cdot}(t) \bm{\xi}_0^{\cdot^{\T}}(t')\rangle = \bm{\Sigma}^{\cdot \cdot}_0\delta(t-t') 
\qquad  
\bm{\xi}_1^{\cdot}(t)= 
\frac{1}{2} (\bm{\Sigma}_1
\bm{\Sigma}_0^{-1}
\bm{\xi}_0
)^\cdot(t)
\qquad \cdot = \rm{s}, \rm{b}
\ee
The expression \eqref{eq:sigmasub} for $\bm{\xi}_1$ is obtained from an expansion of $\sqrt{\bm{\Sigma}}$ in $\delta\bm{x}$ up to first order; it thus preserves the $\sim \delta\bm{x}$ dependence of the noise covariance expected as a leading correction to a Gaussian solution (see Sec.\ II in SM). Note that $\bm{\xi}_1^{\cdot}(t)$, being linearly related to $\bm{\xi}_0^{\cdot}(t)$, does not add stochasticity \emph{per se} but simply corrects the correlation structure by accounting (multiplicatively) for the $\delta\bm{x}$-dependence. In \eqref{eq:chi1}, we have further defined the shorthand
\be
\label{eq:deflam}
\bm{\lambda}^{\T}(t)=\delta\bm{x}^{\rm b^{\T}}(0)\bm{E}_{\rm bb}(t)+\int_{0}^t dt'\bm{\xi}_0^{\rm{b}^{\T}}(t')\bm{E}_{\rm bb}(t-t')
\ee
whose covariance $\langle \bm{\lambda}(t) \bm{\lambda}^{\T}(t') \rangle = \bm{C}^{\rm bb|s}(t,t')$ has the meaning of a bulk-bulk covariance matrix conditioned on knowing the subnetwork trajectory
\be
 \label{eq:bulk_corr}
\bm{C}^{\rm bb|s}(t,t')= \bm{E}_{\rm bb}^{\T}(t)\bm{C}^{\rm bb}(0,0)\bm{E}_{\rm bb}(t') + \int_0^{\text{min}(t,t')} dt''\ebb^{\T}(t-t'')\bm{\Sigma}^{\rm bb}_0\ebb(t'-t'')
\ee
Like for the nonlinear memory, the nonlinear effective noise term in our marginalization approach does not coincide with its analogue in the projection method, the random force; this is true even in the limit $\epsilon \to 0$. In particular, \eqref{eq:chi1} contains a dependence on the whole trajectory $\delta\bm{x}^{\rm s}(t')$. Nonetheless, in the limit $\epsilon \to 0$ of vanishing intrinsic noise the combination of memory and effective noise still gives an approximation \emph{equivalent} to the combination of memory and random force in the projection method~\cite{gaussianvar}, up to $O(\delta\bm{x}^{\rm s \, 2})$. For $\epsilon>0$ the two approximations become different as the random force (\ref{eq:rf0},\ref{eq:rf1}) contains only the extrinsic noise from the uncertainty about the bulk initial state while the effective colored noise (\ref{eq:chi0},\ref{eq:chi1}) contains also the effects of intrinsic fluctuations, i.e.\ the subnetwork intrinsic noise ($\bm{\xi}_0^{\rm{s}}$ and $\bm{\xi}_1^{\rm{s}}$) and bulk intrinsic noise ($\bm{\xi}_0^{\rm{b}}$ and $\bm{\xi}_1^{\rm{b}}$). The 3GA reduced equations thus provide a more complete picture of fluctuations in the subnetwork dynamics: they retain the original subnetwork intrinsic randomness as well as extrinsic (bulk-related) components, i.e.\ the unknown bulk initial conditions and the propagation across the network of bulk intrinsic noise. In the next sections we test the accuracy of the subnetwork equations, depending on which of these contributions to the extrinsic noise is retained in the evaluation.

\section{Accuracy of subnetwork equations with initial uncertainty in the bulk}
\label{sec:accuracyrf}

In the following we assess how well our explicit expressions for extrinsic noise capture the two contributions identified above: (i) initial deviations of the bulk from the steady state and (ii) intrinsic stochastic fluctuations in the bulk dynamics, which are captured only by the 3GA approximation. In this section we address contribution (i). Without intrinsic noise ($\eps\to 0$), the projection and 3GA methods give results at a nonlinear level that agree up to $O(\delta\bm{x}^{\rm s \, 2})$ but not beyond, and only when the combination of memory and effective noise is taken into account \cite{gaussianvar}, thus a comparison between the two methods will also be of interest.

\subsection{Measures of accuracy}
\label{sec:accuracy_measures}
We sample bulk initial conditions from Gaussian distributions centered around zero with variance equal to $\langle\dx_i^2\rangle = \eps_0 y_i$ for $i=1,...,N^{\rm b}$, introducing a parameter $\eps_0$ separate from $\eps$ that sets the amplitude of initial fluctuations. The scaling with $y_i$ corresponds to the Gaussian (large copy number) limit of Poisson fluctuations in the number of molecules from each individual bulk species. The parameter $\eps_0$ is the inverse of an effective volume that converts steady state concentrations $y_i$ to copy numbers $y_i/\eps_0$.

We fix the subnetwork initial conditions, randomly sample bulk initial conditions and then solve both the full reaction equations and the reduced subnetwork equations. Repeating this process, we get at each point in time a distribution of predicted subnetwork concentrations, one according to each of the reduced equations (projection or 3GA) and one according to the full nonlinear reaction equations (here the quantity $\dx$ is simply found by the change of variable $\dx = x - y$ from the full nonlinear solution $x$). Our goal is to assess how close these two distributions are, i.e.\ to quantify how well the reduced equations capture fluctuations due to variation in bulk initial conditions. One possible measure of similarity between the true distribution $p(\dx)$ and the approximated one $q(\dx)$ is the Kullback-Leibler (KL) divergence \cite{kullback}. If we focus on the distributions of a single concentration variable and approximate these as Gaussians $p(\dx) = N(\mu,\sigma^2)$ and $q(\dx) = N(\hat{\mu},\hat{\sigma}^2)$, then 
\begin{equation}
 \label{eq:singlekl}
 \text{KL}(p||q) = \frac{1}{2} \ln\frac{\hat{\sigma}^2}{\sigma^2} + \frac{\sigma^2 -\hat{\sigma}^2 + (\hat{\mu}-\mu)^2}{2\hat{\sigma}^2}
\end{equation}
The Gaussian approximation is exact for the linearized dynamics and proved to be a good approximation in our numerical experiments for the solution of the full nonlinear CLE (see Fig.\ S1a). As a consequence the Gaussian KL estimates are consistent with a non-parametric estimate (see Fig.\ S1b) that makes no assumption on the form of the distributions. As Fig.\ S1b demonstrates, using Gaussian fits for the KL divergence has the advantage of giving substantially less noisy estimates; it is also computationally faster. We note that large KLs (above $O(1)$) indicate that the distributions being compared are quite different, with the integrand in $\text{KL}(p||q)=\int \dx\,p(\dx) \ln[p(\dx)/q(\dx)]$ picking up contributions in the tails of $q(\dx)$ where a Gaussian fit may be insufficient. Our KL estimates in this regime will therefore not be quantitatively accurate and we will only use them for qualitative comparisons, e.g.\ between KL values of different orders of magnitude. We note finally that alternative definitions of distance between steady state distributions in stochastic models of biochemical reaction networks  have been proposed, such as the Wasserstein distance \cite{ocal2019}. We have verified that, when evaluating the Wasserstein distance, the ordering in accuracy among the different approximations we consider and the trends in time are the same as for our Gaussian KL divergence estimates (see Fig.\ S1c). We therefore report results only for the latter.

Returning then to \eqref{eq:singlekl}, to evaluate the Gaussian KL we just need to collect the relevant single-species means and variances. Such a single-species KL is useful to measure how accurately means and variances are captured for individual species, e.g.\ those immediately affected by the interaction with the bulk, which we will call boundary species. At the numerical level, we will work with dimensionless concentrations, namely fractional deviations from the steady state values $\delta\tilde{x}=\delta x / y$ (the definitions of KL is invariant under this change of variables). 

An alternative measure that is sensitive to the accuracy of {\em correlations} between the fluctuations of different subnetwork concentrations at equal times is the normalized mean square error
\begin{equation}
\label{eq:deltaC}
 \Delta C = \frac{\sum_{i,j=1}^{N^{\rm s}}\left(C_{ij}-\hat{C}_{ij}\right)^2/(y_iy_j)^2}{\sum_{i,j=1}^{N^{\rm s}}C_{ij}^2/(y_iy_j)^2}
\end{equation}
where $C_{ij}$ are the elements of the subnetwork correlation matrix $\bm{C}^{\rm ss}(t,t)$ 
from the full reaction equations and $\hat{C}_{ij}$ those of the correlation matrix estimated from a reduction method (3GA or projection). The error $\Delta C$ is constructed in such a way that $\Delta C \sim 0$ when $\hat{C}_{ij}$ approximates well $C_{ij}$ and $\Delta C \sim 1$ when $\hat{C}_{ij}$ significantly underestimates $C_{ij}$. The factors $(y_iy_j)^2$ indicate that we are again looking at fractional deviations from steady states.

\begin{figure}[!ht]
  \centering
  \includegraphics{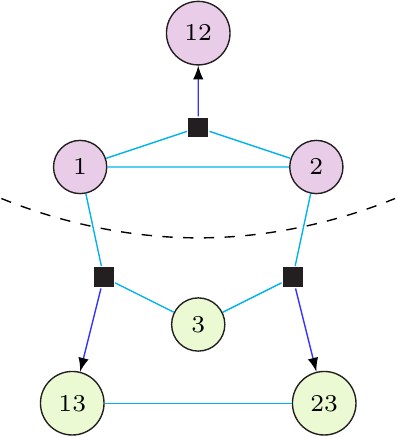}
  \caption[]{\textbf{Simple protein network}. Sketch of a simple protein interaction network composed of 6 species (details specified in Appendix \ref{sec:toy_app}). Species 1, 2, 12 belong to the subnetwork (light red) and species 3, 13, 23 form the bulk (light green). Squares and arrows indicate that protein 1 reacts with protein 2 to form complex 12 and in reverse 12 dissociates into 1 and 2, with analogous reactions between 1 and 3 to form 13, and 2 and 3 to form 23. Species 1 and 2  lie at the boundary of the subnetwork, i.e.\ they directly react with bulk species, while species 12 is internal to the subnetwork.}
  \label{fig:3p3c3rf}
\end{figure}

In order to illustrate the general features of our approach we will often use the small reaction network shown in Fig.~\ref{fig:3p3c3rf}
 
\subsection{Quantitative tests on a simple network model}

\begin{figure}[!ht]
 \includegraphics[width=17cm]{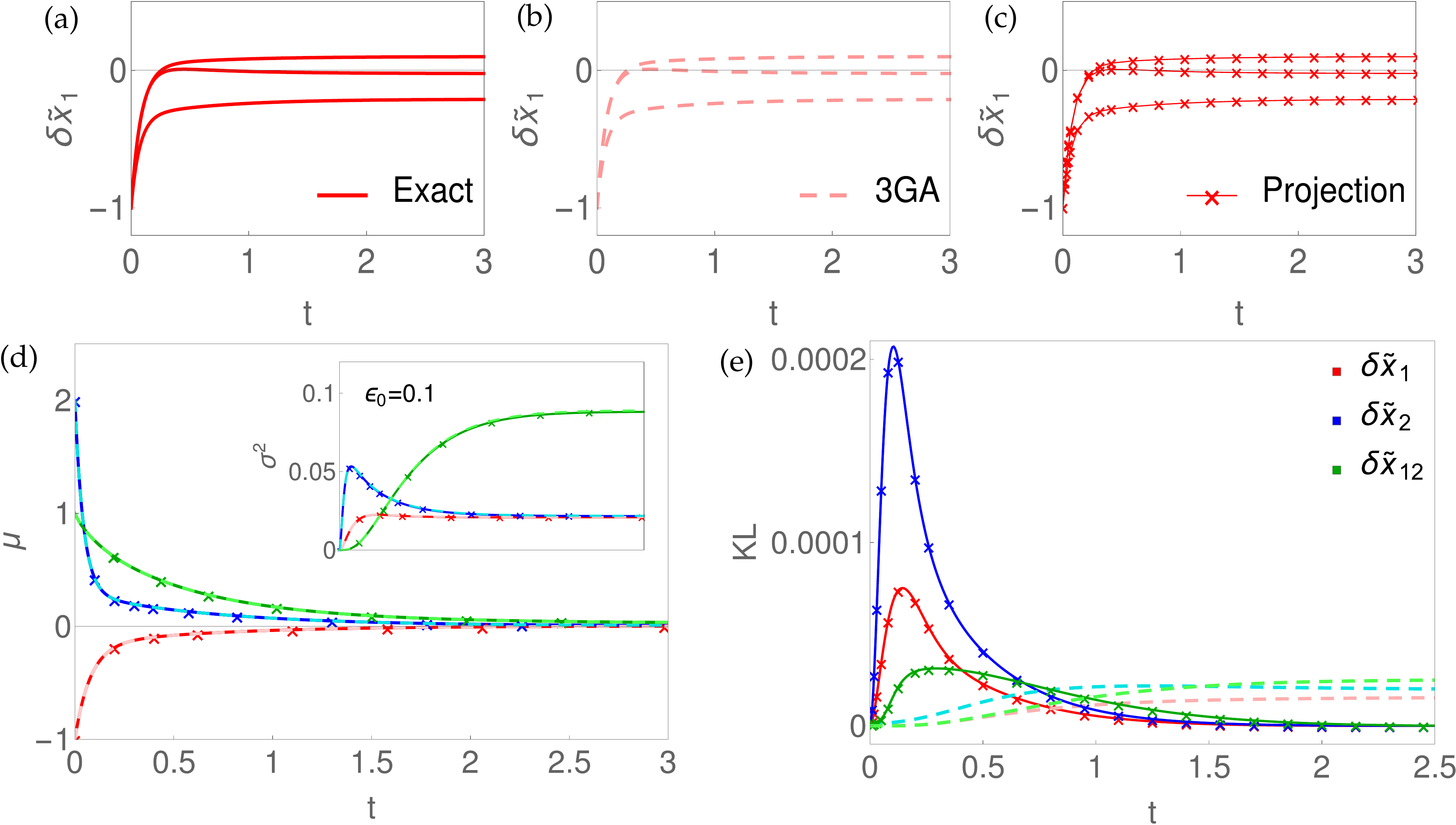}
 \caption[]
 {\textbf{Subnetwork dynamics with extrinsic noise in the simple network model}. Time courses of the subnetwork boundary species $\delta \tilde{x}_1$ of the simple network model of Fig.~\ref{fig:3p3c3rf}: (a) exact, bold lines, (b) 3GA, dashed lines, (c) projection, bold lines with crosses. In each plot, single trajectories correspond to 3 different realizations (the same in a, b, c) of random bulk initial conditions at $\epsilon_0 = 0.1$: their effect is visible both on the transient and the long time behavior. For each of them, the exact trajectory in (a) is accurately reproduced by 3GA (corresponding curve in b) and by projection methods (corresponding curve in c). (d) Exact subnetwork means over 1000 random bulk initial conditions compared to the approximated means by 3GA/projection methods (same line style as in a-c). The inset shows the comparison of variances. (e) Subnetwork KLs for full 3GA and full projected equations. The difference in the long time behavior, induced by random bulk initial conditions, is less well captured by the 3GA, where the errors on means and variance both reach nonzero values at long times, while the initial transient is less well captured by projection methods.}
 \label{fig:3p3c3_full}
\end{figure}

\begin{figure}[!ht]
 \includegraphics[width=15cm]{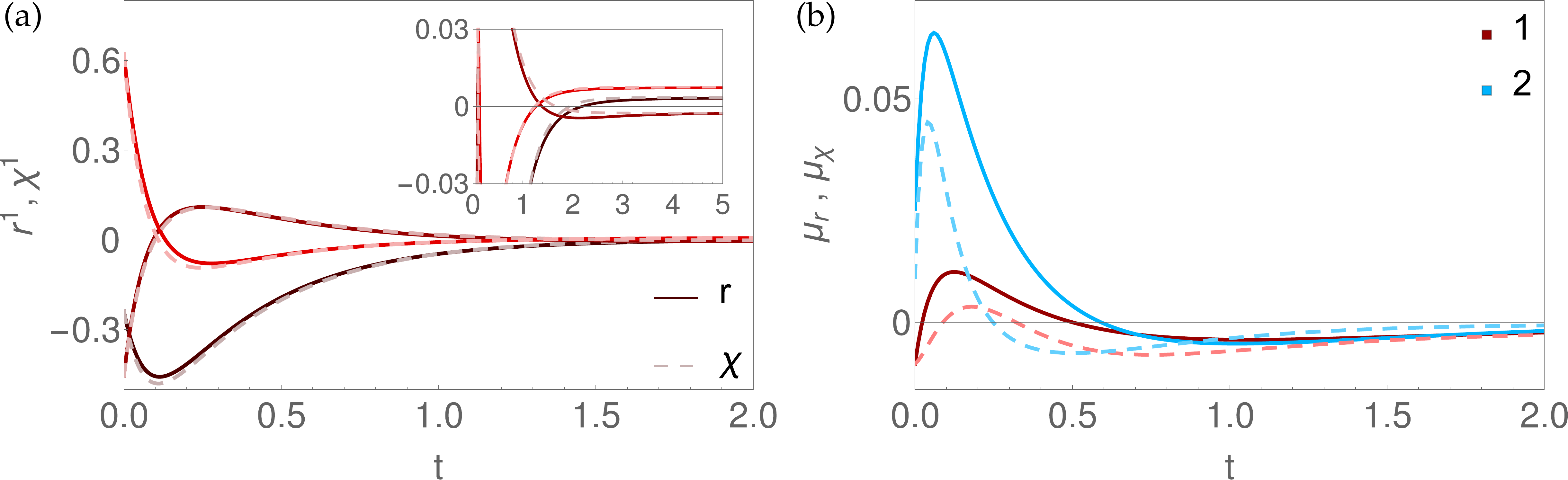}
 \caption[]
 {\textbf{Extrinsic noise from random bulk initial conditions at $\epsilon_0 = 0.1$ in the simple network model}. (a) Random force on $\delta \tilde{x}_1$ (solid lines) and 3GA colored noise (dashed lines) for a sample of three different initial conditions. The zoom in the inset shows that there is a small but nonzero asymptote on each curve. (b) Mean random force (over 1000 bulk initial conditions) on $\delta \tilde{x}_1$ and $\delta \tilde{x}_2$ (solid lines) and mean 3GA colored noise (dashed lines).}
  \label{fig:3p3c3rfplot}
\end{figure}

We sampled 1000 bulk initial conditions at $\epsilon_0 = 0.1$ and then calculated the single-variable KL divergence \eqref{eq:singlekl} between the distributions predicted by the full reaction equations and the reduced (projected or 3GA) equations. We focus here on the two subnetwork boundary species (see Fig.~\ref{fig:3p3c3rf}), which will have extrinsic noise acting on them due to the interaction with the bulk environment. We work with fixed subnetwork initial conditions $\delta\tilde{x}_1(0)=-1$, $\delta\tilde{x}_2(0)= 2$, $\delta\tilde{x}_{12}(0)= 1$, i.e.\ a subnetwork state where the initial deviations from steady state are substantial, ranging from $-100\%$ to $+200\%$ across the three subnetwork species. Solving the subnetwork equations with the full time-dependence of $\bm{r}(t)$ from (\ref{eq:rf0}, \ref{eq:rf1}) and $\bm{\chi}(t)$ from (\ref{eq:chi0}, \ref{eq:chi1}), we see that both model reduction approaches reproduce
the exact dynamics extremely well, both at the level of single trajectories (Fig.~\ref{fig:3p3c3_full}a-c) and of means and variances (Fig.~\ref{fig:3p3c3_full}d). This is confirmed by the low values of the KL divergence (Fig.~\ref{fig:3p3c3_full}e). The small residual KL values show that the two reduction methods perform better in different time regimes (3GA at short times, projection method at long times). Note that even though the projection and 3GA reduction methods are known theoretically to be equivalent only to $O(\delta \bm{x}^{2})$, they agree rather well with each other and with the full dynamics even for the large initial subnetwork deviations from steady state explored in Fig.~\ref{fig:3p3c3_full}, suggesting that higher order corrections in $\delta\bm{x}$, while present in principle, are quantitatively small. Similarly, although the nonlinear random force $\bm{r}(t)$ and nonlinear colored noise $\bm{\chi}(t)$ are not defined in the same way, they both capture the extrinsic noise from initial bulk fluctuations and we observe that structurally they look very similar (Fig.~\ref{fig:3p3c3rfplot}). We note in this context that the means of $\bm{r}(t)$ and $\bm{\chi}(t)$, as evaluated by averaging over many realizations of bulk initial conditions, are different from zero. This is due to terms $\delta\bm{x}^{\rm b}(0)\circ \delta\bm{x}^{\rm b}(0)$ that are present both in $\bm{r}_1(t)$ \eqref{eq:rf1} and $\bm{\chi}_1(t)$ (via $\bm{\lambda}\circ \bm{\lambda}$ in \eqref{eq:chi1}): when averaged, these terms scale as $\bm{C}^{\rm bb}(0,0)\sim \eps_0$ rather than vanishing exactly. The nonzero mean, arising from nonlinear propagation of stochastic effects, can be in principle subtracted and added back as a $t$-dependent piece in the subnetwork equations. However, this would be rather involved without providing any conceptual advantage.

\subsection{``Impulse plus persistent'' extrinsic noise approximation}
\label{sec:approx_rf_main}
In practice it can be difficult to explicitly calculate $\bm{r}(t)$ or $\bm{\chi}(t)$ as functions of time and we therefore explore ways of approximating these extrinsic noise terms. Intuition can be gained from Fig.~\ref{fig:3p3c3rfplot}a, which shows the random force $r^1$ and the colored noise $\chi^1$ acting on $\delta\tilde{x}_{1}$ for three different sets of bulk initial conditions sampled at $\epsilon_0=0.1$. One sees that the extrinsic noise can have a constant piece for large times, as well as contribution that is significant only for short times. The long-time piece arises from the fact that different bulk initial conditions can produce different values of quantities (specifically weighted sums of certain molecule numbers) that are conserved in our example network.

A more systematic analysis, presented in detail in Appendix \ref{sec:approx_rf}, shows that in the projected subnetwork equations one obtains the correct random force-dependent steady state from an approximate form of the random force as long as the approximation preserves both its long-time limit $\rinf$ and the total contribution from the short-time transient $\overline{\Delta\bm{r}} = \int_0^\infty dt'\, [\bm{r}(t')-\bm{r}_\infty]$. This suggests an approximation where the transient part of the random force is replaced by a single impulse at $t=0$, specifically
\begin{equation}
\label{eq:approx_rf}
  \bm{r}(t) \simeq \rinf + \delta(t) \overline{\Delta\bm{r}}
\end{equation}
This can then be approximated further by only taking into account one of the two contributions above. By direct analogy, the extrinsic noise in the 3GA approach can be approximated as
\begin{equation}
\label{eq:approx_chi}
  \bm{\chi}(t) \simeq \cinf + \delta(t) \overline{\Delta\bm{\chi}}
\end{equation}
where $\cinf=\lim_{t\to\infty} \bm{\chi}(t)$ and $\overline{\Delta\bm{\chi}} = \int_0^\infty dt'\, [\bm{\chi}(t')-\bm{\chi}_\infty]$.

Our numerical tests of these approximations show that retaining only the persistent piece of $\bm{r}(t)$ and $\bm{\chi}(t)$ is not a good approximation, resulting in very large KL divergences (data not shown). This makes sense given that in Fig.~\ref{fig:3p3c3rfplot}a one sees that the constant pieces of $\bm{r}(t)$ and $\bm{\chi}(t)$ can be small compared to the transients.

\begin{figure}[!ht]
\includegraphics[width=8.5cm]{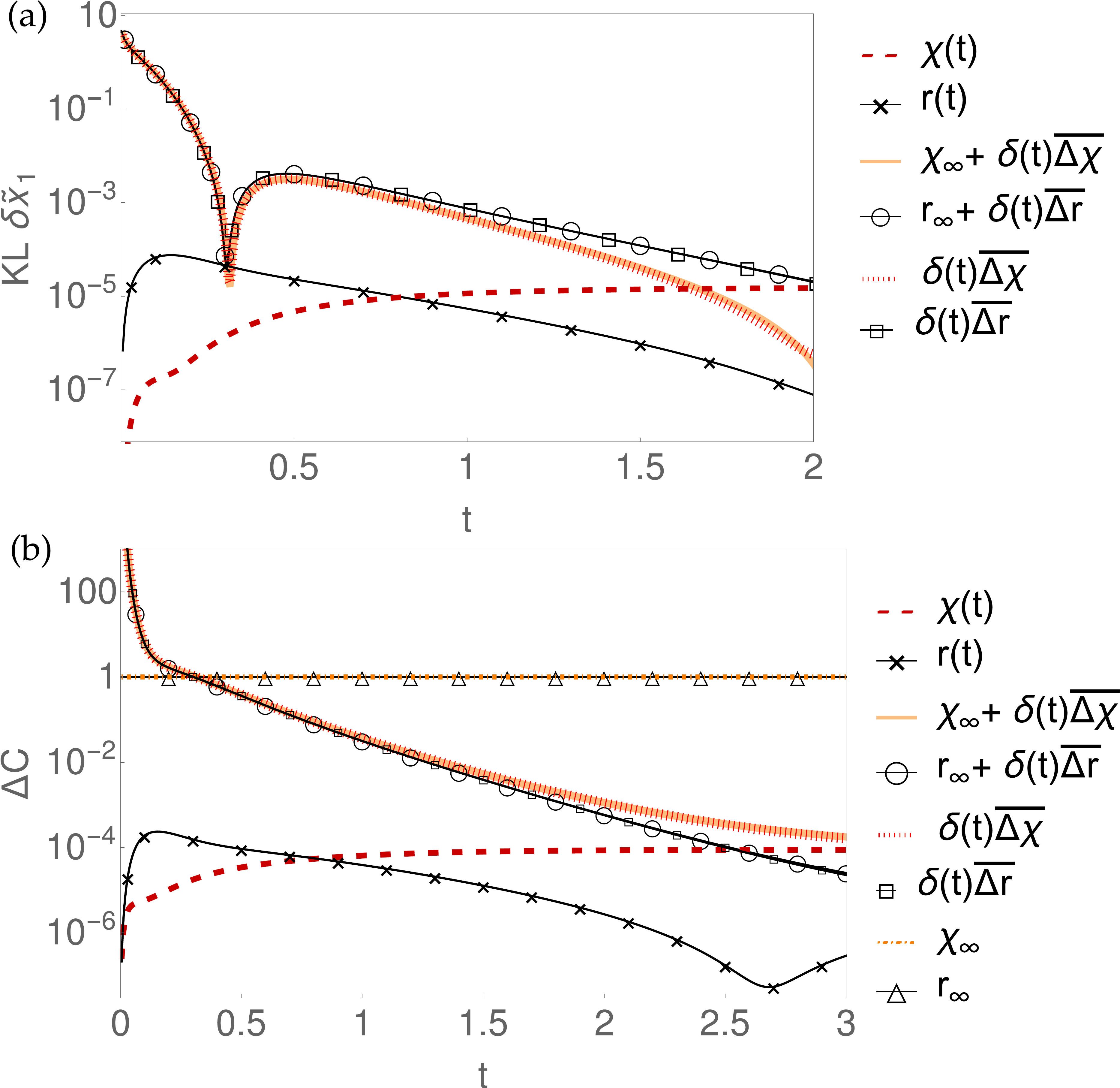}
\caption[]
{\textbf{Accuracy of extrinsic noise approximations at $\epsilon_0 = 0.1$}. (a) Comparison of KLs for $\delta \tilde{x}_1$ for projected equations (solid lines with symbols) and 3GA equations (dashed lines). Lowest two curves: full extrinsic noise; the other four curves show the approximation of the noise by an initial impulse and a persistent piece, and by only an initial impulse as shown in the legend; the KL divergences remain small except in an initial transient. (b) Comparison of errors on $C$, $\Delta C$, for the 3GA and projected equations and for the same approximations of the random force $\bm{r}(t)$ and the colored noise $\bm{\chi}(t)$, respectively. Results are shown in addition for the approximation containing only the persistent piece: here the value $\Delta C \sim 1$ indicates that the true covariances are strongly underestimated. 
}
\label{fig:3p3c3kl}
\end{figure}

Once the transient piece of the extrinsic noise is included approximately as in \eqref{eq:approx_rf} and \eqref{eq:approx_chi}, respectively, much more satisfactory predictions for the subnetwork dynamics are found. The KL divergences (Fig.~\ref{fig:3p3c3kl}a) are small and close to those found previously with the full form of the random force or effective noise, except at very short times -- here the fact that we are not fully capturing the time-dependence of the extrinsic noise transients evidently matters. The deviations $\Delta C$ in the predicted covariances show the same trend, being substantial only for short times and small (but still larger than when using the full form of $\bm{r}(t)$ or $\bm{\chi}(t)$) at long times. In the example shown one can in fact drop the persistent part of the extrinsic noise while still getting very accurate subnetwork dynamics predictions (Fig.~\ref{fig:3p3c3kl}), though we have seen in other tests that this is not generically true (see e.g.\ Sec.\ IV B in SM). Note finally that the properties of the ``impulse plus persistent'' approximation for the extrinsic noise that we have seen so far are also observed in the simpler case of dynamics linearized around the fixed point (data not shown). The only difference is that the subnetwork equations with the full form of the extrinsic noise are exact here and so give zero KL divergences up to sampling errors.

\subsection{Variation with strength $\epsilon_0$ of initial bulk fluctuations}
\label{sec:var_eps0}
So far we have analyzed how accurately the two model reduction methods can capture initial bulk fluctuations of fixed strength $\eps_0$. Now we vary $\epsilon_0$ to assess the impact of the size of these fluctuations. Here it is useful to look at summary measures of accuracy over entire trajectories, so we consider the absolute error in the mean concentrations, averaged over the subnetwork species and over the chosen total time window $T$
\be
\label{eq:deltamuerr}
\overline{\Delta\mu} = \frac{1}{T N^{\rm s}}\int_{0}^{\T} dt \sum_{i=1}^{N^{\rm s}}\frac{1}{y_i}\left|\mu _{i}(t)-\hat{\mu}_{i}(t)\right|
\ee
The normalization by $y_i$ ensures that this quantity is dimensionless, again measuring errors in fractional deviations from steady state values. Similarly we consider the errors in the variances
\be
\label{eq:deltaserr}
\overline{\Delta \sigma^2} = \frac{1}{TN^{\rm s}}\int_{0}^{\T} dt \sum_{i=1}^{N^{\rm s}} \frac{1}{y_i^2}\left|\sigma^2 _{i}(t)-\hat{\sigma}^2_{i}(t)\right|
\ee
and in the full covariance matrix, which includes correlations between species
\be
\label{eq:deltacerr}
\overline{\Delta C} = \frac{1}{T(N^{\rm s})^2}\int_{0}^{\T} dt \sum_{i,j=1}^{N^{\rm s}} \frac{1}{y_i y_j} \left|C_{ij}(t)-\hat{C}_{ij}(t)\right|
\ee
Here, differently from the definition \eqref{eq:deltaC}, we do not take square of the errors $|C_{ij}(t)-\hat{C}_{ij}(t)|$ and we do not normalize them by the exact correlations, in order to avoid having a potentially $\epsilon_0$-dependent normalization constant. To get an understanding of the expected scalings of $\overline{\Delta \mu}$, $\overline{\Delta\sigma^2}$ and $\overline{\Delta C}$ with $\eps_0$ we assume the subnetwork initial conditions to be at steady state, $\delta \bm{x}^{\rm s}(0)=0$. Other initial conditions would generate small errors, of order $\delta\bm{x}^{\rm s}(0)^3$ in $\overline{\Delta\mu}$, even for $\epsilon_0$: these arise from the truncation to quadratic terms in our projection (see \cite{katy} and Sec.\ I in SM). Subnetwork steady state initial conditions therefore allow us to isolate the effects of $\epsilon_0$. (For general initial conditions one broadly expects that the truncation error or the one from $\epsilon_0$ will dominate, depending on which is larger.)

The desired error scaling with $\epsilon_0$ can now be inferred from the construction of the projection and 3GA approximations, which both involve neglecting terms beyond a certain order in the initial bulk fluctuations. As we are still considering $\eps\to 0$ (no intrinsic noise), variances and covariances of subnetwork quantities are only driven by initial bulk fluctuations and must scale with $\eps_0$ to leading order. As explained in Appendix \ref{sec:scalings}, the next terms of $O(\epsilon_0^2)$ are already not treated fully systematically by projection or 3GA, so the errors in means, variances and covariances, $\overline{\Delta\mu}$, $\overline{\Delta\sigma^2}$ and $\overline{\Delta C}$, are expected to scale as $\eps_0^2$. 

\begin{figure}[!ht]
 \includegraphics[width=17cm]{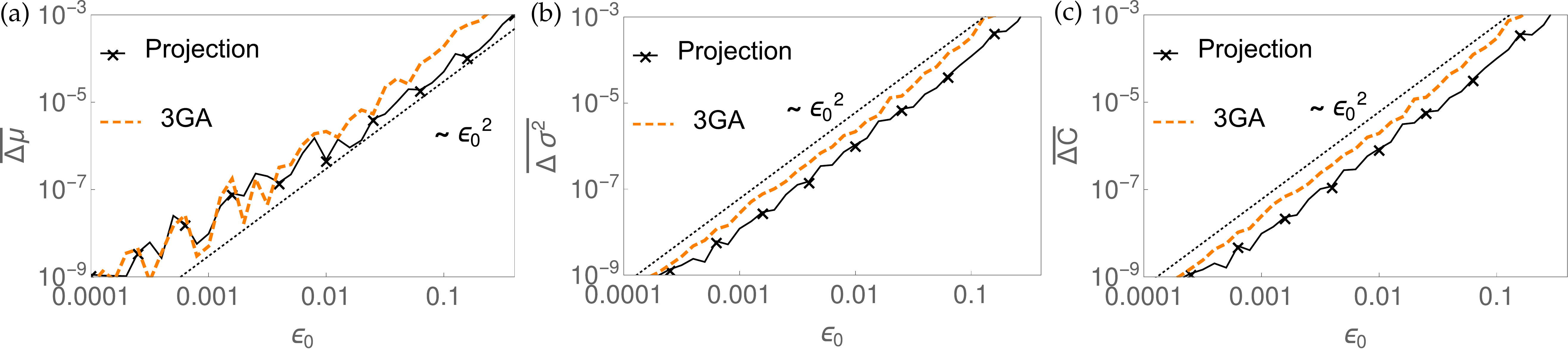}
 \caption[]
 {\textbf{Variation with $\epsilon_0$ of integrated error} on subnetwork means (a), variances (b) and correlations (c) over a sample of 1200 random bulk initial conditions for projected and 3GA equations. Initial conditions in the subnetwork are set to steady state, $\delta \bm{x}^{\rm s}(0)=0$ (see text for discussion). 
 }
 \label{fig:var_eps0}
\end{figure}

These scalings are confirmed numerically in Fig.~\ref{fig:var_eps0} over a wide range of $\eps_0$, with deviations appearing only for very small $\eps_0$ where the errors become too small to estimate reliably. Note that the errors on both means and (co-)variances remain low (below $10^{-3}$) even at the largest $\eps_0 = 0.4$ we used, which corresponds to very substantial initial bulk fluctuations (because steady state copy numbers $y_i/\eps_0$ in our model are then as small as $O(10)$).

\section{Accuracy of subnetwork equations with intrinsic fluctuations}
\label{sec:approx_chi}
\begin{figure}[!ht]
 \includegraphics[width=17cm]{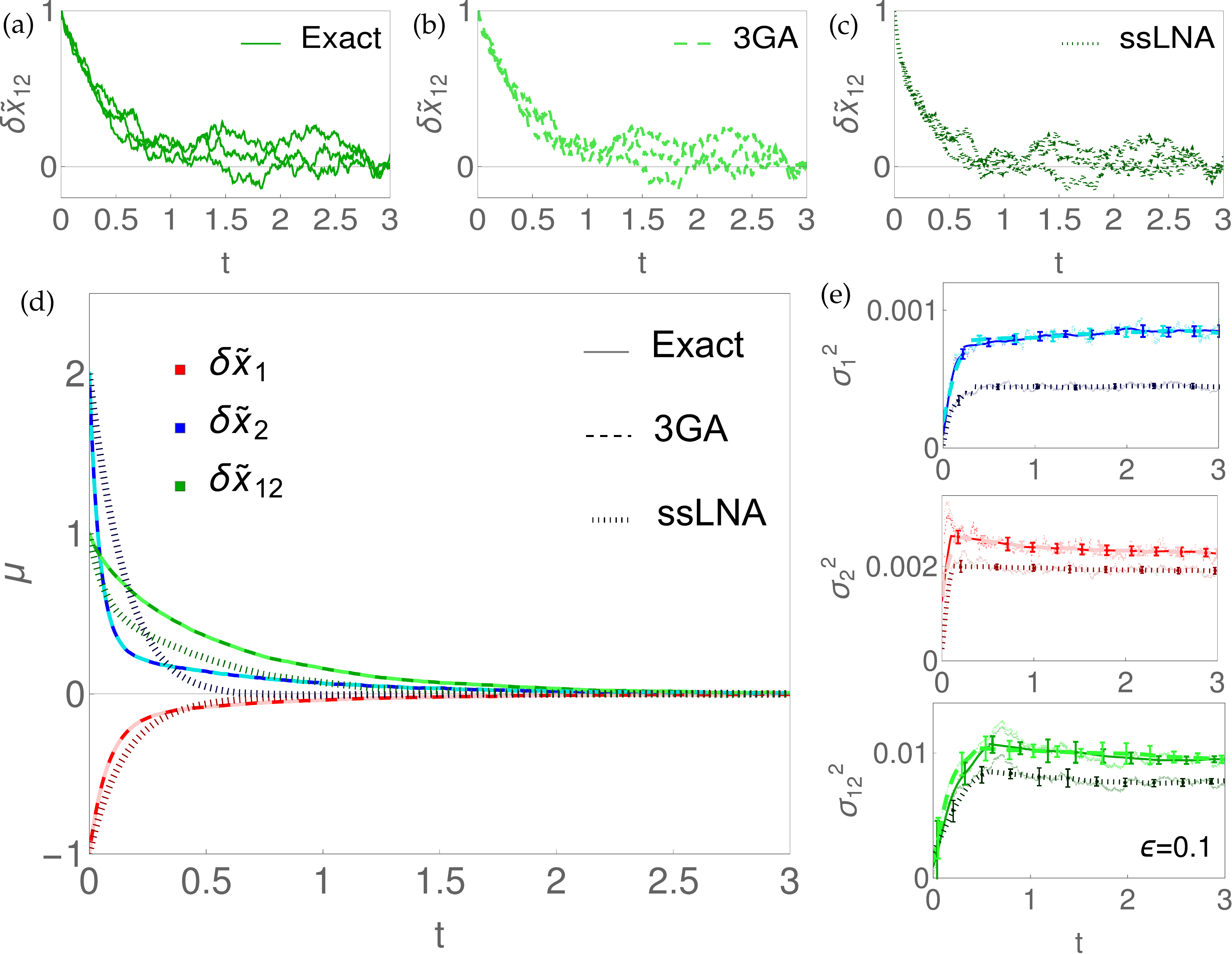}
 \caption[]{\textbf{Subnetwork dynamics with intrinsic noise in the simple network model.} Examples of stochastic trajectories for subnetwork species $\delta \tilde{x}_{12}$: (a) exact dynamics (bold), (b) 3GA model reduction (dashed), (c) ssLNA (dotted). The different trajectories in each plot correspond to 3 different realizations of white noise with $\epsilon = 0.1$ and the same 3 noise realizations were used in each case to emphasize the differences arising from the distinct nature of the approximations. The bulk is initially taken at steady state, i.e.\ $\epsilon_0 = 0$. Each exact trajectory in (a) is accurately reproduced by 3GA (corresponding curve in b) but less well by ssLNA (corresponding curve in c). Next we compare exact means (d) and variances (e), calculated from a sample of 1000 noise realizations, to the ones given by 3GA and ssLNA (same line styles as in a-c). The numerical results for the variances (points in e) have been smoothed in time for better visibility, with a representative set of error bars shown that have been estimated from the smoothing bins.}
 \label{fig:3p3c3eps}
\end{figure}

So far we have restricted ourselves to the limit of noiseless dynamics, i.e.\ $\epsilon \to 0$. As a next step, we consider the intrinsic (white Gaussian) noise $\bm{\xi}(t)$ from \eqref{eq:steq} for all subnetwork and bulk species while setting $\epsilon_0 = 0$. Examples of single trajectories for different realizations of the noise are shown in Fig.~\ref{fig:3p3c3eps}a-c for the simple protein interaction network of Fig.~\ref{fig:3p3c3rf}.

As projection methods provide only trajectories conditionally averaged over the noise, we need to use the 3GA to analyze the nonlinear dynamics in the presence of such fluctuations\footnote{The projection method can in principle be applied for $\epsilon > 0$ to derive reduced dynamical equations also for the {\em fluctuating} subnetwork concentrations, but the resulting expressions (e.g.\ for the memory functions and random force) cannot be evaluated in closed form without further approximation \cite{herrera-delgado2020,chorin2000}}. We will compare its performance to the Steady State Linear Noise Approximation (ssLNA) \cite{thomas,thomasproj}, which is based on the assumption that the bulk dynamics are fast enough to reach equilibrium with respect to the instantaneous subnetwork state, and to intermediate approximations that retain only some features of the effective noise $\bm{\chi}(t)$ on the subnetwork dynamics, as summarized schematically in Tab.\ \ref{table:tableapp}. Building such intermediate approximations allows us to assess the impact of the different extrinsic noise contributions that arise in the process of model reduction, such as nonlinearity and time correlations.

The first intermediate approximation we explore is ``linear noise'', where in the 3GA-reduced dynamics \eqref{eq:nonlin3GA} we keep the nonlinear memory (i.e.\ the integrals containing $\bm{M}^{\rm ss}(t,t')$ and $\bm{M}_{\rm 3GA}^{\rm ss,s}(t,t',t'')$) but we replace the nonlinear (hence multiplicative) and colored effective noise $\bm{\chi}(t)=\bm{\chi}_0(t)+\bm{\chi}_1(t)$ with its \emph{linear} (hence additive), but still time-correlated, part $\bm{\chi}_0(t)$ 
from \eqref{eq:chi0}. We refer to this approximation as 3GA-$\bm{\chi}_0$.

We further study a ``white linear noise'' version of the 3GA-$\bm{\chi}_0$ where, in analogy to the ssLNA, we assume that the bulk dynamics is fast w.r.t.\ the subnetwork dynamics so that we can ignore the time correlations it induces in the effective noise. We thus approximate $\bm{\chi}_0(t)$ by a white (uncorrelated in time) noise $\bm{\chi}_0^{w}(t)$ with a time-dependent covariance matrix. The later is determined such that the covariance of the {\em cumulative} noise $\bm{\chi}_0^{w}(t)$ over a short time interval $t\ldots t+dt$ is the same as for the linear effective noise $\bm{\chi}_0(t)$; here the time interval $dt$ should be short compared to the dynamics of the subnetwork but larger than the timescale on which bulk fluctuations are correlated. Under these assumptions one finds explicitly
\be
\label{eq:chi0int}
\bm{\chi}_0^{w}{}^{\T}(t)=\bm{\xi}_0^{\rm s^{\T}}(t) -\bm{\xi}_0^{\rm b^{\T}}(t)\bm{L}^{\rm bb \,-1}\bm{L}^{\rm bs}
\ee
The resulting approximation is Markovian at the level of the stochastic part but stays non-Markovian in its deterministic part as it retains the full nonlinear memory of the 3GA; we will refer to it as 3GA-$\bm{\chi}_0^{w}$. As an aside we note that $\bm{\chi}_0^{w}(t)$ is identical to the ssLNA effective noise evaluated at the subnetwork steady state rather than the transient subnetwork concentrations used in the ssLNA, see \cite{thomas,thomasproj}.

\begin{table}
\begin{center}
\begin{tabular}{|c|c|c|c|c|} 
\hline
Approximation & Effective Noise: from $\bm{\Sigma}$ & from $\bm{C}^{\rm{bb}}(0,0)$ & Memory & Terms \\ [0.5ex]
\hline\hline
Projection & \xmark & \cmark & \cmark & Eq.\,\eqref{eq:nonlinprojeq} with \eqref{eq:rf0} and \eqref{eq:rf1}\\
3GA & Nonlinear, Multiplicative, Colored & \cmark & \cmark & Eq.\,\eqref{eq:nonlin3GA} with \eqref{eq:chi0} and \eqref{eq:chi1}\\
3GA\small{, additive n.s.} & Nonlinear, Additive, Colored & \cmark & \cmark & Eq.\,\eqref{eq:nonlin3GA} with \eqref{eq:chi0} and \eqref{eq:chi1}; $\bm{\xi}_1^{\cdot}$ dropped from \eqref{eq:chi1}\\
3GA-$\bm{\chi}_0$\small{, multiplicative n.s.} & Linear, Multiplicative, Colored & \cmark & \cmark & Eq.\,\eqref{eq:nonlin3GA} with \eqref{eq:chi0};  $\bm{\xi}_0^{\cdot} \rightarrow \bm{\xi}_0^{\cdot} + \bm{\xi}_1^{\cdot}$ ($\bm{\xi}_1^{\cdot}$ from eq. \ref{eq:sigmasub})\\
3GA-$\bm{\chi}_0$ & Linear, Additive, Colored & \cmark & \cmark & Eq.\,\eqref{eq:nonlin3GA} with \eqref{eq:chi0}\\
3GA-$\bm{\chi}_0^{w}$ & Linear, Additive, White & \cmark & \cmark & Eq.\,\eqref{eq:nonlin3GA} with \eqref{eq:chi0} approximated as white\\
ssLNA & Linear, Multiplicative, White & \xmark & \xmark & see Ref.\ \cite{thomas,thomasproj} \\
\hline
\end{tabular}
\end{center}
\caption{Comparison of features of projection methods, 3GA and the different approximate model reduction approaches to dynamics with intrinsic noise discussed in Sec.\ \ref{sec:approx_chi}, listed in descending order of sophistication. ``n.s.'' is shorthand for ``noise source''.}
\label{table:tableapp}
\end{table}

\subsection{Quantitative tests on a simple network model}
We can assess the accuracy of the 3GA compared to the ssLNA and the intermediate approximations introduced above in terms of the KL divergence \eqref{eq:singlekl} and the covariance accuracy $\Delta C$ \eqref{eq:deltaC}. To estimate the required statistics we use a sample of 1000 realizations of intrinsic noise with $\eps = 0.1$ (we set $\eps_0 = 0$ to focus on the $\eps$ dependence). When comparing the four approximations discussed so far, the difference between 3GA and 3GA-$\bm{\chi}_0$ quantifies the importance of nonlinearity of noise; the difference between 3GA-$\bm{\chi}_0$ and 3GA-$\bm{\chi}_0^{w}$ shows the impact of time correlations (``color'' ) in the noise, and the comparison to the ssLNA provides information on the importance of memory.

\begin{figure}[!ht]
\includegraphics[width=8.5cm]{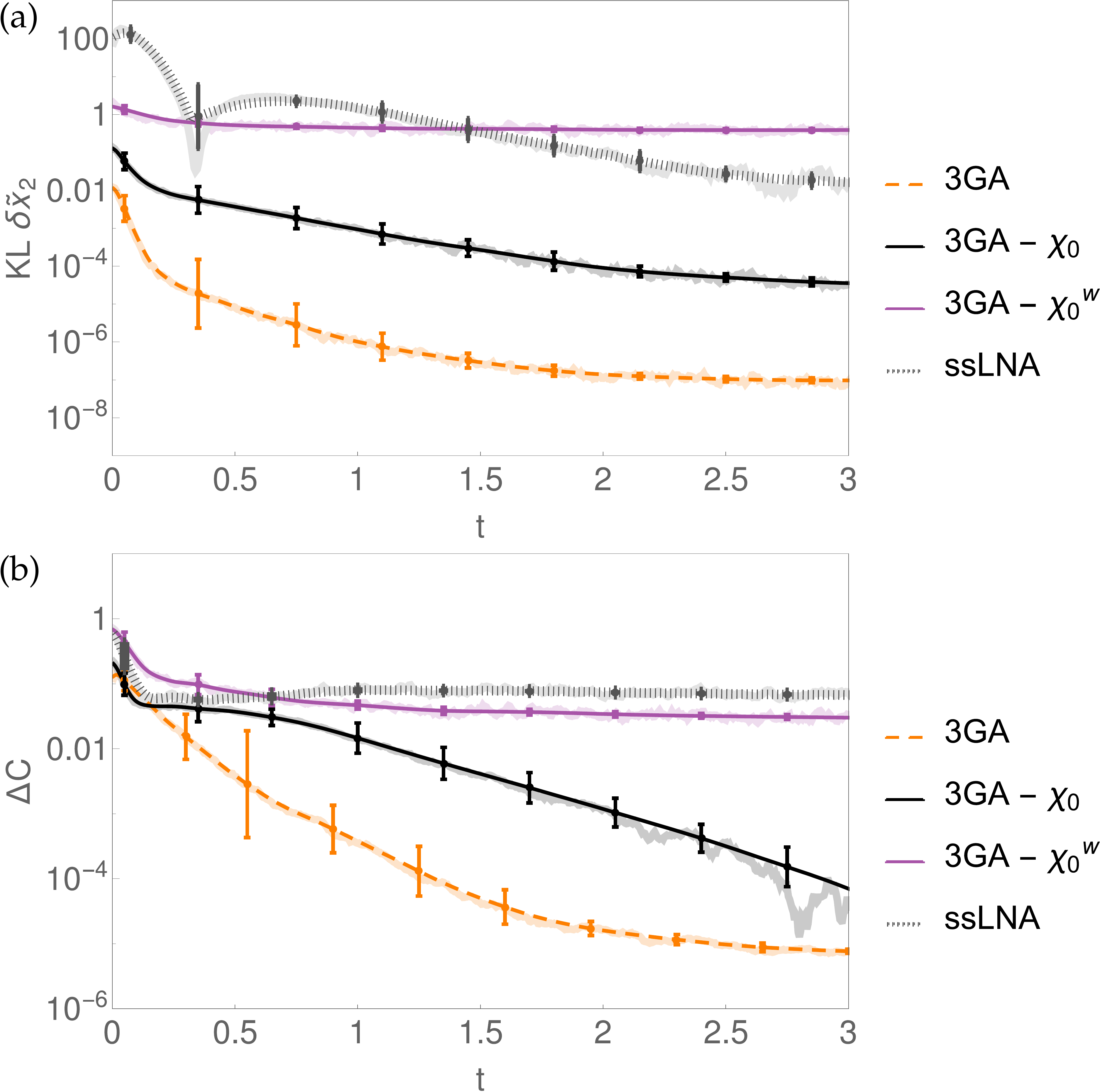}
\caption[]{\textbf{Accuracy of effective subnetwork noise approximations at $\epsilon = 0.1$}. (a) KL of subnetwork boundary species $\delta \tilde{x}_2$ and (b) error $\Delta C$, for 3GA, ssLNA and the intermediate approximations (3GA-$\bm{\chi}_0$ and 3GA-$\bm{\chi}^{w}_0$), estimated from a sample of 1000 noise realizations ($\epsilon = 0.1$). Smoothed curves with corresponding error bars are shown in dark color in addition to the raw numerical data (light color). The full 3GA reduction works best, though neglecting nonlinear noise effects (3GA-$\chi_0$) still gives a good approximation (small KL values).}
\label{fig:3p3c3kleps}
\end{figure}

By way of a first orientation, Fig.~\ref{fig:3p3c3eps} shows that the full 3GA model reduction gives results that are very close to the full dynamics both for individual trajectories (Fig.~\ref{fig:3p3c3eps}a,b) and for time-dependent mean trajectories and variances (Fig.~\ref{fig:3p3c3eps}d,e). The ssLNA (Fig.~\ref{fig:3p3c3eps}c) shows significant deviations, which can be attributed both to its neglect of memory (which is important for the means, Fig.~\ref{fig:3p3c3eps}d) and of time correlations in the extrinsic noise, resulting in variances being underestimated (Fig.~\ref{fig:3p3c3eps}e). The quantification of these deviations in terms of single species KL-divergence and covariance matrix accuracy (Fig.~\ref{fig:3p3c3kleps}a,b) confirms this. The 3GA-$\bm{\chi}_0^{w}$ approximation gives similar accuracies as the ssLNA but is significantly better for short times as is visible in the KL-divergence values in Fig.~\ref{fig:3p3c3kleps}a. This is due to its inclusion of memory effects, which in previous studies \cite{katy,gaussianvar} were shown to give approximation errors that were orders of magnitude lower than for memoryless approximations like the ssLNA. At long times, where the system has essentially reached its steady state, memory becomes unimportant and the 3GA-$\bm{\chi}_0^{w}$ and ssLNA perform similarly.

The 3GA-$\bm{\chi}_0$ approximation, which keeps track of time correlations in the extrinsic noise, is significantly more accurate than the ``white'' approximations (ssLNA and 3GA-$\bm{\chi}_0^{w}$). Its errors as shown in Fig.~\ref{fig:3p3c3kleps} are clearly still larger than those of the full 3GA but remain small in quantitative terms; for a comparison with the corresponding Wasserstein distances, which lead to the same conclusion, see Fig.\ S1c). We can thus summarize by saying that accurate subnetwork reductions require at least memory terms to make reasonable predictions for average time courses; for quantitative fluctuation estimates, time correlations in the extrinsic noise are also important while nonlinear (multiplicative) features of this noise can be neglected without seriously impairing accuracy.

We also tested the accuracy of the various approximations when noise and random initial conditions in the bulk are both present, see Fig.~S6. The initial bulk uncertainty can then -- and does in our simple network example -- become a significant driver of fluctuations in the dynamics. This makes the different treatment of the stochastic noise in the 3GA, 3GA-$\bm{\chi}_0$ and 3GA-$\bm{\chi}_0^{w}$ approximations less important, with all three methods giving low errors. Only the ssLNA continues to perform poorly because it assumes the bulk is always at steady state so cannot capture initial bulk fluctuations.
 
\subsection{Variation with strength $\epsilon$ of stochastic fluctuations}
\label{sec:var_eps}
\begin{figure}[!ht]
\includegraphics[width=16cm]{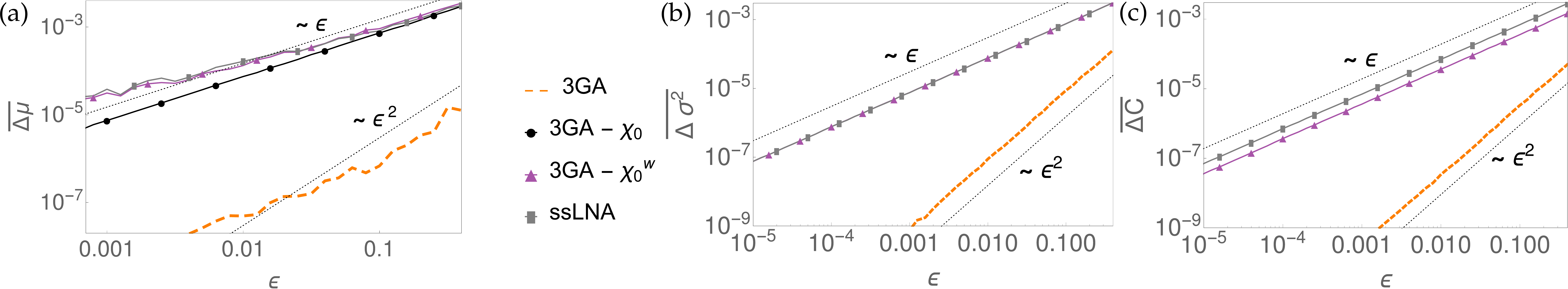}
\caption[]
{\textbf{Variation with $\epsilon$ of integrated error}: subnetwork means (a), variances (b) and correlations (c) over a sample of 2500 noise realizations for 3GA, ssLNA and intermediate approximations of 3GA. Dotted lines show the theoretically expected power laws (see Appendix \ref{sec:scalings}). Initial conditions in the subnetwork are at steady state, $\delta \bm{x}^{\rm s}(0)=0$. For the means in (a) we use 3500 realizations of stochastic trajectories to suppress sampling errors, which are more significant here and mean we cannot meaningfully access noise levels $\epsilon < 0.001 \sim 1/n$. (The deviations that are visible at low $\epsilon$ arise from finite sample size errors on the means, which scale linearly with $\epsilon$.)
For the same reason we do not show results for the 3GA-$\bm{\chi}_0$ approximation in (b,c).}
\label{fig:var_eps}
\end{figure}

The parameter $\epsilon$ we have introduced sets the amplitude of stochastic fluctuations and plays a role complementary to $\epsilon_0$, which controls the amplitude of deviations from the steady state in the bulk at the initial time. As a complement to the analysis in Sec.~\ref{sec:var_eps0}, where we varied $\epsilon_0$ at $\epsilon=0$, we now assess how the accuracy of the 3GA, ssLNA, 3GA-$\bm{\chi}_0$ and 3GA-$\bm{\chi}_0^{w}$ varies with $\epsilon$ when we set $\epsilon_0 = 0$. As before we set the subnetwork initial conditions to be at steady state to avoid competing error contributions. For each approximation, the error on the means scales as the first order in $\epsilon$ not systematically accounted for, thus one has (as discussed in Appendix \ref{sec:scalings}) $\Delta \mu \sim \epsilon^{2}$ for 3GA, $\Delta \mu \sim \epsilon$ for 3GA-$\bm{\chi}_0$, 3GA-$\bm{\chi}_0^{w}$ and ssLNA. For errors on covariances one finds $\overline{\Delta C} \sim \epsilon^{2}$ (3GA, 3GA-$\bm{\chi}_0$) and $\overline{\Delta C} \sim \epsilon$ (3GA-$\bm{\chi}_0^{w}$ and ssLNA); for $\overline{\Delta \sigma^2}$ the same scalings hold.
We confirm the validity of these power laws for our simple network model in Fig.~\ref{fig:var_eps}. The errors of the ssLNA in particular are highest as expected, though in quantitative terms they remain moderate. This arises from our use of steady state subnetwork initial conditions, which keep the exact mean concentrations close to steady state and hence to the ssLNA prediction (remaining exactly at steady state). In the general case the ssLNA makes larger errors on the means as shown by the large KL divergences in Fig.~\ref{fig:3p3c3kleps}a.

\section{Epidermal Growth Factor Receptor signalling network}
\label{sec:egfr}
The insights established by the previous analysis allow us to understand how to approximate the extrinsic noise, stemming both from random initial conditions in the unobserved bulk and the propagation of stochastic fluctuations, in biochemical networks. We now apply these insights to a key network around the Epidermal Growth Factor Receptor (EGFR). 
In this network, copy numbers of the relevant molecular species are large \cite{Beck2011, Nagaraj2011} and stochastic fluctuations therefore modest \emph{per se}. As a consequence of this, attention has focused on the role of the cellular context in shaping heterogeneous outcomes of EGFR signalling. For instance, the observed heterogeneity of EGF dose-response curves for downstream processes such as the translocation of Extracellular signal-Regulated Kinases (ERK) to the nucleus \cite{iwamoto} and Ras signalling \cite{kim} has been proposed to arise from cell-to-cell variation of signalling proteins like EGFR and of other species involved in the signalling pathway. 

\begin{figure}[!ht]
 \centering
 \includegraphics[width=8.5cm]{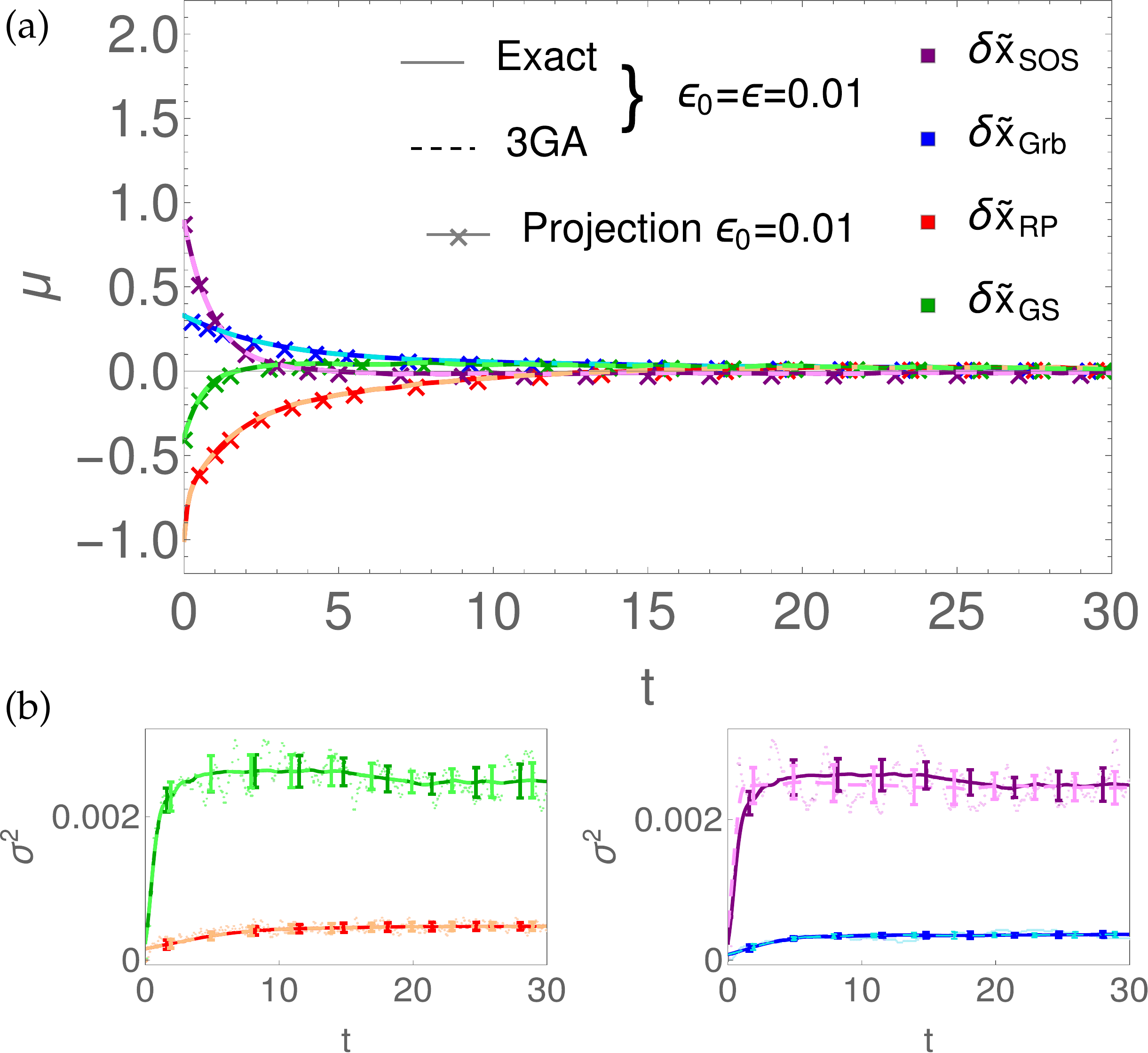}
 \caption[]
 {\textbf{Subnetwork dynamics in the EGFR network with intrinsic and extrinsic noise}. (a) Temporal evolution of mean concentrations (normalized deviations from steady state) over 200 realizations of bulk random initial conditions and stochastic noise with $\epsilon = \epsilon_0 = 0.01$ (bold lines: exact, dashed lines: 3GA). Trajectories by projection methods represent conditional means for $\epsilon \to 0$ and their average over the same random initial conditions (lines with crosses) is added as a comparison. (b) Comparison of variances, exact and 3GA, for the boundary species SOS, Grb, RP, GS. Smoothing (with corresponding error bars) is applied and shown in dark color on top of numerical data (in light color). Here and in all EGFR plots, time is expressed in seconds.}
 \label{fig:egfr}
\end{figure}

We consider the model by Kholodenko et al.\ \cite{kholodenko}, which describes the time evolution of 29 species involved in the EGFR signalling pathway. Following \cite{katy}, we choose a subnetwork of 19 species, including Epidermal Growth Factor (EGF) and EGFR, which trigger the signalling reactions, and a bulk of 10 species, namely Shc (Src homology and collagen domain protein) and all its complexes. The 4 subnetwork boundary species, directly affected by the interaction with the bulk, are SOS (Son of Sevenless homolog protein), Grb2 (Growth factor receptor-binding protein 2), RP (Tyrosine phosphorylated EGFR) and GS (Grb2-SOS complex); example results for their average time evolution and their variance around this are shown in Fig.~\ref{fig:egfr}.

In the EGFR network as modeled in \cite{kholodenko}, steady state concentrations $y_i$ range from $\sim 0.03$ to $\sim 600$ nMol. If we estimate cells to have a diameter of $20 \rm{\mu m}$ \cite{milo2010} and hence a volume of order $(20 \rm{\mu m})^3$, this gives absolute steady state number of molecules $Vy_i$ in a reaction compartment of volume $V$ in the range $\sim 140$ to $\sim 3 \cdot 10^6$, satisfying the criterion $Vy_i \gg 1$ for our Gaussian initial condition approximation and the treatment by a CLE. We use units where the smallest $y_i$ is order unity and so choose a fluctuation amplitude $\epsilon=0.01$ that correctly reproduces the smallest copy numbers $y_i/\epsilon\sim 100$.

We first study the accuracy of subnetwork equations with initial bulk uncertainty only, where we can apply both the path integral and projection approaches. We set $\epsilon_0 = 0.01$ as just estimated. As the computational cost of the EGFR network is higher than for the simple network, we sample here 200 realizations of bulk initial conditions for the bulk and measure the accuracy of the full random force and 3GA effective noise and the different approximations to them. The quantitative results for the time-dependent KL divergences and $\Delta C$ are shown in Fig.\ S2 and confirm the general trends observed in the simple network: including the transient part $\overline{\Delta\bm{r}}$ of the extrinsic noise but simplifying it to an impulse at $t=0$ gives reasonable approximations. The persistent part $\bm{r}_{\infty}$ on its own yields low accuracy and adding it to the transient $\overline{\Delta\bm{r}}$ does not make a significant difference to predictions at short times. Looking at specific combinations of concentrations that feature in conservation laws does however reveal a non-negligible role of $\bm{r}_{\infty}$ in reproducing correlated fluctuations in the long-time steady state (see Appendix \ref{sec:approx_rf} and Sec.~IV B in SM).

\begin{figure}[!ht]
\centering
\includegraphics[width=16cm]{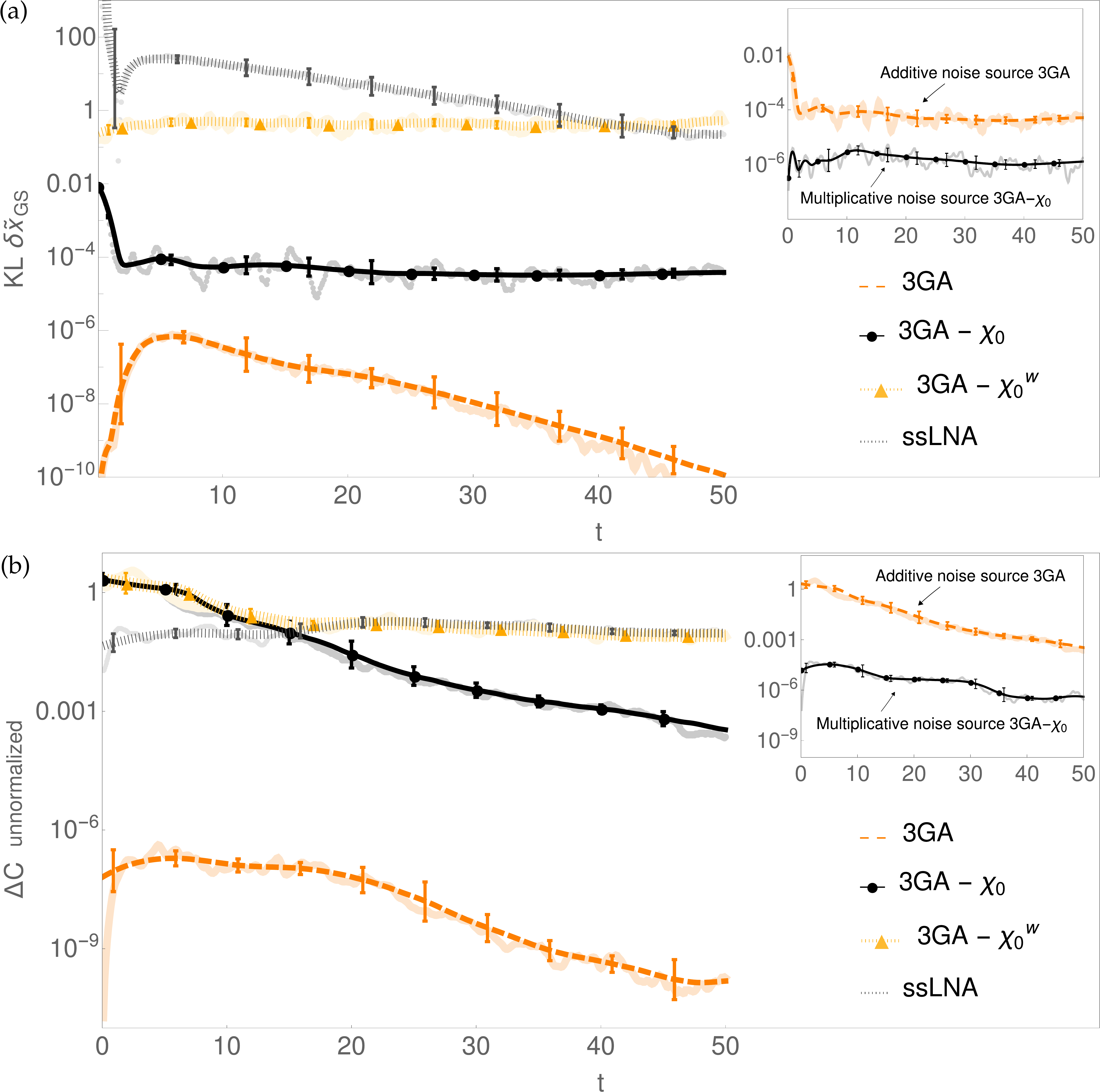}
\caption[]{\textbf{Accuracy of effective subnetwork noise approximations at $\epsilon = 0.01$}. (a) KL for the boundary species GS and (b) covariance matrix error estimated on a sample of 200 realizations of white noise with $\epsilon = 0.01$ for 3GA, intermediate approximations of 3GA (3GA-$\bm{\chi}_0$ and 3GA-$\bm{\chi}_0^w$) and ssLNA; the bulk is initially at steady state. The inset in (a) shows that the KL for the 3GA with additive noise source is very similar to the one of 3GA-$\bm{\chi}_0$; the inset in (b) shows that the same approximation gives a covariance error of $O(1)$, like the other approximations where the noise is additive (3GA-$\bm{\chi}_0$ and 3GA-$\bm{\chi}_0^w$); these approaches all underestimate the covariance on several subnetwork species. The insets in (a) and (b) also show that an improvement with respect to additive noise approximations is achieved by a linear noise approximation similar to 3GA-$\bm{\chi}_0$, if we neglect only the nonlinear propagation of effective noise but keep the multiplicative noise source; see Tab.\ \ref{table:tableapp} for details of this approximation. The covariance error in (b) was calculated \emph{without} the normalization by steady state concentrations, to avoid it becoming dominated by a few interior subnetwork species that have extremely low steady state concentrations and initially large deviations from this steady state (an example is shown in Fig.\ S5). For comparison, the version with normalized $\Delta C$ (as defined by \eqref{eq:deltaC}) is provided in SM, Fig.\ S4.}
\label{fig:egfr_eps}
\end{figure}

We next assess the performance of 3GA compared to the ssLNA and the linear-noise approximations of Sec.~\ref{sec:approx_chi} (and Tab.\ \ref{table:tableapp}) in the presence of (only) stochastic fluctuations of amplitude $\epsilon = 0.01$; results are summarized in Fig.~\ref{fig:egfr_eps}. In all these quantitative tests, we have chosen challenging subnetwork initial conditions as in Ref.~\cite{katy} that maximize nonlinear effects in the subnetwork. This results in large initial values of the components of $\delta\bm{\tilde{x}}$ for some species. The linearization of $\sqrt{\bm{\Sigma}}$ in  $\delta\bm{x}$ that we have used so far (see Sec.\ II in SM) then no longer makes sense, so we work with the full multiplicative noise $\bm{\xi}^{\cdot}(t)$ with covariance (\ref{eq:fluct},\ref{eq:fluctdx}) and define its nonlinear part as $\bm{\xi}_1^{\cdot}(t)=\bm{\xi}^{\cdot}(t)-\bm{\xi}_0^{\cdot}(t)$, $\cdot=$ s,b. For more typical initial conditions closer to the steady state we have verified that our standard linearization of $\sqrt{\bm{\Sigma}}$ can be applied to the EGFR system without difficulties and gives sensible results.

The KL divergences resulting from the various approximations follow similar trends for the EGFR network model as in our simple network model; Fig.~\ref{fig:egfr_eps}a shows data for one of the boundary species in the subnetwork. The same is true of the error on the covariance (Fig.~\ref{fig:egfr_eps}b), with one exception: the ssLNA has lower errors in the initial transient here than the 3GA-$\bm{\chi}_0$ and 3GA-$\bm{\chi}_0^{w}$. The reason is the accuracy with which correlations of interior subnetwork species are described, an effect that is not picked up by our single-species KL divergences (which have other benefits as explained above, specifically allowing us to focus on the memory effects that only affect boundary species). Specifically, some interior subnetwork species (mainly complexes formed by RP) are subject to strong relative fluctuations, as they have low steady state concentrations but start from much larger initial values (see for example Fig.\ S5). The ssLNA, whose noise properties depend on the deterministic, fully nonlinear temporal trajectory is able to reproduce these large fluctuations, while its rather rough treatment of effects from the bulk has limited effect on interior subnetwork species. In contrast, our approximations with purely additive noise (3GA-$\bm{\chi}_0$ and 3GA-$\bm{\chi}_0^{w}$) underestimate these large interior subnetwork correlations, giving an error $\Delta C$ that is close to unity initially. 

This behavior allows us to further appreciate the role of the different terms that make up the nonlinear (multiplicative) contribution to the effective noise $\bm{\chi}_1(t)$ \eqref{eq:chi1} as predicted by the 3GA. It consists of (i) a multiplicative noise generated by nonlinear subnetwork-bulk reactions (i.e.\ the $\bm{\lambda}$-dependent terms of \eqref{eq:chi1}) and (ii) a multiplicative noise \emph{source} (i.e.\ terms $\bm{\xi}^{\rm s}_1$ and $\bm{\xi}^{\rm b}_1$ of \eqref{eq:chi1}). We have tested what accuracy is achieved if we neglect either (i) or (ii): in the former case, we are effectively implementing a 3GA-$\bm{\chi}_0$ with multiplicative noise source while in the later we obtain a 3GA approximation with additive noise source. Neglecting (i) has an impact only on boundary species (expected to make up a small fraction of large networks, e.g.\ here we have a subnetwork of 19 species, of which only 4 are boundary species) and we have found that indeed this keeps errors very small, as shown by the curves labelled ``multiplicative noise source 3GA-$\bm{\chi}_0$'' in the insets of Fig.~\ref{fig:egfr_eps}. On the other hand, as soon as (ii) is neglected, which affects how well the fluctuations of every subnetwork species are captured, the accuracy of the 3GA becomes essentially the same as the 3GA-$\bm{\chi}_0$ (curves labelled ``additive noise source 3GA''). From these numerical tests we can conclude that the improvement in accuracy of the 3GA over the 3GA-$\bm{\chi}_0$ in modeling the EGFR system results mainly from accounting for the multiplicative nature of the noise sources.

Finally we consider the full reaction equations including all stochastic terms, i.e.\ both intrinsic noise of amplitude $\epsilon$ and fluctuating bulk initial conditions of amplitude $\epsilon_0$. As both types of fluctuations arise from the same finite reaction volume we take them equal, $\eps=\eps_0 =0.01$. Averaging over multiple realizations of noise and randomly sampled bulk initial conditions one obtains true conditional (on the initial subnetwork state) averages of subnetwork concentrations, which can be compared to the predictions from the projected equations for the same bulk initial conditions. These are approximate not only in their treatment of the random force (by truncating after terms of order $(\delta \bm{x}^{\rm b})^2$, but also via the fact that the memory functions used are calculated for $\eps\to 0$, see \cite{katy}). Nonetheless, the agreement with the exact averages and those predicted by the 3GA method is excellent, see Fig.~\ref{fig:egfr}a. Total variances can be estimated only in the 3GA approach, which yields extremely accurate results as shown in Fig.~\ref{fig:egfr}b. In the EGFR network, initial bulk uncertainty is not the main contribution to the overall effective noise as it was for the simple network model, hence the comparison of performance between 3GA and the other approximations from Tab.\ \ref{table:tableapp} is essentially the same as in the case where only stochastic noise is present, as we show in Fig.~S7.

\section{Discussion}

In this paper we have considered stochastic effects in mathematical models of biochemical subnetworks embedded in an environment (the bulk). This setting allows us to  study systematically the emergence of {\em extrinsic noise} acting on the subnetwork due to the presence of the bulk, and to analyze the structure of the various contributions to this noise. The general strategy here is to remove, i.e.\ integrate out or ``marginalize'', the bulk degrees of freedom to arrive at a reduced subnetwork description. For dynamics linearized around a fixed point, this can be done explicitly, showing that the time evolution of subnetwork boundary species -- those that interact directly with the bulk -- acquires both memory terms and an extrinsic noise that is colored (time-correlated). The extrinsic noise consists of two contributions: the first arises from the propagation of stochastic fluctuations in the bulk. The second contribution, which we refer to as random force, comes from the unknown bulk initial conditions. This fundamental structure can be generalized to nonlinear equations by two methods of systematic model reduction,  namely, projection methods \cite{katy} and a path-based marginalization over the bulk \cite{gaussianvar} (3GA). While in previous works \cite{katy,gaussianvar} we focused on memory terms, by assuming the bulk to be initially at steady state and consider vanishing intrinsic noise, we have in this study investigated the effects of the environment when it is not initially at steady state and one has to account also for its intrinsic fluctuations.

We saw that when only randomness in bulk initial conditions is present, it is possible to approximate the random force appearing in the projected equations as the sum of a short time piece and a constant piece that capture its transient and persistent parts, respectively. The first contribution is thus simplified to an impulse at $t=0$ (a $\delta$-function) with amplitude given by the integral of the short time random force. It has the simple effect of perturbing the subnetwork initial conditions randomly. Adding the time-persistent piece gives what we called the ``impulse plus persistent'' approximation. It gives accurate results, with the transient piece being essential and the persistent piece giving smaller corrections. Analogous approximations for the colored noise appearing in the path integral 3GA approach lead to the same trends.

We next considered the fully stochastic case. To include intrinsic noise one needs to step away from the projection approach, with its focus on conditional averages; we then use instead the path integral technique 3GA, which can track all sources of fluctuations in the time evolution. Accordingly there are two separate variances, one ($\eps_0$) for the initial bulk fluctuations and another ($\eps$) for the stochastic noise affecting all chemical reactions. We analyzed the scalings of errors with both parameters and found them to follow similar power laws, though their relative impact on the accuracy of predictions for the subnetwork dynamics depends on details of the reaction network such as reaction rates. (Note that for $\eps>0$, relative concentration fluctuations cannot in general be made smaller than $O(\eps)$, even for mean field-like situations where each molecular species takes part in many reactions; approaches based on linearization in fluctuations therefore always remain approximate to some extent except in the joint limit of $\eps\to 0$ and $\eps_0\to 0$.)

We compared the 3GA approach with simplifications treating the extrinsic noise as linear (additive) or white linear (additive and without time correlations), and with the steady-state linear noise approximation (ssLNA); the latter treats all bulk dynamics as effectively instantaneous. We assessed the accuracy of these approximations for a simple network model and for the EGFR network and found that retaining in the subnetwork-reduced dynamics the time correlations (``color'') of the extrinsic noise and its multiplicative (subnetwork-state dependent) nature is more important than nonlinear corrections.
The time correlations in particular proved to be essential for properly reproducing the amplitude and temporal variation of correlations induced by extrinsic noise. The best choice of approximation for the extrinsic noise may be system-dependent: if there are transients where some concentrations that affect the colored noise covariance are far from steady state, as for the EGFR network, then any approximation which disregards the multiplicative nature of the noise source will be poor, regardless of the nonlinearities it accounts for, while relatively simple approximations like the ssLNA can perform well. By contrast, if there are other dominant sources of variation such as the initial bulk uncertainty in our simple network model, then the focus needs to be on capturing these properly, while for the stochastic fluctuations one can use approximations that are simpler than the full 3GA. An important question that would deserve future investigation is whether one can find a systematic strategy to assess the expected magnitude of the different contributions to the reduced dynamics starting from the rates and structure of the reaction network.

Several recent studies have concentrated on distinguishing sources of noise by conditioning on some part of the network in such a way as to split the fluctuations of some biochemical species into intrinsic contributions and those coming from either a static \cite{swain,bowsher} or dynamic \cite{paulsson} environment, which also contribute with memory-like terms \cite{zechner2014a, falk2019}. Our expression for the colored noise (\ref{eq:chi0}, \ref{eq:chi1}) shows explicitly the decomposition of stochasticity into different contributions, first into noise intrinsic to the subnetwork and extrinsic noise, due to the interaction with the environment \cite{elowitz,swain,shahrezaei}. A key aspect in our result is  that we can further decompose the extrinsic noise into separate contributions coming from the random fluctuations inherent to biochemical reactions in the bulk and the uncertainty in its initial state. Our derivation of this decomposition is general (for unary and binary reactions, up to the second order) and fully retains dynamical features such as the transients in the time courses before a steady state is reached. Our result therefore generalizes previous steady state analyses \cite{Paulsson2004, shibata} to dynamics, without any a priori assumption on the timescale of environmental fluctuations; the fast bulk approximation is recovered as a particular case. The tools to include other types of reactions in this framework, e.g.\ Michaelis-Menten enzyme reactions or gene regulation, are already available \cite{rubinMM, herrera-delgado}. Our approach to stochastic effects in subnetwork modeling allows us to understand to what extent variation in the system of interest is determined by fluctuations in the environment. If processes in the bulk can be parametrized and their rates can be estimated, at least within certain ranges, then our method enables one to put forward hypotheses on the sources of observed variability in the subnetwork. In this way the properties of extrinsic noise can be {\em deduced} from statistical assumptions about the bulk dynamics and these assumptions can then be refined by comparing the predicted and the measured effects on the subnetwork. For the sake of simplicity we have limited the analysis to the Gaussian limit of Poisson fluctuation statistics for the bulk initial conditions fluctuations, but the generality of \eqref{eq:chi0} and \eqref{eq:chi1} leaves open the possibility of considering different choices of fluctuating initial conditions. In addition, the strategy we have discussed in principle also applies to heterogeneity in bulk reaction rate constants. For mass-action kinetics this can be treated on the same footing as fluctuations in bulk species, by including the relevant rates into the state-space of the process as ``virtual'' bulk variables. This equivalence has been exploited to capture within the same mathematical framework extrinsic factors consisting either of heterogeneous, time-invariant reaction parameters -- akin to reaction rates in the bulk \cite{zechner2012,zechner2014} -- or of temporally fluctuating variables -- akin to bulk species \cite{zechner2014a}.

We have taken the EGFR network, whose structure and rates are well-characterized, as an illustrative example of how our mathematical methods can capture the dynamic propagation of noise as well as the persistent and the transient contributions. Here we have chosen the bulk to be the protein Src homology and collagen domain protein (Shc) and any complexes that include Shc, but one could straightforwardly repeat this analysis for other network splits, e.g.\ allocating to the bulk the extracellular stimuli that induce EGF binding, EGFR conformational changes and phosporylation events, with in the subnetwork a selected set of reactions that are activated downstream. Among the targets, one could consider molecular species that are crucial to cell-fate decisions, such as Sos, Grb2 as well as Ras, MAPK, Erk; recent investigations highlight these as molecular species that are sensitive to cellular context variability \cite{iwamoto, santos, kim}. In this way our mathematical framework could be applied to study how the variability in external signals and in the abundance of EGFR (induced e.g.\ by downregulation, internalization and subsequent degradation or recycling \cite{starbuck}) is conveyed by protein reaction cascades along the pathway to ultimately affect cellular responses to signalling. The linear scaling of 3GA reduction with the size of the bulk (see discussion in Appendix \ref{sec:eff_solver_3GA}) enables one to deal with the high combinatorial complexity of protein interaction networks more efficiently than projection methods (which scale quadratically with bulk size). This would be important e.g.\ for studying a more recent version of Kholodenko’s EGFR signalling model \cite{faeder} that contains a significantly increased number of molecular species (over 300) by incorporating more details on protein domains and on feasible biochemical reactions. The example of heterogeneity in EGFR response hence sets the stage for a systematic use of our model reduction schemes, which can be flexibly adjusted to the theoretical or experimental question under investigation.

\section*{Supplementary Material}
\textbf{I. Memory and random force in projection methods}\\
\textbf{II. Path integral approach 3GA}\\
\textbf{III. KL divergence estimation}\\
Figure S1: Numerical tests for KL divergence estimation.\\
\textbf{IV. Epidermal Growth Factor Receptor}\\
Figure S2: Accuracy of approximations of the extrinsic noise from random initial conditions at $\epsilon_0 = 0.01$.\\
Figure S3: Accuracy of prediction with conservation laws.\\
Figure S4: Accuracy of effective subnetwork noise approximations at $\epsilon = 0.01$.\\
Figure S5: Fluctuations magnitude for intra-subnetwork species RPL.\\
\textbf{V. Joint sampling of extrinsic and intrinsic noise}\\
Figure S6: Accuracy of approximations of effective subnetwork noise in the simple network model at $\epsilon = \epsilon_0 = 0.1$.\\
Figure S7: Accuracy of approximations of effective subnetwork noise in the EGFR network at $\epsilon = \epsilon_0= 0.01$.\\

\section*{Acknowledgements}
The authors acknowledge helpful discussions with Ramon Grima, Edgar Herrera-Delgado and Brian Munsky.

\section*{Data Availability Statement}
The data that support the findings of this study are available from the corresponding author upon reasonable request.

\begin{appendices}

\appendix
\section{Approximation ``impulse plus persistent'' of the random force}
\label{sec:approx_rf}
To understand in more detail what properties the random force in the full projection approach should have, we go back to the subnetwork equations for the concentration deviations in the subnetwork, denoted collectively by $\bdx^{\rm{s}}$. 
For simplicity we shall focus on the case of linearized dynamics at $\epsilon \to 0$, written in general form as 
\be
\label{eq:projlin}
\frac{d}{dt}\bdx^{\rm{s}^{\T}} = \bdx^{\rm{s}^{\T}}\bm{L}^{\rm{ss}} +\int_0^t dt'\,\bdx^{\rm{s}^{\T}}\bm{M}^{\rm{ss}}(t-t') + 
\bmrz^{\T}
\ee
where $\bmrz(t)$ is given by \eqref{eq:rf0}. To write down the solution of the projected equations in the presence of nonzero random forces, we can introduce the Green's function $\bu(t)$ for the system \eqref{eq:projlin}, which is determined by the rate matrix and memory function. In terms of this Green's function, the subnetwork concentrations at time $t$ are 
simply
\begin{equation}
  \label{eq:projectedeqns_soln}
  \bdx^{\rm{s}^{\T}}(t) = \int_0^t dt'\,\bmrz^{\T}(t') \bu(t-t') + \bdx^{\rm{s}^{\T}}(0) \bu(t)
\end{equation}
The second term depends only on the subnetwork initial conditions and not the random force, so we will ignore it in the following discussion. The convolution structure of $\bdx^{\rm{s}^{\T}}(t)$ suggests that for further analysis one should split both $\bm{r}$ and 
$\bm{U}$ into two parts, one transient and one permanent, i.e.\ surviving to infinite time. Given again the convolution structure, it is convenient to work with the Laplace Transform
\begin{equation}
\delta\hat{\bm{x}}^{\rm{s}^{\T}}(z) = \rfhatz(z)^{\T}\uhat(z)
\end{equation}
where $\uhat(z)$ is given by
\begin{equation}
\label{eq:uz}
\uhat(z) = \left( z - \bm{L}^{\rm ss} - \bm{\hat{M}}^{\rm ss}(z) \right)^{-1}
\end{equation}
Decomposing $\uhat(z)$ and $\rfhat(z)$ into the transient and permanent part, we write 
\begin{equation}
 \begin{split}
\rfhatz(z) &= \frac{1}{z}\rinfz(z)+\Delta\rfhatz(z)\\
\uhat(z) &= \frac{1}{z}\uinf + \Delta\uhat(z)
 \end{split}
\end{equation}
The variation of the steady state due to the random force is then found as
\begin{equation}
\label{eq:expandssrf}
\begin{split}
  \bdx^{\rm{s}^{\T}}(t\rightarrow\infty) &= \lim_{z\to 0} z\,\rfhatz^{\T}(z)\uhat(z)\\
  &=\lim_{z\to 0} z \left(\frac{1}{z}\rinfz(z)+\Delta\rfhatz(z)\right)^{\T}\left(\frac{1}{z}\uinf + \Delta\uhat(z)\right)
\\
&=\lim_{z\to 0}\left(\frac{1}{z}\rinfz^{\T}\uinf + \rinfz^{\T}\Delta\uhat(z) + \Delta\rfhatz^{\T}(z)\uinf + z
\Delta\rfhatz^{\T}(z) \Delta\uhat(z) + \order(z) \right)
\end{split}
\end{equation}
We show below that the apparently divergent first term does not contribute while the last term vanishes for $z\to 0$. 
In the two remaining finite contributions, the small $z$-limit gives
\begin{equation}
\Delta\rfhatz(z\to 0) = 
\int_0^\infty dt'\,\Delta\bmrz(t') = 
\int_0^\infty dt'\, [\bmrz(t')-\rinfz] \equiv \overline{\Delta\bmrz}
\end{equation}
and similarly $\Delta\uhat(z\to 0)=\overline{\Delta\bu}$, so that
\begin{equation}
\begin{split}
  \bdx^{\rm{s}^{\T}}(t\rightarrow\infty) &= 
\rinfz^{\T}\overline{\Delta\bu} + \overline{\Delta{\bmrz}}^{\T}\uinf
\end{split}
\end{equation}
This decomposition has two contributions, which are respectively the transient response to the persistent part of the random force, and the persistent response to the transient part of the random force.

The analysis so far tells us that we will get the correct random force-dependent steady state from an approximate form of the random force as long as the approximation preserves both $\rinfz$ and $\overline{\Delta \bmrz}$. Our proposal is then to construct the approximation by ``concentrating'' the transient part of the random force into a spike at $t=0$, specifically
\begin{equation}
\label{approx_rfapp}
  \bmrz(t) = \rinfz+\Delta\bmrz(t)
  \simeq \rinfz + \delta(t) \overline{\Delta\bmrz}
\end{equation}
Accordingly we then approximate the linearized projected equations as
\begin{equation}
\label{eq:rfapprox}
  \frac{d}{dt}\bdx^{\rm{s}^{\T}} = \bdx^{\rm{s}^{\T}}\bm{L}^{\rm{ss}} +\int_0^t dt'\,\bdx^{\rm{s}^{\T}}\bm{M}^{\rm{ss}}(t-t') + \rinfz^{\T} + \overline{\Delta\bmrz}^{\T}\delta(t)
\end{equation}
Integrating a small time interval around $t=0$ then shows that the effect of the $\delta$-piece of the random force is equivalent to a random change in the initial subnetwork conditions by an amount $\overline{\Delta\bm{r}_0}$.

To see how the above approximation is implemented in practice, we recall that in terms of the subblocks of the matrix $\bm{L}$ representing the adjoint Fokker-Planck matrix operator the random force is written as \eqref{eq:rf0}, i.e.\
\begin{equation}
  \bmrz^{\T}(t) =  \bdx^{\rm{b}}(0)^{\T}e^{\bm{L}^{\rm{bb}t}}\bm{L}^{\rm{bs}}
\end{equation}

If $\bar{\bm{\rho}}$ and $\bar{\bm{l}}$ are the right and left eigenvectors and $\mu$ the eigenvalues of the matrix $\bm{L}^{\rm{bb}}$ we can write the persistent and transient components of the random force explicitly as
\begin{equation}
\label{eq:rsums}
  \begin{split}
    \rinfz^{\T} &= \bdx^{\rm{b}^{\T}}(0)\sum_{\beta,\mu_\beta = 0} \bar{\bm{\rho}}_\beta \bar{\bm{l}}_\beta^{\T}\bm{L}^{\rm{bs}}\\
\overline{\Delta{\bmrz}}^{\T} &= - \bdx^{\rm{b}^{\T}}(0)\sum_{\beta,\mu_\beta\neq 0} \frac{1}{\mu_\beta}\bar{\bm{\rho}}_\beta \bar{\bm{l}}_\beta^{\T}
\bm{L}^{\rm{bs}}
  \end{split}
\end{equation}

We can now use the eigenvector structure of $\rinfz$ and $\uinf$ to show that the first term in \eqref{eq:expandssrf} vanishes. Let $\bm{\rho}^{\rm{s}}$ and $\bm{l}^{\rm{s}}$ be the subnetwork components of the 
right and left eigenvectors of the full matrix $\bm{L}$, and $\lambda$ the corresponding eigenvalues. Then we can decompose $\uhat(z)$ \eqref{eq:uz} into
\be
\uhat = \sum_{\alpha, \lambda_{\alpha}=0} \frac{1}{z} \bm{\rho}_{\alpha}^{\rm{s}} \bm{l}_{\alpha}^{\rm{s}\,\rm{T}}+
\sum_{\alpha, \lambda_{\alpha} \neq 0} \frac{1}{z- \lambda_{\alpha}}  \bm{\rho}_{\alpha}^{\rm{s}} \bm{l}_{\alpha}^{\rm{s}\,\rm{T}}
\ee
from which one reads off 
\begin{equation}
\label{eq:uinf}
\begin{split}
\uinf & =  \sum_{\alpha, \lambda_{\alpha}=0} 
\bm{\rho}_{\alpha}^{\rm{s}} \bm{l}_{\alpha}^{\rm{s}\,\rm{T}}\\
\Delta \uhat &= \sum_{\alpha, \lambda_{\alpha} \neq 0} \frac{1}{z- \lambda_{\alpha}}  \bm{\rho}_{\alpha}^{\rm{s}} \bm{l}_{\alpha}^{\rm{s}\,\rm{T}}
\end{split}
\end{equation}
Then we can write $\rinfz^{\T}\uinf$ using \eqref{eq:rsums} and \eqref{eq:uinf} as
\begin{equation}
\label{eq:uinf0}
  \begin{split}
    \rinfz^{\T}\uinf &= \bdx^{\rm{b}^{\T}}(0)\sum_{\beta,\mu_\beta = 0}\sum_{\alpha,\lambda_\alpha = 0} \bar{\bm{\rho}}_\beta \bar{\bm{l}}_\beta^{\T}\bm{L}^{\rm{bs}}\bm{\rho}_{\alpha}^{\rm{s}}\bm{l}_{\alpha}^{\rm{s}^{\T}}\\
&= -\bdx^{\rm{b}^{\T}}(0)\sum_{\beta,\mu_\beta = 0}\sum_{\alpha,\lambda_\alpha = 0} \bar{\bm{\rho}}_\beta \bar{\bm{l}}_\beta^{\T}\bm{L}^{\rm{bb}}\bm{\rho}_{\alpha}^{\rm{b}}\bm{l}_{\alpha}^{\rm{s}^{\T}}\\
&=0
  \end{split}
\end{equation}
where the last line follows because $\bar{\bm{l}}_\beta^{\T}\bm{L}^{\rm{bb}}=0$ when $\bar{\bm{l}}_\beta$ is an eigenvector with zero eigenvalue of $\bm{L}^{\rm{bb}}$. We have also used the fact that $\bm{L}^{\rm{bs}}\bm{\rho}_\alpha^{\rm{s}} + \bm{L}^{\rm{bb}}\bm{\rho}_\alpha^{\rm{b}} = 0$
for any right eigenvector of $\bm{L}$ with eigenvalue $\lambda_{\alpha} = 0$; this identity allows one to eliminate $\bm{L}^{\rm{bs}}$ in favour of $\bm{L}^{\rm{bb}}$ in \eqref{eq:uinf0}). Note $(-\bm{L}^{\rm{ss}} - \bm{M}^{\rm{ss}})$ becomes non-invertible when we consider conserved quantities and this occurs precisely when the response has a persistent part $\bm{U}_{\infty}$: therefore the persistent part of the random force originates from conservation laws (see Sec.~IV B in SM for an illustration in the context of the EGFR network). 

Motivated by the above analysis, we extend the approximation \eqref{approx_rfapp} to the random force $\bm{r}(t) = \bm{r}_0(t) + \bm{r}_1(t)$ of the nonlinear dynamics (to get \eqref{eq:approx_rf}) as well as to the colored noise $\bm{\chi}(t) = \bm{\chi}_0(t) + \bm{\chi}_1(t)$ of the 3GA-equations in the limit $\epsilon \to 0$  \eqref{eq:approx_chi}. 
Disentangling the purely $\delta\bm{x}^{\rm b}(0)$-dependent terms from the memory is not completely obvious from the mathematical derivation itself of 
the 3GA (see Sec.\ II in SM) and is easiest to understand from a numerical point of view, as explained in Appendix \ref{sec:eff_solver_3GA_chi}.

\cleardoublepage

\section{Scalings of statistics with $\epsilon_0$ and $\epsilon$}
\label{sec:scalings}
In this appendix, we assess how the errors (\ref{eq:deltamuerr},\ref{eq:deltaserr},\ref{eq:deltacerr}) scale with $\epsilon_0$, i.e.\ with the variance of the initial bulk fluctuations, and with $\epsilon$, the variance of stochastic fluctuations in the dynamics. To estimate this theoretically, we need to expand the solution of 3GA and projection equations in $\epsilon_0$ and $\epsilon$, which enter via $\delta \bm{x}^{\rm b}(0) \sim \sqrt{\epsilon_0}$ and $\bm{\xi} \sim \sqrt{\epsilon}$, and see to what order the 3GA and projection methods are systematic.

We first focus on the 3GA path integral approach, which includes both dependencies. The subnetwork-reduced equation of motion \eqref{eq:nonlin3GA}, accounting for $\delta \bm{x}^{\rm b}(0)$ and $\bm{\xi}$ through $\bm{\chi}$ (\ref{eq:chi0}, \ref{eq:chi1}), has the schematic form
\begin{equation}
\label{eq:schemeeq}
\frac{d}{d t}\bdx^{\rm{s}} = \bdx^{\rm{s}} + \bdx^{\rm{s}\,2} + \bm{\xi} + \bdx^{\rm{s}}\bm{\xi} + O(\bdx^{\rm{s}\,3},\bdx^{\rm{s}\,2}\bm{\xi})
\end{equation}
where all terms are in general convolutions (e.g.\ the memory terms) but for simplicity we do not write these explicitly. To start with we shall also assume zero subnetwork initial conditions, $\delta\bm{x}^{\rm s}(0)=0$. The logic of the 3GA (Sec.\ II in SM) consists of keeping only the linear term on the r.h.s.\ of (\ref{eq:schemeeq}) initially, solving with Green's functions and then treating the other terms perturbatively. In this way, $\bdx^{\rm{s}}$ is expressed as an expansion in $\delta \bm{x}^{\rm b}(0)$ and $\bm{\xi}$ up to \emph{second} order (again involving time convolutions that we leave implicit), giving schematically a solution
\be
\label{eq:scal}
\bdx^{\rm{s}} = \bm{\xi} + \bm{\xi}^2 + \delta\bm{x}^{\rm b}(0) + \delta\bm{x}^{\rm b\,2}(0) + \bm{\xi}\delta\bm{x}^{\rm b}(0)
\ee
This solution will be accurate up to the order written, i.e.\ all third order terms are unsystematic. To disentangle the dependence on $\epsilon$ from the one on $\epsilon_0$, let us set $\delta \bm{x}^{\rm b}(0)=0$. If we consider that pure $\bm{\xi}^3$-terms average out as $\bm{\xi}$ is drawn from a Gaussian with zero mean (see Eq.\ \eqref{eq:fluct}), \eqref{eq:scal} gives an estimate of the absolute errors on the means as $\overline{\Delta \mu} \sim \bm{\xi}^4 \sim \epsilon^{2}$. For the second moments, we take the square of \eqref{eq:scal}, showing that the first terms not systematically included are fourth order ones in $\bm{\xi}$. The errors on correlations $\overline{\Delta C}$ and on the variance $\overline{\Delta \sigma^2}$ are thus also expected to scale as $\epsilon^{2}$ in leading order. 

We next look at errors within the simper approximations. The 3GA-$\bm{\chi}_0$, which neglects $\bm{\chi}_1(t)$, can be systematic only up to $O(\sqrt{\epsilon})$, i.e.\ up to $\bm{\xi}$ in the perturbative expansion \eqref{eq:scal}. This gives errors on means that are now $\overline{\Delta \mu} \sim \bm{\xi}^2 \sim \epsilon$, while from the square of \eqref{eq:scal} one estimates errors on second moments $\overline{\Delta C} \sim \overline{\Delta \sigma^2} \sim \epsilon^{2}$. Approximations that take the noise as white, like the 3GA-$\bm{\chi}_0^{w}$ and ssLNA, are no longer systematic at $O(\sqrt{\epsilon})$ as they change the stochastic noise properties already at leading order; but as terms $O(\sqrt{\epsilon})$ average out for Gaussian noise one again gets $\overline{\Delta \mu} \sim \epsilon $.
The correlations and variances will be incorrect already to leading order, on the other hand, i.e.\ have errors $\sim \epsilon$.
The ssLNA, which retains nonlinearities only at the level of the deterministic dynamics and so actually gives an $\epsilon$-independent prediction of the means, has the same error scalings as the 3GA-$\bm{\chi}_0^{w}$.

The same line of reasoning as above is valid for the scalings with $\epsilon_0$, given the dependence on $\delta\bm{x}^{\rm b}(0)$ in \eqref{eq:scal}. The dependence of $\bdx^{\rm{s}}$ from projection approach on $\delta\bm{x}^{\rm b}(0)$ is also the same given that it is determined through the random force. (We recall that the projection method, as developed in \cite{katy} and summarized in Sec.\ I of SM, can predict the linear and quadratic contributions to the random force in the initial values $\delta\bm{x}^{\rm b}(0)$ but not terms $O(\delta\bm{x}^{\rm b}(0)^3)$). As a result one obtains, for both 3GA and projection methods, the scalings $\overline{\Delta \mu} \sim \overline{\Delta C} \sim \overline{\Delta \sigma^2} \sim \epsilon_0^{2}$.

Finally, the effect of nonzero initial conditions $\delta\bm{x}^{\rm s}(0)$ in the subnetwork would be to generate additional deviations. For the mean values these will scale as $O(\delta\bm{x}^{\rm s}(0)^3)$ and will dominate over the sources of error discussed above once $\epsilon$ and $\epsilon_0$ are sufficiently small.

\section{Simple network model}
\label{sec:toy_app}
The simple network model of Fig.~\ref{fig:3p3c3rf} is composed of the following reactions \\
\ba
1 + 2 \mathrel{\mathop{\rightleftarrows}^{k^+_{12}}_{k^-_{12}}} 12 \notag\\
1 + 3 \mathrel{\mathop{\rightleftarrows}^{k^+_{13}}_{k^-_{13}}} 13 \notag\\
2 + 3 \mathrel{\mathop{\rightleftarrows}^{k^+_{23}}_{k^-_{23}}} 23 \notag\\
1 \mathrel{\mathop{\rightleftarrows}^{\lambda_{12}}_{\lambda_{21}}} 2\notag\\
13 \mathrel{\mathop{\rightleftarrows}^{\lambda_{13}}_{\lambda_{23}}} 23\notag\\
\ea
which can be described by mass-action kinetics (\ref{eq:steq},\ref{eq:sf}). $k^+_{12}$, $k^+_{13}$, $k^+_{23}$ are complex formation rate constants, $k^-_{12}$,  
$k^-_{13}$, $k^-_{23}$ are complex dissociation rates and $\lambda_{12}$, $\lambda_{21}$, $\lambda_{13}$, $\lambda_{23}$ are unary reaction rates. For our numerical examples their values are set to $k^+_{12} = 1$, $k^-_{12} = 2$, $k^+_{13} = 1$, $k^-_{13}= 2$, $k^+_{23} = 2$, $k^-_{23} = 3$, $\lambda_{12}=0.5$, $\lambda_{21}=1$, $\lambda_{13}=0.5$, $\lambda_{23}=1$, which gives the steady state values $y_1 = 0.5$, $y_2 = 0.2$, $y_{12} = 0.05$, $y_3 = 9.5$, $y_{13} = 2.4$, $y_{23} = 1.25$.

\section{Effective 3GA equations solver with effective noise}
\label{sec:eff_solver_3GA}
In this appendix, we discuss the numerical solution of the 3GA subnetwork equations with effective noise from the bulk. For an analogous discussion on the numerical implementation of projected equations we refer to \cite{rubin_thesis}. The numerical solution of the stochastic integro-differential equations \eqref{eq:nonlin3GA} can be simplified by mapping every integral term into the solution of ordinary stochastic differential equations. The subnetwork reduced dynamics has the form
\be
\label{eq:eqsub}
\frac{d}{dt} \delta\bm{x}^{\rm s^{\T}}(t) = \delta\bm{x}^{\rm s^{\T}}(t)\bm{L}^{\rm ss} + 
\big(\delta\bm{x}^{\rm s^{\T}}(t)\circ\delta\bm{x}^{\rm s^{\T}}(t))\bm{L}^{\rm ss,s}
+\mathcal{\bm{M}}^{\T}(t) + \bm{\chi}^{\T}(t)
\ee
In the derivation of the effective noise $\bm{\chi}(t)$ we have rewritten this quantity (see Eq.~\eqref{eq:chi1}) in terms of $\bm{\lambda}(t)=\bm{E}_{\rm bb}^{\T}(t)\delta\bm{x}^{\rm b}(0)+\int_{0}^t dt'\,\bm{E}_{\rm bb}^{\T}(t-t')\bm{\xi}_0^{\rm{b}}(t')$ (Eq.~\eqref{eq:deflam}), which is the overall linear noise coming out of the bulk, both from initial bulk fluctuations $\delta\bm{x}^{\rm b}(0)$ and stochastic fluctuations $\bm{\xi}^{\rm{b}}(t)$. Thus, to implement the colored noise, we need auxiliary equations also for $\bm{\lambda}(t)$. In this process it is expedient to combine $\bm{\lambda}(t)$ with the contribution needed to implement the memory, $ \bm{\p}(t)$, see \cite{gaussianvar} and Sec.\ II in SM. We define $\bm{\g}_0(t)$ as $\bm{\g}_0^{\T}(t) = 
\int_{0}^{t} dt' \big[\delta \bm{x}^{\rm s^{\T}}(t')\bm{L}^{\rm sb}+\bm{\xi}_0^{\rm b^{\T}}(t')]\bm{E}_{\rm bb}(t-t')
+\delta\bm{x}^{\rm b^{\T}}(0)\ebb(t) = \bm{\p}(t) + \bm{\lambda}(t)$ and $\bm{\g}_1(t)$ as $\bm{\g}_1^{\T}(t) =
\int_{0}^{t} dt'\bm{\xi}_1^{\rm b^{\T}}(t')\bm{E}_{\rm bb}(t-t')$. Their sum $\bm{\g}(t) = \bm{\g}_0(t)+\bm{\g}_1(t) = \bm{\p}(t) + \bm{\lambda}(t) + \bm{\g}_1(t)$ is a vector of size $N^{\rm b}$ (number of bulk species) which can be obtained by solving the $N^{\rm b}$ differential equations
\be
\label{eqg}
\frac{d\bm{\g}^{\T}(t)}{dt}= \bm{\g}^{\T}(t)\bm{L}^{\rm bb} + \delta\bm{x}^{\rm s^{\T}}\bm{L}^{\rm sb}(t)
+\bm{\xi}_0^{\rm b^{\T}}(t) + \bm{\xi}_1^{\rm b^{\T}}(t)
\ee
with initial condition $\bm{\g}(0)=\delta\bm{x}^{\rm b}(0)$.
The evolution of $\bm{g}$ is then the linearized bulk dynamics with an additive part $\langle \bm{\xi}_0^{\rm{b}}(t) \bm{\xi}_0^{\rm{b}^{\T}}(t')\rangle = \bm{\Sigma}^{\rm bb}_0\delta(t-t')$ and a multiplicative part $\langle \bm{\xi}_1^{\rm{b}}(t) \bm{\xi}_1^{\rm{b}^{\T}}(t')\rangle = 
\bm{\Sigma}^{\rm bb}_1(\delta\bm{x}^{\rm s}, \delta\bm{x}^{\rm b}=\bm{g})\delta(t-t')$. The non-Markovian part of the effective subnetwork equation \eqref{eq:eqsub} can be written as
\be
\label{intimpl}
\begin{split}
\mathcal{\bm{M}}^{\T}(t)+ \bm{\chi}^{\T}(t) = &\,\bm{g}^{\T}(t)\bm{L}^{\rm bs} + \bm{\xi}_0^{\rm s^{\T}}(t) + \bm{\xi}_1^{\rm s^{\T}}(t)+
\big(\delta\bm{x}^{\rm s}(t)\circ\bm{g}_0(t)\big)^{\T}\bm{L}^{\rm sb,s}+\\
&\int_0^t dt'\left[\big(\delta\bm{x}^{\rm s}(t')\circ\bm{g}_0(t')\big)^{\T}\bm{L}^{\rm sb,b}+
\big(\bm{g}_0(t')\circ\bm{g}_0(t')\big)^{\T}\bm{L}^{\rm bb,b}\right]\ebb(t-t')\bm{L}^{\rm bs}
\end{split}
\ee 
The $\bm{\p}(t)$-part of $\bm{\g}_0(t)$ contributes to the memory $\mathcal{\bm{M}}(t)$ (as explained in \cite{gaussianvar}),
while the contributions from $\bm{\lambda}(t)$ and
$\bm{\g}_1(t)$, along with the additional terms in $\bm{\xi}_0^{\rm s^{\T}}(t)$ and $\bm{\xi}_1^{\rm s^{\T}}(t)$, give rise to the subnetwork colored noise $\bm{\chi}(t)$. The time integral left in the last term of \eqref{intimpl} can be evaluated by introducing \emph{ad hoc} auxiliary variables, 
similarly to the strategy adopted in \cite{katy,gaussianvar}.
We diagonalize $\bm{L}^{\rm bb}$ and decompose the exponential kernel $\ebb(t-t') = e^{\bm{L}^{\rm bb}(t-t')}$ into a superposition of pure exponentials
\be
\begin{split}
&\int_0^t dt'\left[\big(\delta\bm{x}^{\rm s}(t')\circ\bm{g}_0(t')\big)^{\T}\bm{L}^{\rm sb,b}+
\big(\bm{g}_0(t')\circ\bm{g}_0(t')\big)^{\T}\bm{L}^{\rm bb,b}\right]e^{\bm{L}^{\rm bb}(t-t')}\bm{L}^{\rm bs}=\\
&=\int_0^t dt' \left[\big(\delta\bm{x}^{\rm s}(t')\circ\bm{g}_0(t')\big)^{\T}\bm{L}^{\rm sb,b}+
\big(\bm{g}_0(t')\circ\bm{g}_0(t')\big)^{\T}\bm{L}^{\rm bb,b}\right] \sum_{\beta=1}^{N^{\rm b}}\bar{\bm{\rho}}_\beta \bar{\bm{l}}_\beta^{\T}
e^{\mu_\beta(t-t')}\bm{L}^{\rm bs}
\end{split} 
 \ee 
where the $\mu_\beta$ are the $N^{\rm b}$ eigenvalues of $\bm{L}^{\rm bb}$, $\bar{\bm{\rho}}_\beta$ and $\bar{\bm{l}}_\beta$ are respectively its right and left eigenvectors and give the coefficients of this decomposition. 
To evaluate the time  integral we can now solve a differential equations for each component of the vector $\bm{z}(t)=\lbrace z^\beta(t) \rbrace$, $\beta=1,...,N^{\rm b}$
\be
\label{eqz}
\frac{d}{dt}z^{\beta}(t) = \mu_\beta \,z^{\beta}(t) + \left[\big(\delta\bm{x}^{\rm s}(t')\circ\bm{g}_0(t')\big)^{\T}\bm{L}^{\rm sb,b}+
\big(\bm{g}_0(t')\circ\bm{g}_0(t')\big)^{\T}\bm{L}^{\rm bb,b}\right]\bar{\bm{\rho}}_\beta \qquad z^{\beta}(0)=0
\ee
In terms of these auxiliary variables we can finally rewrite  \eqref{eq:eqsub} as
\be
\label{eq:eqsub_fin}
\begin{split}
\frac{d}{dt} \delta\bm{x}^{\rm s^{\T}}(t) =& \, \delta\bm{x}^{\rm s^{\T}}(t)\bm{L}^{\rm ss} + \big(\delta\bm{x}^{\rm s}(t)
\circ\delta\bm{x}^{\rm s}(t)\big)^{\T}\bm{L}^{\rm ss,s}+
\bm{g}^{\T}(t)\bm{L}^{\rm bs}+ \bm{\xi}_0^{\rm s^{\T}}(t) + \bm{\xi}_1^{\rm s^{\T}}(t) \\
&+\big(\delta\bm{x}^{\rm s}(t)\circ\bm{g}_0(t)\big)^{\T}\bm{L}^{\rm sb,s} + 
 \sum_{\beta=1}^{N^{\rm b}}z^{\beta}(t) \bar{\bm{l}}_\beta^{\T}\bm{L}^{\rm bs} 
 \end{split}
\ee 
As in \cite{gaussianvar}, solving the $N^{\rm s}$ subnetwork equations is equivalent to solving a system with $2 N^{\rm b}$ additional equations, $N^{\rm b}$ for $\bm{g}(t)$ (see Eq.~\eqref{eqg}) and $N^{\rm b}$ for $\bm{z}(t)$ (see Eq.~\eqref{eqz}). For projection methods \cite{katy}, on the other hand, the analogous reduction from memory integrals to differential equations requires a number of additional variables equal to the dimension of the bulk variable space with all the possible concentration products,  which scales as $(N^{\rm b}\times N^{\rm b})+(N^{\rm s}\times N^{\rm b})$.

\subsection{$\epsilon_0$-component of $\bm{\chi}(t)$.}
\label{sec:eff_solver_3GA_chi}
In the regime $\epsilon \to 0$, i.e.\ where the extrinsic noise arises only from the uncertainty on bulk initial conditions, one uses only the $\delta\bm{x}^{\rm b}(0)$ part of $\bm{\lambda}(t)$ and sets $\bm{g}_1=0$ (so $\bm{g}=\bm{g}_0$). Next, the resulting $\bm{\chi}(t)$ can then be further approximated by short-time and long-time parts \eqref{eq:approx_chi}, after having isolated the 3GA colored noise $\bm{\chi}(t)$ from the memory. To do so, we solve the r.h.s. of \eqref{intimpl} either with the full $\bm{g}(t)$ or with $\bm{g}(t)=\bm{\p}(t)$ (i.e.\ without the sources of randomness $\delta\bm{x}^{\rm b}(0)$ appearing in $\bm{\lambda}$) and we take as $\bm{\chi}(t)$ the difference between these two solutions; note that in both of these solutions, $\delta \bm{x}^{\rm s}(t)$ is evaluated at the nonzero $\delta\bm{x}^{\rm b}(0)$.

\subsection{$\epsilon$-component of $\bm{\chi}(t)$.}
\label{sec:eff_solver_3GA_eps} 
In the presence of stochastic fluctuations 
one notices that the variables $\bm{g}$ and $\bm{g}_0$ appearing in  \eqref{eq:eqsub_fin} become distinct. To avoid having to calculate these variables separately, one can approximate in one of two ways: (i) use $\bm{g}(t)$ everywhere; this gives the full 3GA approximation with multiplicative noise source but implicitly adds within the colored noise $\bm{\chi}_1(t)$ some terms $\sim (\delta\bm{x}^{\rm s})^{3/2}$ of higher order than we have kept; (ii) use $\bm{g}_0(t)$ everywhere. The later option is equivalent to dropping the multiplicative noise terms $\bm{\xi}_1^{\rm b}$ in \eqref{eqg}. One should then consistently also neglect $\bm{\xi}_1^{\rm s}$ in \eqref{eq:eqsub_fin} and this defines what we have called the ``3GA with additive noise source'' in Tab.~\ref{table:tableapp}.

The noise can be further approximated by either 3GA-$\bm{\chi}_0$ or 3GA-$\bm{\chi}_0^{w}$ (see Sec.~\ref{sec:approx_chi} and Tab.~\ref{table:tableapp}). To implement the former, we neglect $\bm{\xi}_1^{\rm s}$ in \eqref{eq:eqsub_fin} and introduce a new variable $\bm{g}_0^{*}$ evolving according to the deterministic version of \eqref{eqg}; we then replace $\bm{g}_0$ by $\bm{g}_0^{*}$ in the nonlinear terms in \eqref{eq:eqsub_fin} with coefficients $\bm{L}^{\rm sb,s}$ and similarly in \eqref{eqz} with coefficients $\bm{L}^{\rm sb,b}$, $\bm{L}^{\rm bb,b}$. This ensures that the nonlinear memory is preserved but the effective noise appears only through linear terms $\bm{L}^{\rm bs}$. For the 3GA-$\bm{\chi}_0^{w}$ approximation we use the same set of equations where $\bm{g}_0^{*}$ appears also in the term multiplied by $\bm{L}^{\rm bs}$ in \eqref{eq:eqsub_fin}, so that propagation of noise from the bulk is excluded also in this linear term; instead we explicitly add the linear white noise $\bm{\chi}_0^{w}$ from \eqref{eq:chi0int} to \eqref{eq:eqsub_fin}. Without initial bulk fluctuations, one needs simply to drop the corresponding contributions to $\bm{\lambda}$ and $\bm{g}_0$. 

\end{appendices}

\cleardoublepage
\bibliography{mybib}
\cleardoublepage

\input{main_SI}

\end{document}

%% file: main_SI.tex
\setcounter{page}{0}
\setcounter{section}{0}
\pagenumbering{arabic}

\begin{center}\LARGE{Supplementary Material: \vspace{0.5cm} \\
Systematic model reduction captures the dynamics of extrinsic noise in biochemical subnetworks}\end{center}
\vspace{0.5cm}
\begin{center} Barbara Bravi$^1$, Katy J. Rubin$^2$, Peter Sollich$^{2,3}$ \end{center}

\vspace{0.5cm}
\begin{center}
\emph{$^1$ Institute of Physics, \'Ecole Polytechnique F\'ed\'erale de Lausanne, CH-1015 Lausanne, Switzerland\\
Current affiliation: Laboratoire de Physique de l'Ecole Normale Sup\'erieure, ENS, Universit\'e PSL, CNRS, Sorbonne Universit\'e, Universit\'e de Paris, F-75005 Paris, France}\\
\emph{$^2$Department of Mathematics, King's College London, Strand, London WC2R 2LS, UK}\\
\emph{$^3$Institute for Theoretical Physics, Georg-August-University G\"ottingen, Friedrich-Hund-Platz 1, 37077 G\"ottingen, Germany}
\end{center}

\cleardoublepage

\beginsupplement

\addtocounter{equation}{-1}

\section{Memory and random force in projection methods}
\label{sec:derivation_PM}

In this section we explain the model reduction approach to unary and binary biochemical reactions based on projection methods derived in \cite{katy}. The starting point is the Fokker-Planck equation associated to the Langevin dynamics \eqref{eq:steq} in the main text describing the time evolution of $P(\bm{x}, t)$, the probability of a certain set of concentrations $\bm{x}$ at time $t$
\be
\label{eq:fp}
\frac{\partial}{\partial t} P(\bm{x}, t) = - \frac{\partial}{\partial \bm{x}} \left( \bm{S} \bm{f} P \right) 
+ \frac{1}{2}\frac{\partial^2}{\partial \bm{x}^2} \left( \bm{\Sigma} P \right) = \mathcal{L}^{\rm{T}} P(\bm{x}, t)
\ee
which can be written in terms of $\mathcal{L}^{\rm T}$, the so-called adjoint Fokker-Planck operator.

For a linearized dynamics as \eqref{eq:lin} in the main text, one finds \cite{katy} that projection operators as well as $\mathcal{L}$ can be represented by matrices with a simple block structure; for $\mathcal{L}$ the blocks are the coefficient matrices featuring in (\ref{eqdxs}, \ref{eqdxb}) of the main text. If one applies the projection approach to linearized dynamics one obtains \eqref{eq:linearreduced} (main text), suitably averaged so that the noise term disappears. 

For nonlinear projected equations, one projects the dynamics onto a space of concentrations and products of concentrations; let us call $z_{\alpha}$ the vector formed by concatenating these observables. The adjoint Fokker-Planck operator $\mathcal{L}$ can then be written in matrix form such that \be
\label{eq:matrep}
\frac{d}{dt} z_{\alpha} = \sum_{\beta} z_{\beta} \bm{L}_{\beta \alpha} + \delta x^3 + \mathcal{O}(\epsilon) 
\ee
i.e.\ $\bm{L}$, multiplied by the vector of observables from the left, determines their time evolution. We shall emphasize that this matrix representation captures neither cubic terms nor ones of $O(\epsilon)$; the latter can be dropped in any case by restricting attention to the limit $\epsilon \to 0$.
$\bm{L}_{\beta \alpha}$ here is a matrix in an ``enlarged'' space, i.e.\ $\beta$ and $\alpha$ can refer to both single concentrations or products, and again has a block structure determined by whether $\beta$ and $\alpha$ belong to the subnetwork or the bulk. It reads explicitly as follows
\[\bm{L}=\begin{pmatrix}
\bm{L}^{\rm ss} & \bm{L}^{\rm sb}  & \mathbb{0}  & \mathbb{0}  & \mathbb{0}  \\
\bm{L}^{\rm bs} & \bm{L}^{\rm bb}  & \mathbb{0}  & \mathbb{0}  & \mathbb{0}  \\
\bm{L}^{\rm ss,s} & \bm{L}^{\rm ss,b}  & \bm{L}^{\rm ss,\rm ss}  & \bm{L}^{\rm ss,\rm sb}  & \bm{L}^{\rm ss,\rm bb}  \\
\bm{L}^{\rm sb,s} & \bm{L}^{\rm sb,b}  & \bm{L}^{\rm sb,\rm ss}  & \bm{L}^{\rm sb,\rm sb}  & \bm{L}^{\rm sb,\rm bb}  \\
\bm{L}^{\rm bb,s} & \bm{L}^{\rm bb,b}  & \bm{L}^{\rm bb,\rm ss}  & \bm{L}^{\rm bb,\rm sb}  & \bm{L}^{\rm bb,\rm bb}  \\
\end{pmatrix}\]
i.e.\ it consists of 5 block rows and columns, referring to linear subnetwork concentrations ($\rm s$), subnetwork concentration products ($\rm ss$), mixed subnetwork-bulk products ($\rm sb$), which are taken as part of the bulk subspace, and finally bulk products ($\rm bb$). The dynamics for the products is encoded in the bottom right blocks ($\bm{L}^{\rm ss,\rm ss}$, $\bm{L}^{\rm ss,\rm sb}$ etc.) and its coefficients are taken from the linearized dynamics via
\be
\frac{d}{dt}(\delta x_i \delta x_j)= \delta x_j \frac{d}{dt} \delta x_i + \delta x_i \frac{d}{dt} \delta x_j 
\ee
As a consequence, only products of concentrations (rather than linear concentration variables) feature in the evolution of a product, hence the top right block has all zero elements; also $\bm{L}^{\rm ss,\rm bb}=\bm{L}^{\rm bb,\rm ss} = 0$. Genuine nonlinearities appear in the bottom left blocks ($\bm{L}^{\rm ss,s}$, $\bm{L}^{\rm ss,b}$ etc.), containing the coefficients multiplying products in the equations for linear subnetwork variables. Via these blocks, nonlinearities are represented as a linear coupling between a concentration and products of two concentrations. The assumptions we made on what reactions between the bulk and the subnetwork can occur (see Sec.~\ref{sec:setup} in the main text) imply that $\bm{L}^{\rm ss,b}=\bm{L}^{\rm bb,s}=\mathbb{0}$.

Another essential ingredient of the projection approach is the definition of two operators, $\bm{P}$ and $\bm{Q}$, with $\bm{P}$ projecting onto the subnetwork subspace and $\bm{Q}$ the orthogonal space. These projection operators read in block form
 \[\bm{P}=\begin{pmatrix}
\mathbb{1}   & \mathbb{0}  & \mathbb{0}  & \mathbb{0}  & \mathbb{0}  \\
\mathbb{0}   & \mathbb{0}  & \mathbb{0}  & \mathbb{0}  & \mathbb{0} \\
\mathbb{0}  & \mathbb{0}  &\mathbb{1}   & \mathbb{0}  & \mathbb{0} \\
\mathbb{0}   & \mathbb{0}  & \mathbb{0}  & \mathbb{0}  & \mathbb{0} \\
\mathbb{0}   & \mathbb{0}  & \mathbb{0}  & \mathbb{0}  & \mathbb{0} \\
\end{pmatrix} \qquad \bm{Q}=\begin{pmatrix}
\mathbb{0}   & \mathbb{0}  & \mathbb{0}  & \mathbb{0}  & \mathbb{0}  \\
\mathbb{0}   & \mathbb{1}  & \mathbb{0}  & \mathbb{0}  & \mathbb{0} \\
\mathbb{0}  & \mathbb{0}  &\mathbb{0}   & \mathbb{0}  & \mathbb{0} \\
\mathbb{0}   & \mathbb{0}  & \mathbb{0}  & \mathbb{1}  & \mathbb{0} \\
\mathbb{0}   & \mathbb{0}  & \mathbb{0}  & \mathbb{0}  & \mathbb{1} \\
\end{pmatrix}\]

The goal of the projection method is to find explicit expressions for the memory function and the random force for the full nonlinear dynamics, by
exploiting the above representation of the dynamics in terms of matrices. In particular one needs \cite{katy} the combinations of matrices $\bm{PL}$, $\bm{QLQ}$ and $\bm{QL}$. $\bm{QLQ}$ has a lower triangular block structure, thus $\bm{E}(t)=e^{\bm{QLQ}t}$ has the same structure with diagonal blocks that are the exponentials of those in $\bm{QLQ}$, i.e.\ for example $[e^{\bm{QLQ}t}]_{\rm bb}=\ebb(t)=e^{\bm{L}^{\rm bb}t}$. \red{Given the action of the bulk projector $\bm{Q}$ both on the right and on the left, the non-zero elements of $\bm{E}(t)$ are only the ones with both indices referring to bulk species}. The intra-subnetwork dynamics is extracted from the top left block (the one associated with subnetwork indices only) of $\bm{PL}$ and consists of a linear part with coefficients $\bm{L}^{\rm ss}$ and a nonlinear part $\bm{L}^{\rm ss,s}$.

The memory functions come from the $\rm ss$ blocks of $\bm{P}\bm{L}\bm{Q} \,e^{\bm{Q}\bm{L}\bm{Q} t} \bm{Q}\bm{L}$; the linear memory function is given by \eqref{eq:mem0} (main text) while the nonlinear one reads
 \begin{equation}
  \label{eq:memProj}
 \bm{M}_{\rm P}^{\rm ss,s}(t-t')=\bm{L}^{\rm ss,\rm sb}\bm{E}_{\rm sb,b}(t-t')\bm{L}^{\rm bs} + \bm{L}^{\rm ss,sb} \bm{E}_{\rm sb,\rm sb}(t-t') 
 \bm{L}^{\rm sb,s}  
 \end{equation}
Finally the random force is given by the $\rm s$-entries of $\bm{z}^{\rm b^{\T}}(0) e^{\bm{Q}\bm{L}\bm{Q} t} \bm{Q}\bm{L}$. The components of $\bm{z}^{\rm b^{\T}}(0)$ that relate to single bulk concentrations give the contribution $\bm{r}_0(t)$ (see \eqref{eq:rf0} in the main text), while the bulk product observables in $\bm{z}^{\rm b^{\T}}(0)$ give the nonlinear part $\bm{r}_1(t)$ of the random force (see \eqref{eq:rf1} in the main text). The nonlinear deterministic part, i.e.\ memory function and subnetwork drift, in contrast to the non-linear random force $\bm{r}_1(t)$, can be calculated in closed form. This is because (only) the two former quantities are defined by products of operators with a projector $\bm{P}$ on the left, which then removes the $O(\delta x^3)$ terms in \eqref{eq:matrep}.

\cleardoublepage
\section{Path integral approach 3GA}
\label{sec:Gau_app}

In this section we extend the 3GA approach developed in \cite{gaussianvar} to the presence of multiplicative noise; for the sake of completeness, we give an overview of the entire derivation. The Langevin dynamics \eqref{eq:steq} (main text) can be rewritten vectorially as 
\begin{equation}
\label{eq:steqvec}
\frac{d}{dt}\bm{x}(t) = \bm{S}\bm{f}(\bm{x}(t)) +\bm{\xi}(t) 
\end{equation}
We can study the time-dependent statistics of protein concentrations by appeal to the Martin--Siggia--Rose--Janssen--De Dominicis (MSRJD)
formalism \cite{martin,janssen,dedominicis,kamenev,PathMethods,rabello}. The starting point is to work formally with $P(\bm{x})$, the normalized probability distribution of protein 
concentration {\em paths}, i.e.\ the distribution over the set of stochastic temporal trajectories $\bm{x}=\lbrace x_i(t) \rbrace$ generated by \eqref{eq:steqvec}. (The notation here is different from that in the previous section, where $\bm{x}$ referred to a concentration vector at a single time.) The essence of the formalism is to introduce a set of ``response'' variables $\hat{\bm{x}}$ that 
allow one to write a joint measure over trajectories $P(\bm{x}, \hat{\bm{x}})$ as
\be
\label{msrjd}
P(\bm{x}, \hat{\bm{x}})=\int D\bm{\xi} P (\bm{\xi}) e^{\text{i}\hat{\bm{x}}^{\T}(d\bm{x}/dt - \bm{Sf}-\bm{\xi})} = \frac{e^{\mathcal{H}}}{Z}
\ee
where $Z \equiv 1$ by construction (see \cite{gaussianvar}) and $P(\bm{\xi})$ is the Gaussian distribution of the noise, with zero mean and covariance \eqref{eq:fluct} given in the main text. The integration over this Gaussian noise can be performed by appealing to
the Hubbard-Stratonovich identity
\begin{equation}
\label{hs}
\int D\bm{\xi} P (\bm{\xi}) e^{\pm\text{i} \hat{\bm{x}}^{\T} \bm{\xi}} =
e^{-\frac{1}{2}\bm{\hat{x}}^{\T}\bm{\Sigma}\bm{\hat{x}}}
\end{equation}
and the integration measure should be understood as $D\bm{\xi}= \prod_{i}d\xi_i(t)$. As highlighted in \eqref{msrjd}, in this way the
probability over paths is recast as the exponential of an ``action'' $\mathcal{H}$
\be
\label{action}
\mathcal{H}= \int_0^{T} dt\Bigg\lbrace \sum_{i}\text{i}\hat{x}_i(t)\bigg(\frac{d}{dt} x_i(t)-[\bm{S}\bm{f}(\bm{x}(t))]_i\bigg)
+\frac{1}{2}\sum_{ij} \text{i} \hat{x}_i(t)\bm{\Sigma}_{ij}(\bm{x}(t)) \text{i} \hat{x}_j(t) \Bigg\rbrace 
\ee
which makes it amenable to the application of field-theoretic perturbation theory \cite{kleinert}. We observe a simple ``rule'' in the MSRJD formalism: 
whatever multiplies $\text{i} \hat{x}_i(t)$ gives the deterministic part of the dynamics (drift term) while 
the quadratic part in $\hat{x}$ (and containing $\bm{\Sigma}$) defines the noise or diffusion term.

The above representation of the dynamics is convenient for integrating out the entire time 
evolution of the bulk via a path integral. This process yields the marginal distribution over subnetwork 
variables only, $\bm{x}^{\rm s}$ and $\hat{\bm{x}}^{\rm s}$, which encodes the subnetwork dynamics via an ``effective'' (i.e.\ reduced to the subnetwork) action $\mathcal{H}_{\text{eff}}$ defined by
\be
\int D\bm{x}^{\rm b}D\bm{\hat{x}}^{\rm b} e^{\mathcal{H}} = e^{\mathcal{H}_{\text{eff}}}
\ee
The notation for the path integral $D\bm{x}^{\rm b}\,D\bm{\hat{x}}^{\rm b}$ is a shorthand for $\prod_{it}dx_i(t)d\hat{x}_i(t)/(2\pi)$ 
and means that formally the integration is over all possible trajectories for $\bm{x}^{\rm b}$ and $\hat{\bm{x}}^{\rm b}$. We resort to perturbation theory to make this marginalization analytically tractable. 
Let us decompose the action into a ``non-interacting'' part $\mathcal{H}_0$, \red{containing only terms quadratic in all the variables (hence drift and noise covariance of the linearized dynamics, \eqref{eq:lin}-\eqref{eq:effdiffusiony} in the main text)} and an interacting one, containing higher powers, which will be treated as a perturbation $\Delta\mathcal{H}$
\begin{equation}
\label{eq:act_pert}
 \mathcal{H}=\mathcal{H}_0+\Delta\mathcal{H}
\end{equation}
The effective action for the subnetwork dynamics becomes then, via a standard perturbation expansion,
\begin{eqnarray}
\label{eq:effactpert}
&&\mathcal{H}_{\text{eff}}=\ln {\int D\bm{x}^{\rm b}D\bm{\hat{x}}^{\rm b} e^{\mathcal{H}_0 + \Delta \mathcal{H}}} \\
&&=\ln \left(\int D\bm{x}^{\rm b}D\bm{\hat{x}}^{\rm b} e^{\mathcal{H}_0}\left(1+\frac{\int D\bm{x}^{\rm b}D\bm{\hat{x}}^{\rm b} 
e^{\mathcal{H}_0}\Delta \mathcal{H}}{\int D\bm{x}^{\rm b}D\bm{\hat{x}}^{\rm b} e^{\mathcal{H}_0}}+ 
\mathcal{O}(\Delta \mathcal{H})^2\right)\right)\notag\\
&&=\left(\ln\int D\bm{x}^{\rm b}D\bm{\hat{x}}^{\rm b} e^{\mathcal{H}_0}\right)+\int D\bm{x}^{\rm b}D\bm{\hat{x}}^{\rm b}
Q_0(\bm{x}^{\rm b},\bm{\hat{x}}^{\rm b}|\bm{x}^{\rm s},\bm{\hat{x}}^{\rm s})\Delta \mathcal{H}+
\mathcal{O}(\Delta \mathcal{H})^2
\notag
\end{eqnarray}
The average in the last line is over  $Q_0(\bm{x}^{\rm b},\bm{\hat{x}}^{\rm b}|\bm{x}^{\rm s},\bm{\hat{x}}^{\rm s}) =
e^{\mathcal{H}_0}/\int D\bm{x}^{\rm b}D\bm{\hat{x}}^{\rm b} e^{\mathcal{H}_0}$, the probability distribution of bulk trajectories conditioned on a fixed 
subnetwork time evolution. This distribution is the ratio of $e^{\mathcal{H}_0}=Q_0(\bm{x}^{\rm b},\bm{\hat{x}}^{\rm b},\bm{x}^{\rm s}, \bm{\hat{x}}^{\rm s})$ and 
its marginal $\int  D\bm{x}^{\rm b}D\bm{\hat{x}}^{\rm b} e^{\mathcal{H}_0}=Q_0(\bm{x}^{\rm s},\bm{\hat{x}}^{\rm s})$, both of which are Gaussian, so is Gaussian itself. 
These Gaussian distributions can be chosen in such a way as to retrieve the linearized dynamics of Sec.~\ref{sec:lin} in the main text, centered 
around the deterministic trajectory at steady state. The perturbative expansion then allows us to go beyond the linear dynamics (Gaussian Approximation) in a controlled way up to e.g.\ cubic terms; this is what we define as the 3rd order Gaussian Approximation, 3GA. To evaluate the required perturbative correction we need to consider subnetwork variables as fixed and work out the corresponding conditional bulk probabilities. The perturbation terms in the action that we keep are the third order terms in $\Delta \mathcal{H}$, which explicitly read
\begin{equation}
\label{eq:deltah}
\begin{split}
\fl \Delta \mathcal{H} = \int_0^{T} dt\Bigg\lbrace&\sum_{i}\text{i}\hat{x}_i(t) \Big[\sum_{j,l,j \neq l} k^{+}_{ij,l}
\delta x_i(t) \delta x_j(t)-\frac{1}{2}\sum_{j,l,j\neq l}k^{+}_{jl,i}\delta x_j(t)\delta x_l(t) + \\
&\fl \sum_{l}k^{+}_{ii,l}\delta x_i(t) \delta x_i(t) -\frac{1}{2}\sum_{j}k^{+}_{jj,i}\delta x_j(t)\delta x_j(t)\Big]
+ \frac{1}{2}\sum_{ij} \text{i} \hat{x}_i(t)\bm{\Sigma}_{ij}(\delta\bm{x}(t)) \text{i} \hat{x}_j(t)\Bigg\rbrace
\end{split}
\end{equation}
For ease of analysis, we divide \eqref{eq:deltah} into 
\be
\label{eq:deltah2}
\Delta \mathcal{H} = \Delta \mathcal{H}^D + \Delta \mathcal{H}^{\Sigma}
\ee
where $\Delta \mathcal{H}^D$ contains the piece contributing to the drift and $\Delta \mathcal{H}^{\Sigma}$ the piece for the noise. We switch to the vectorial notation of the main text (which allows us to separate species belonging to bulk and subnetwork)
 \begin{eqnarray}
\label{deltaH}
&& \fl \Delta\mathcal{H}^D= \int_0^{T} dt\bigg\lbrace\bigg[\big(\delta\bm{x}^{\rm s}\circ \delta\bm{x}^{\rm s}\big)^{\T}\bm{L}^{\rm  ss, s}+ 
\big(\delta\bm{x}^{\rm s}\circ\delta\bm{x}^{\rm b}\big)^{\T}\bm{L}^{\rm  sb, s}
+\big(\delta\bm{x}^{\rm b}\circ \delta\bm{x}^{\rm b}\big)^{\T}\bm{L}^{\rm  bb, s}\bigg]\text{i}\hat{\bm{x}}^{\rm s}+\notag\\
&&\fl \qquad\qquad\qquad\bigg[\big(\delta\bm{x}^{\rm s}\circ \delta\bm{x}^{\rm s}\big)^{\T}\bm{L}^{\rm  ss, b}+
\big(\delta\bm{x}^{\rm s}\circ\delta\bm{x}^{\rm b}\big)^{\T}\bm{L}^{\rm  sb, b}+\big(\delta\bm{x}^{\rm b}\circ
\delta\bm{x}^{\rm b}\big)\bm{L}^{\rm  bb, b}\bigg]\text{i}\hat{\bm{x}}^{\rm b}\bigg\rbrace 
 \end{eqnarray}
where $\circ$ indicates a ``flattened'' outer product and $\bm{L}^{\rm sb,s}$, $\bm{L}^{\rm sb,b}$, $\bm{L}^{\rm bb,b}, \bm{L}^{\rm  ss,b}, \bm{L}^{\rm  bb,s}$ are matrices collecting the relevant nonlinear couplings (see Sec.~\ref{sec:proj} of the main text). As explained in Sec.~\ref{sec:setup} of the main text, we do not consider all the possible reactions occurring between subnetwork and bulk, thus we set $\bm{L}^{\rm ss, b}=\bm{L}^{\rm bb,s}\equiv 0$. The variables $\delta\bm{x}^{\rm s}$ and $\delta\bm{x}^{\rm b}$ are defined as differences w.r.t.\ the marginal Gaussian means $\bm{\mu}^{\rm s}(t)$ and $\bm{\mu}^{\rm b}(t)$ while we have dropped $\delta$ from $\hat{\bm{x}}^{\rm s}$ and $\hat{\bm{x}}^{\rm b}$ as the marginal auxiliary means $\bm{\hat{\mu}}^{\rm s}=\bm{\hat{\mu}}^{\rm b} \equiv 0$, as a consequence of the normalization condition \cite{PathMethods,rabello}. Note that the normalization condition implies also that correlations of auxiliary variables $\langle \hat{\bm{x}}^{\rm T, \cdot} \hat{\bm{x}}^{\cdot}\rangle \equiv 0$ ($\cdot$ = s,b); in addition, one has vanishing equal time responses $\bm{R}^{\cdot \cdot}(t,t)  = \langle \hat{\bm{x}}^{\rm T, \cdot} \bm{x}^{\cdot}\rangle$ as a consequence of It\^{o} discretization \cite{PathMethods, vankampenito}. 
Note that the second term in \eqref{eq:deltah2} was disregarded in our previous work \cite{gaussianvar} by taking the limit $\epsilon \to 0$; in the present work we extend the derivation to include it. It reads
\begin{eqnarray}
\label{deltaHn}
&& \Delta\mathcal{H}^{\Sigma}= \frac{1}{2}\int_0^{T} dt\bigg\lbrace \text{i} \hat{\bm{x}}^{\rm s^{\T}} \bm{\Sigma}_1^{\rm ss} \text{i} \hat{\bm{x}}^{\rm s} +
\text{i} \hat{\bm{x}}^{\rm s^{\T}} \bm{\Sigma}_1^{\rm sb}\text{i} \hat{\bm{x}}^{\rm b} +
\text{i} \hat{\bm{x}}^{\rm b^{\T}} \bm{\Sigma}_1^{\rm bs} \text{i} \hat{\bm{x}}^{\rm s} +
\text{i} \hat{\bm{x}}^{\rm b^{\T}} \bm{\Sigma}_1^{\rm bb}\text{i} \hat{\bm{x}}^{\rm b}\bigg\rbrace
 \end{eqnarray}
where $\bm{\Sigma}_1^{\cdot \cdot}$, $\cdot = \rm{s},\rm{b}$ denotes the $\delta\bm{x}-$dependent part of noise correlations, see \eqref{eq:fluctdx} in the main text.

The first order correction in the effective action, as given by the second term in \eqref{eq:effactpert}, reduces to an average w.r.t.\ the bulk distribution conditional on the subnetwork $Q_0(\bm{x}^{\rm b},\bm{\hat{x}}^{\rm b}|\bm{x}^{\rm s},\bm{\hat{x}}^{\rm s})$. 
This average is equivalent to replacing combinations of bulk variables in \eqref{eq:deltah} by conditional Gaussian moments which, by Wick's theorem, can always be expressed in terms of means and correlations. The relevant statistics are here the conditional means $\delta\bm{\mu}^{\rm b|s}=\bm{\mu}^{\rm b|s} (t)-\bm{\mu}^{\rm b}(t)$, $\hat{\bm{\mu}}^{\rm b|s}$ and the conditional correlations $\bm{C}^{\rm bb|s}(t,t)$ given by \eqref{eq:bulk_corr} in the main text, hence
\begin{eqnarray}
\label{dsCompact}
\langle\Delta \mathcal{H}^D\rangle &=&
 \int_0^{T} dt \bigg\lbrace
 \bigg[\big(\delta\bm{x}^{\rm s}\circ\delta\bm{x}^{\rm s}\big)^{\T}\bm{L}^{\rm ss,s}
 +\big(\delta\bm{x}^{\rm s}\circ\delta\bm{\mu}^{\rm b|s}\big)^{\T}\bm{L}^{\rm sb,s}\bigg]\text{i}\hat{\bm{x}}^{\rm{s}}+\\
&&\quad \quad \quad\quad
\bigg[\big(\delta\bm{x}^{\rm s}\circ\delta\bm{\mu}^{\rm b|s}\big)^{\T}\bm{L}^{\rm sb,b}+
\big(\delta\bm{\mu}^{\rm b|s}\circ\delta\bm{\mu}^{\rm b|s}+\bm{C}^{\rm bb|s}(t,t)\big)^{\T}\bm{L}^{\rm bb,b}\bigg]
\text{i}\hat{\bm{\mu}}^{\rm b|s}\bigg\rbrace \notag
\end{eqnarray}
where one uses that the conditional average $\langle\delta \bm{x}^{\rm b}(t)\delta \bm{x}^{\rm b\, \it{T}}(t)\rangle=
\delta \bm{\mu}^{\rm b|s}(t)\delta \bm{\mu}^{\rm b|s\, \it{T}}(t)+\bm{C}^{\rm bb|s}(t,t)$. Furthermore
\begin{eqnarray}
\label{dsCompactn}
\langle\Delta \mathcal{H}^{\Sigma}\rangle =&
\frac{1}{2} \int_0^{T} dt \bigg\lbrace \text{i}\hat{\bm{x}}^{\rm s^{\T}}\bm{\Sigma}_1^{\rm ss}(\delta\bm{x}^{\rm s},\delta\bm{\mu}^{\rm b|s})\text{i}\hat{\bm{x}}^{\rm s} +
\text{i}\hat{\bm{x}}^{\rm s^{\T}}\bm{\Sigma}_1^{\rm sb}(\delta\bm{x}^{\rm s},\delta\bm{\mu}^{\rm b|s})\text{i}\hat{\bm{\mu}}^{\rm b|s}\notag\\
 &+\text{i}\hat{\bm{\mu}}^{\rm b|s^{\T}} \bm{\Sigma}_1^{\rm bs}(\delta\bm{x}^{\rm s},\delta\bm{\mu}^{\rm b|s})\text{i} \hat{\bm{x}}^{\rm s} +
\text{i} \hat{\bm{\mu}}^{\rm b|s^{\T}} \bm{\Sigma}_1^{\rm bb}(\delta\bm{x}^{\rm s},\delta\bm{\mu}^{\rm b|s})\text{i}\hat{\bm{\mu}}^{\rm b|s}
\bigg\rbrace
\end{eqnarray}
Here we have highlighted that, by integrating over the bulk, $\bm{\Sigma}_1$ is calculated in either $\delta\bm{x}^{\rm s}$ or in the bulk conditional means $\delta\bm{\mu}^{\rm b|s}$. In applying Wick's theorem we have used that the conditional correlator of auxiliary variables and the equal-time conditional response both vanish \cite{gaussianvar}, as in the full dynamics. We next substitute the expression for $\text{i}\hat{\bm{\mu}}^{\rm b|s}$
\begin{equation}
\label{condaux}
\ii\hat{\bm{\mu}}^{\rm b|s}(t)= \int_{t}^{T} dt' \,\bm{E}_{\rm bb}(t'-t)\bm{L}^{\rm bs}\text{i}\hat{\bm{x}}^{\rm s}(t') 
\end{equation}
and for the conditional mean $\delta\bm{\mu}^{\rm b|s}(t) = \bm{\nu}(t)+\bm{\hat{\nu}}(t)$
\be
\label{bulkmean}
\bm{\nu}(t) = \int_0^t dt' \ebb^{\T}(t-t')(\bm{L}^{\rm sb})^{\T}\delta\bm{x}^{\rm s}(t') \qquad
\bm{\hat{\nu}}(t) = -
\int_0^{T}dt'\bm{C}^{\rm bb|s}(t,t')\bm{L}^{\rm bs}\text{i}\bm{\hat{x}}^{\rm s}(t')
\ee
The reduced dynamics for $\delta\bm{x}^{\rm s}(t)$ (Eq.\ \eqref{eq:nonlin3GA} in the main text) includes a memory $\bm{M}^{\rm ss}(t,t')$ and a colored noise $\bm{\chi}(t)$ with covariance 
$\langle\bm{\chi}(t)\bm{\chi}^{\T}(t')\rangle=\bm{N}^{\rm ss}(t,t')$ to be determined. The deterministic part of such dynamics can be extracted 
from the action \eqref{eq:act_pert} by considering the terms multiplying $\text{i}\hat{\bm{x}}^{\rm s}(t)$, while the noise always enters the MSRJD formalism via quadratic terms in the 
auxiliary variables $\sim\hat{\bm{x}}^{\rm s}$, as made evident by \eqref{hs}. One can write $\bm{N}^{\rm ss}(t,t')=\bm{N}^{\rm ss}_{0}(t,t')+ \bm{N}^{\rm ss}_{1}(t,t')$ as the sum of a purely Gaussian term and a first order correction. Similarly the memory comprises two terms: $\bm{M}^{\rm ss}(t-t')$, which like
$\bm{N}^{\rm ss}_{0}(t,t')$ can be calculated starting from the quadratic part of the action and $\bm{M}^{\rm s,ss}(t,t',t'')$, which like 
$\bm{N}^{\rm ss}_{1}(t,t')$ contains the contributions of cubic terms treated by means of perturbation theory. A schematic summary of these construction steps is provided in Tab.\ \ref{table:tabledx}. Expressions for linear and nonlinear memory were analyzed in \cite{gaussianvar}. $\bm{N}^{\rm ss}_{0}(t,t')$ is the covariance of noise in the linearized dynamics $\bm{\chi}_0$ (see \eqref{eq:chi0} in the main text) and can be read off from the part of the action that is quadratic in $\hat{\bm{x}}^{\rm s}$, i.e.\ $\mathcal{H}_0$
\begin{eqnarray}
\label{eq:covL}
 \fl \bm{N}_0^{\rm ss}(t,t'')&=&\bm{\Sigma}_0^{\rm ss}\delta(t-t'') +
 \bm{\Sigma}_0^{\rm sb}\bm{L}^{\rm bs}\bm{E}_{\rm bb}(t-t'') +(\bm{L}^{\rm bs})^{\T}\bm{E}_{\rm bb}^{\T}(t-t'')\bm{\Sigma}_0^{\rm bs} +\\
 \fl && (\bm{L}^{\rm bs})^{\T}\bm{E}_{\rm bb}^{\T}(t)\bm{C}^{\rm bb}(0,0)\bm{E}_{\rm bb}(t'')\bm{L}^{\rm bs}+
\int_0^t dt'(\bm{L}^{\rm bs})^{\T}\bm{E}_{\rm bb}^{\T}(t-t')\bm{\Sigma}_0^{\rm bb}\bm{E}_{\rm bb}(t''-t')\bm{L}^{\rm bs}\notag
\end{eqnarray}
where bulk-subnetwork correlations in the initial conditions are assumed to be zero $\bm{C}^{\rm bs}(0,0)= \bm{C}^{\rm sb}(0,0) \equiv 0$ and $t<t''$. To 
illustrate the procedure for $\bm{N}^{\rm ss}_{1}(t,t')$, let us start from $\mathcal{H}^D$ and focus on the first term in the second line of \eqref{dsCompact}, which becomes
\begin{eqnarray}
\label{dsCompact1}
\fl  &&\int_0^{T}dt'\int_{t'}^{T} dt \,\big(\delta\bm{x}^{\rm s}(t')\circ\delta\bm{\mu}^{\rm b|s}(t')\big)^{\T} 
\bm{L}^{\rm sb,b}\bm{E}_{\rm bb}(t-t')\bm{L}^{\rm bs}\text{i}\hat{\bm{x}}^{\rm s}(t)=\\
\fl \qquad&&\int_0^{T}dt'\int_{t'}^{\T} dt\,
\bigg(\delta\bm{x}^{\rm s}(t')\circ \int_0^{t'} dt'' \ebb^{\T}(t'-t'')(\bm{L}^{\rm sb})^{\T}\delta\bm{x}^{\rm s}(t'') \notag\\
&&\qquad \qquad \qquad - \delta\bm{x}^{\rm s}(t')\circ \int_0^{T}dt''\bm{C}^{\rm bb|s}(t',t'')\bm{L}^{\rm bs}\text{i}
\bm{\hat{x}}^{\rm s}(t'')\bigg)^{\T} \bm{L}^{\rm sb,b}\bm{E}_{\rm bb}(t-t')\bm{L}^{\rm bs}\text{i}\hat{\bm{x}}^{\rm s}(t) \notag
\end{eqnarray}
The first of the resulting two terms contributes to the reduced subnetwork dynamics via a temporal integral defining a nonlinear memory term
$\bm{M}^{\rm s,ss}(t,t',t'')$ (see \cite{gaussianvar}). In the second term of \eqref{dsCompact1}, the factor multiplying $\text{i}\hat{\bm{x}}^{\rm s\,\it{T}}(t'')$ and $\text{i}\hat{\bm{x}}^{\rm s}(t)$, which is
\be
\int_0^t dt'\big(\delta\bm{x}^{\rm s}(t')\circ\bm{C}^{\rm bb|s}(t',t'')\bm{L}^{\rm bs}\big)^{\T}\bm{L}^{\rm sb,b}\bm{E}_{\rm bb}(t-t')\bm{L}^{\rm bs}\ ,
\ee
contributes to the covariance of the effective noise $\bm{N}^{\rm ss}_{1}(t,t')$ with an additional $\delta \bm{x}^{\rm s}$-dependence. Note that we have extended our 
circle product notation here to products of vectors and matrices: a term of the form
$(\bm{v}^{\rm s} \circ \bm{A}^{\rm bs})^{\T}\bm{L}^{\rm sb,b}$, where $\bm{v}^{\rm s}$ is a $N^{\rm s}-$ dimensional vector and
$\bm{A}^{\rm bs}$ a $N^{\rm b}\times N^{\rm s}$ matrix, is to be read as a matrix with $ji$-element 
$\sum_{k, l} \big(v_k A_{lj}\big)\bm{L}_{kl,i}$.

Treating the other terms in \eqref{dsCompact} in the same way and dropping third order terms $\sim \bm{\hat{x}}^{\rm s}\bm{\hat{x}}^{\rm s}\bm{\hat{x}}^{\rm s}$ that encode non-vanishing higher cumulants of the noise distribution one obtains the perturbative contribution to $\bm{N}^{\rm ss}_{1}(t,t')$ from the terms $\sim \bm{\hat{x}}^{\rm s}\bm{\hat{x}}^{\rm s}\delta\bm{x}^{\rm s}$. For extracting this type of terms from the diffusive 
part of the action \eqref{dsCompactn} one 
needs to consider only the $\bm{\p}$ part of the conditional mean \eqref{bulkmean}. One finds for e.g.\ $t<t''$
\begin{eqnarray}
\label{eq:covNL}
\fl && \bm{N}_1^{\rm ss}(t,t'')=\bm{\Sigma}_1^{\rm ss}(\delta\bm{x}^{\rm s}(t),\bm{\p}(t))\delta(t-t'') + \bm{\Sigma}_1^{\rm sb}(\delta\bm{x}^{\rm s}(t),\bm{\p}(t))\ebb(t''-t) \bm{L}^{\rm bs}\\
\fl &&+ (\bm{L}^{\rm bs})^{\T}\ebb^{\T}(t''-t)\bm{\Sigma}_1^{\rm bs}(\delta\bm{x}^{\rm s}(t),\bm{\p}(t)) +\int_0^t dt'(\bm{L}^{\rm bs})^{\T}\ebb^{\T}(t-t')\bm{\Sigma}^{\rm bb}(\delta\bm{x}^{\rm s}(t'),\bm{\p}(t'))\ebb(t''-t') 
\bm{L}^{\rm bs} \notag\\
\fl &&+\Bigg\lbrace\big((\bm{L}^{\rm bs})^{\T}\bm{C}^{\rm bb|s}(t,t'')\circ\delta\bm{x}^{\rm s^{\T}}(t)\big)\bm{L}^{\rm sb, s}+\int_0^t dt' 
\big((\bm{L}^{\rm bs})^{\T}\bm{C}^{\rm bb|s}(t',t'')\circ\delta\bm{x}^{\rm s^{\T}}(t')\big)\bm{L}^{\rm sb, \rm b}\bm{E}_{\rm bb}(t-t')\bm{L}^{\rm bs}\notag\\
\fl &&+\int_0^t ds\int_0^{s}dt'\bigg[\delta\bm{x}^{\rm s^{\T}}(t') \bm{L}^{\rm sb}\bm{E}_{\rm bb}(s-t')\circ(\bm{L}^{\rm bs})^{\T}
\bm{C}^{\rm bb|s}(s,t'')+\notag\\
\fl &&\qquad \qquad\qquad \quad(\bm{L}^{\rm bs})^{\T}\bm{C}^{\rm bb|s}(s,t'')\circ\delta\bm{x}^{\rm s^{\T}}(t')
\bm{L}^{\rm sb}\bm{E}_{\rm bb}(s-t')\bigg]\bm{L}^{\rm bb, \rm b}
\bm{E}_{\rm bb}(t-s)\bm{L}^{\rm bs}\Bigg\rbrace+\lbrace\ldots\rbrace^{\T}\notag
\end{eqnarray}
where $\lbrace\ldots\rbrace^{\T}$ indicates the transpose of all preceding terms in curly brackets, producing a symmetric matrix overall. Here $\bm{\Sigma}_1^{\cdot \cdot}$, $\cdot = \rm{s}, \rm{b}$, depends linearly on $\delta\bm{x}^{\rm s}$ either directly or via the $\bm{\nu}$ part of the bulk conditional means \eqref{bulkmean}. Interestingly, also the deterministic part of the action \eqref{dsCompact} contributes to effective noise in the subnetwork-reduced dynamics, in particular as concerns the connection to random initial bulk conditions via $\bm{C}^{\rm bb|s}$. Equations \eqref{eq:covL} and \eqref{eq:covNL} imply that the effective noise statistics, including temporal correlations, can be derived from the statistics of the initial states of the bulk and the one of white noise.

In this expansion, which is systematic to order $\delta\bm{x}$, the intrinsic noise source has covariance $\bm{\Sigma} \sim \bm{\Sigma}_0+\bm{\Sigma}_1$, 
but there is no guarantee that this matrix is positive definite in such a way that one can effectively work with an intrinsic noise defined accordingly. 
To overcome this issue, we directly expand the square root $\bm{D} = \sqrt{\bm{\Sigma}} \sim \bm{D}_0 + \bm{D}_1$ where $\bm{D}_0 = \sqrt{\bm{\Sigma_0}}$ and 
$\bm{D}_1 = \frac{1}{2}\bm{\Sigma_1}\bm{\Sigma_0}^{-1/2}$ (such that $\bm{D} \bm{D}^{\T}$ gives $\bm{\Sigma}$ order by order). 
In this way, the effective intrinsic noise $\bm{\xi}^{\cdot}(t)$ (for $\cdot$ =s,b) that can be defined from this expansion is composed of two terms
\be
\bm{\xi}^{\cdot}(t) 
= \bm{\xi}^{\cdot}_0(t) + \bm{\xi}^{\cdot}_1(t) = (\bm{D}_0 +\bm{D}_1)\bm{\Gamma}(t)
\ee
which are characterized by Eq.\ \eqref{eq:sigmasub} (main text). $\bm{\Gamma}(t)$ is a $R-$dimensional (one component per reaction) Gaussian white and isotropic noise with $\langle \bm{\Gamma}(t)\rangle = 0$ and $\langle \bm{\Gamma}(t) \bm{\Gamma}(t')\rangle=\mathbb{1}\delta(t-t')$:  
$\bm{\xi}_0^{\rm s}(t)$ and $\bm{\xi}_0^{\rm b}(t)$ are correlated by being constructed from the same vector $\bm{\Gamma}(t)$, $\bm{\xi}_1^{\rm s}(t)$ and $\bm{\xi}_1^{\rm b}(t)$ can then be calculated directly from the same white noise trajectory, just by linearizing $\sqrt{\bm{\Sigma}}$ in $\delta\bm{x}$ without further randomness.

We can now determine $\bm{\chi}_1(t)$, the nonlinear effective noise whose covariance is $\bm{N}_1^{\rm ss}$ up to $O(\delta \bm{x}^{\rm s \, 2})$
\begin{eqnarray}    
 \label{chi1a}   
  \fl &&\bm{\chi}_1^{\T}(t)=\\
  \fl && \bm{\xi}_1^{\rm s^{\T}}(t) + \int_{-\infty}^t dt'\,\bm{\xi}_1^{\rm b^{\T}}(t')\bm{E}_{\rm bb}(t-t')\bm{L}^{\rm bs}+
     \big(\delta\bm{x}^{\rm{s}}(t) \circ \bm{\lambda}(t) \big)^{\T}\bm{L}^{\rm sb,s}+\int_0^t dt' 
    \big(\delta\bm{x}^{\rm{s}}(t')\circ \bm{\lambda}(t')\big)^{\T}\bm{L}^{\rm sb,b}\bm{E}_{\rm bb}(t-t')
     \bm{L}^{\rm bs}\notag \\ 
 \fl &&+\int_0^t ds\int_0^{s}dt'\bigg[\big(\bm{\lambda}(s)\circ \bm{E}^{\T}_{\rm bb}(s-t')(\bm{L}^{\rm sb})^{\T}\delta\bm{x}^{\rm{s}}(t')
 +\bm{E}^{\T}_{\rm bb}(s-t')(\bm{L}^{\rm sb})^{\T}\delta\bm{x}^{\rm{s}}(t')\circ \bm{\lambda}(s)\big)^{\T}\bm{L}^{\rm bb,b}\bigg]\bm{E}_{\rm bb}(t-s)\bm{L}^{\rm bs}  \notag  
\end{eqnarray}
where $\bm{\lambda}(t)$ was defined as in \eqref{eq:deflam} of the main text such that
$\langle \bm{\lambda}(t) \bm{\lambda}^{\T}(t') \rangle = \bm{C}^{\rm bb|s}(t,t')$. $\bm{\chi}_1(t)$ is just the $\delta\bm{x}^{\rm s}(t)$-dependent part of the effective colored noise. From expression \eqref{dsCompact}, it can be seen that the Gaussian effective dynamics of $\delta\bm{x}^{\rm s}(t)$ exhibits 
also the $\bm{x}^{\rm s}$-independent term
\begin{equation}
 \label{addTerm}
\int_0^t dt' \bm{C}^{\rm bb|s}(t',t')\bm{L}^{\rm bb, \rm b}\bm{E}_{\rm bb}(t-t')\bm{L}^{\rm bs}=\langle\bm{\psi}_1^{\T}(t)\rangle
\end{equation}
where all the elements of $\bm{C}^{\rm bb|s}(t',t')$ are meant as arranged into a $N^{\rm b}(N^{\rm b}+1)/2$-dimensional vector.
\eqref{addTerm} can be seen as the conditional average of a vector $\bm{\psi}_1(t)$
\be
\label{psi1}
\bm{\psi}_1^{\T}(t) =
\int_0^t dt'  \big(\bm{\lambda}(t')\circ\,\bm{\lambda}(t')\big)^{\T}\bm{L}^{\rm bb,b}\bm{E}_{\rm bb}(t-t')\bm{L}^{\rm bs}
\ee

The conditional average of $\bm{\psi}_1$ produces in the effective action a cubic term 
$\sim \hat{\bm{x}}^{s}\delta\bm{x}^{\rm b} \delta\bm{x}^{\rm b}$, an additional source of extrinsic noise. Thus we redefine $\bm{\chi}_1(t)$ in such a way as 
to include it, i.e.\ $\bm{\chi}_1(t) \leftarrow \bm{\chi}_1(t) + \bm{\psi}_1(t)$ and we obtain \eqref{eq:chi1} of the main text.
\begin{table}
\begin{center}
 \begin{tabular}{|c|c|} 
 \hline
 \textbf{Term in effective action} & \textbf{Role in $\delta\bm{x}^{\rm s}$ dynamics} \\ [0.5ex] 
 \hline\hline
 $\sim \bm{\hat{x}}^{\rm s}(t)\delta\bm{x}^{\rm s}(t)$ & Linear intrasubnetwork dynamics \hspace{0.2cm} \tikz\draw[red,fill=red] (0,0) circle (.4ex);\\ 
 \hline
 $\sim \bm{\hat{x}}^{\rm s}(t)\delta\bm{x}^{\rm s}(t)\delta\bm{x}^{\rm s}(t)$ & Nonlinear intrasubnetwork dynamics \hspace{0.2cm} \tikz\draw[red,fill=red] (0,0) circle (.4ex);\\
 \hline
 $\sim \bm{\hat{x}}^{\rm s}(t) \bm{\hat{x}}^{\rm s}(t)$ & Linear intrinsic noise\hspace{0.2cm} \tikz\draw[red,fill=red] (0,0) circle (.4ex);\\
 \hline
 $\sim \bm{\hat{x}}^{\rm s}(t)\delta\bm{x}^{\rm s}(t)\bm{\hat{x}}^{\rm s}(t)$ & Nonlinear intrinsic noise\hspace{0.2cm} \tikz\draw[red,fill=red] (0,0) circle (.4ex);\\
 \hline
$\sim \bm{\hat{x}}^{\rm s}(t)\delta\bm{x}^{\rm s}(t')$ & Linear memory \hspace{0.2cm} \tikz\draw[darkpastelgreen,fill=darkpastelgreen] (0,0) circle (.4ex);\\
 \hline
 $\sim \bm{\hat{x}}^{\rm s}(t)\delta\bm{x}^{\rm s}(t') \delta\bm{x}^{\rm s}(t'')$ & Nonlinear memory \hspace{0.2cm} \tikz\draw[darkpastelgreen,fill=darkpastelgreen] (0,0) circle (.4ex); \\
  \hline
  $\sim \bm{\hat{x}}^{\rm s}(t) \bm{\hat{x}}^{\rm s}(t')$ & Linear extrinsic noise \hspace{0.2cm} \tikz\draw[darkpastelgreen,fill=darkpastelgreen] (0,0) circle (.4ex); \\
  \hline
 $\sim \bm{\hat{x}}^{\rm s}(t)\delta\bm{x}^{\rm s}(t') \bm{\hat{x}}^{\rm s}(t'')$ & Nonlinear extrinsic noise  \hspace{0.2cm} \tikz\draw[darkpastelgreen,fill=darkpastelgreen] (0,0) circle (.4ex); \\
 \hline
\end{tabular}
\end{center}
\caption{Construction of the stochastic subnetwork dynamics via the path integral approach 3GA. Terms arising from the bulk (green filled circle) differ from subnetwork ones (red filled circle) as they couple different time steps. i.e.\ they are non-Markovian.}
\label{table:tabledx}
\end{table}
\cleardoublepage

\section{KL divergence estimation}
\begin{figure}[!ht]
\centering
\includegraphics[width=\textwidth]{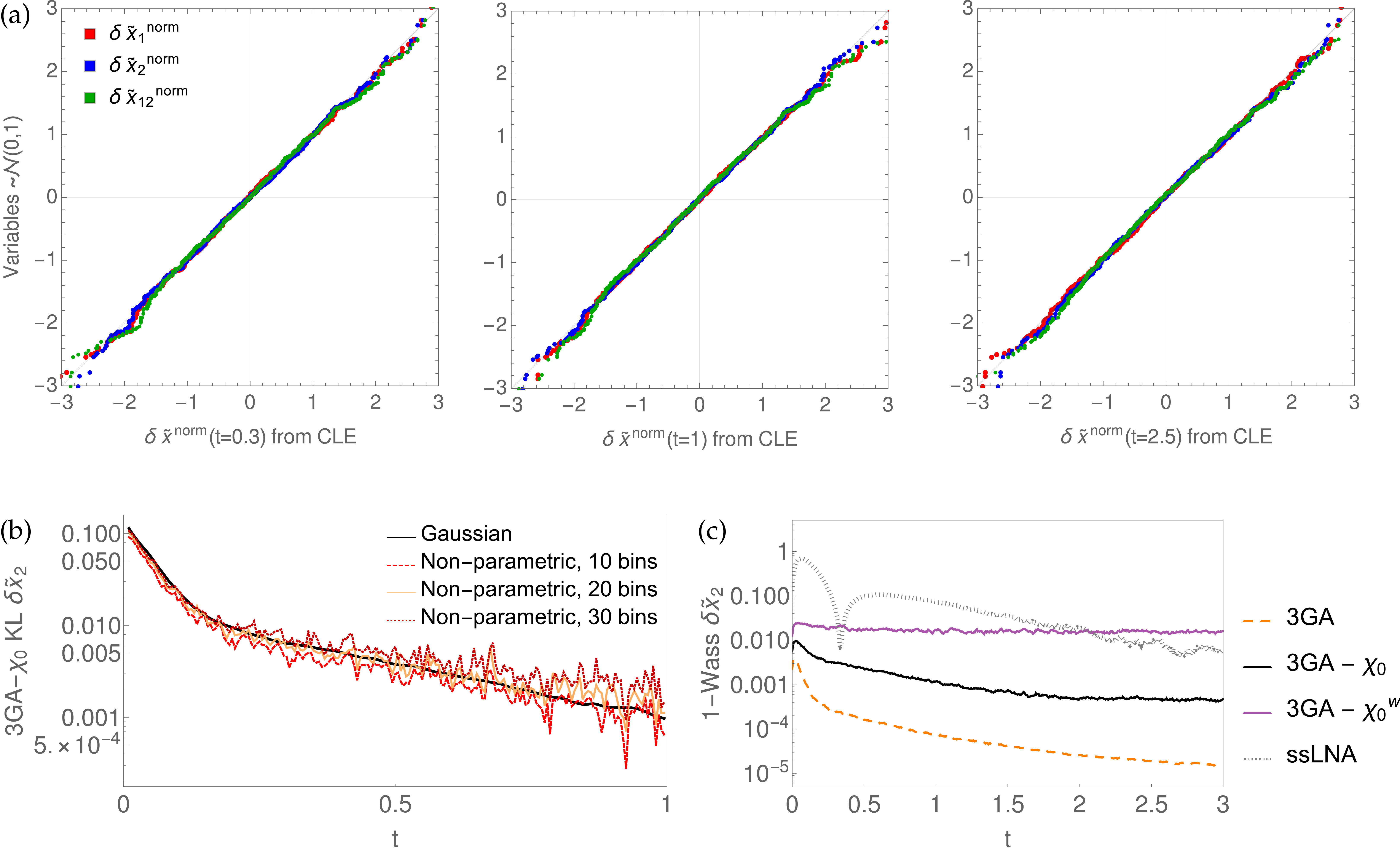}
\caption[]{\textbf{Numerical tests for KL divergence estimation}. (a) Quantile-quantile plots: each plot contains 1000 realizations of the stochastic trajectories ($\epsilon = 0.1$) for the subnetwork species in the simple network model at time steps $t=0.3, 1, 2.5$ (same trajectories as used in Figs.~\ref{fig:3p3c3eps}-\ref{fig:3p3c3kleps}). The notation $\delta \tilde{x}^{\rm norm}$ indicates that we subtracted from $\delta \tilde{x}$ its mean and divided by the standard deviation so that its distribution can be directly compared to samples from a zero mean, unit variance Gaussian. (b) KL divergence for $\delta \tilde{x}_{2}$ trajectories in the simple network model for the 3GA-$\chi_0$ approximation. Estimates are shown from Gaussian distribution fits and from histogram estimators with 10, 20, 30 bins respectively. To improve the accuracy of density estimation from histograms, we have focused on a shorter time range ($t=0$ to 1) for which we can increase the number of realizations to $10^4$. (c) $1$-Wasserstein distance between $\delta\tilde{x}_2$ exact and approximated trajectories (from 1000 noise realizations at $\epsilon =0.1$) as a function of time, for 3GA, 3GA-$\bm{\chi}_0$, 3GA-$\bm{\chi}^{w}_0$ and ssLNA.}
\label{fig:wasser}
\end{figure}

In this section we discuss the numerical tests supporting our choice of estimating KL divergences by a Gaussian fitting [expression \eqref{eq:singlekl} in the main text]. We have first tested that the concentration distributions over time produced by the full nonlinear CLE are well approximated by Gaussians. We do this using the graphical method of quantile-quantile plots, see Fig.\ \ref{fig:wasser}(a). In this test, the data under consideration (here concentrations from the realizations of the subnetwork dynamics at a fixed time point) are sorted and plotted vs.\ data sampled (and again sorted) from the theoretically expected distribution, here a Gaussian distribution. This produces a diagonal line when the distributions match, and Fig.\ \ref{fig:wasser}(a) shows that this holds true to very good accuracy.

To quantify the impact of the Gaussian approximation on the actual values of the KL, we have next compared the parametric (Gaussian) KL estimate to a non-parametric one. For the latter we use a histogram estimator that simply derives probability estimates from counts and plugs these into the integral defining the KL divergence. We perform the comparison for the 3GA-$\chi_0$, as its values lie in the small KL-regime, and obtain excellent agreement [Fig.\ \ref{fig:wasser}(b)]. Large KL divergences (as for 3GA-$\chi_0^w$ and ssLNA) pick up contributions mainly from ranges where one of the two distributions is very small, hence they are more difficult to estimate numerically; as explained in the main text we only use these to establish qualitative trends therefore. For small KL values (as for the 3GA), i.e.\ where our approximations perform well, quantative values are much more important. There the Gaussian KL estimators are advantageous because the non-parametric estimators are much more strongly affected by sampling errors as well as the effects of histogram discretization.

Finally, we have measured an alternative distance between distributions, the $p$-Wasserstein distance with exponent $p=1$ \cite{villani2009}, see Fig.\ \ref{fig:wasser}(c). The performance assessment resulting from the trends there is consistent with the one resulting from the KL divergence for the same time courses [Fig.\ \ref{fig:3p3c3kleps}]. The 1-Wasserstein distance was evaluated using the implementation in the SciPy 1.0 package \cite{virtanen2020}.

\cleardoublepage
\section{Epidermal Growth Factor Receptor}
\label{sec:egfr_si}
\subsection{Accuracy with initial uncertainty}

\begin{figure}[!ht]
 \centering
\includegraphics[width=\textwidth]{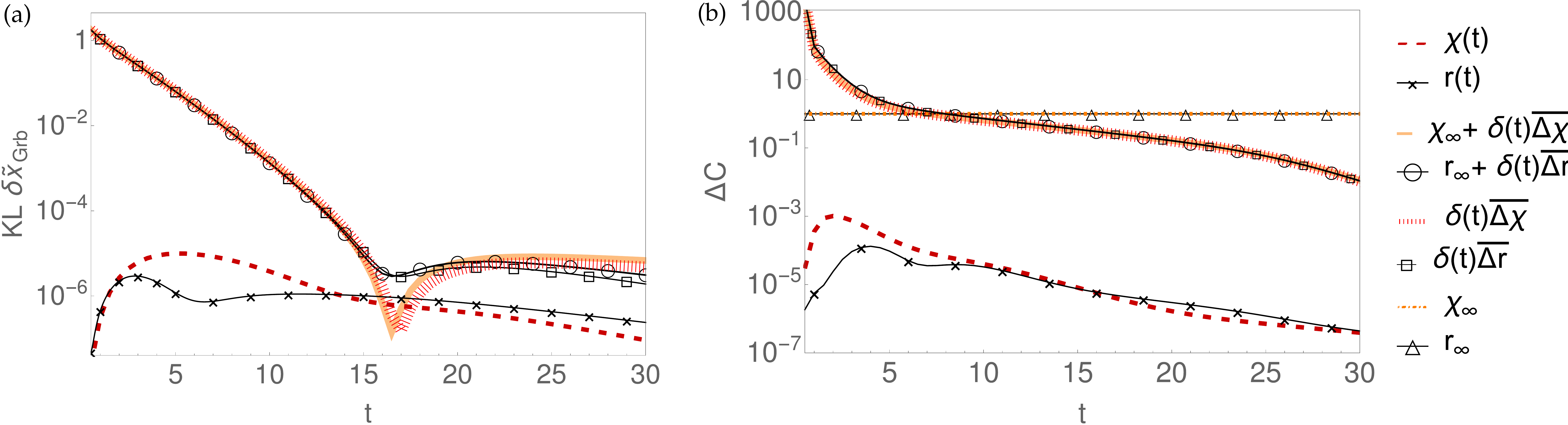}
\caption[]{\textbf{Accuracy of approximations of the extrinsic noise from random initial conditions at $\epsilon_0 = 0.01$}. (a) KL divergence of boundary species Grb in the EGFR network for projected and 3GA-reduced equations with either the full time-dependent extrinsic noise or its approximation by $\delta$-piece (impulse) plus constant (persistent piece) and by $\delta$-piece only from a sample of 200 random bulk initial conditions at $\epsilon_0 = 0.01$; there is no intrinsic noise ($\epsilon = 0$). All KLs are close to zero and have slightly larger values only during an initial period of around 10 s. The KL for the constant piece-only is not shown as its values are above $10^3$. (b) Error $\Delta C$ for the same approximations as in (a) and for the approximation of extrinsic noise by the constant piece only.}
 \label{fig:egfr_eps0}
\end{figure}

\subsection{Conservation laws}
\label{sec:cons_law}
The analysis of the EGFR network offers a useful context in which to understand in what situations the $\bm{r}_{\infty}$-part of the random force might prove important for proper prediction of subnetwork concentrations, i.e.\ for the long-time behavior of species appearing in conservation laws. 

In the EGFR system of equations from \cite{kholodenko}, there are 9 conservation laws and all the subnetwork species participate in at least one of them. As an illustrative example, we focus on the following conservation law, written in terms of the fractional deviations from steady state $\delta \tilde{x}$
\be
y_{\text{GS}}\delta \tilde{x}_{\text{GS}} + y_{\text{RGS}}\delta \tilde{x}_{\text{RGS}} + y_{\text{RShGS}}\delta \tilde{x}_{\text{RShGS}} + y_{\text{ShGS}}\delta \tilde{x}_{\text{ShGS}} + y_{\text{SOS}}\delta \tilde{x}_{\text{SOS}} =0
\ee
It involves a combination of subnetwork species (SOS and GS, which are on the boundary, and RGS, which is internal to the subnetwork) and bulk species (RShGS and ShGS). The time evolution of the boundary species SOS and GS contains an explicit random force term, which can be approximated as a sum of impulse and persistent pieces as described in Sec.~\ref{sec:approx_rf_main} of the main text. 

We consider the long time ($t=200$) value of $\delta \tilde{x}_{\text{SOS}}$ for different, randomly sampled, bulk initial conditions as given by 
projected equations with either full or approximated random force. We subtract and divide by the exact solution $\delta \hat{x}_{\text{SOS}}$ to get
\be
\rho_{\text{SOS}} = \frac{\delta \tilde{x}_{\text{SOS}} - \delta \hat{x}_{\text{SOS}}}{\delta \hat{x}_{\text{SOS}}}
\ee
in such a way that $\rho_{\text{SOS}}=0$ corresponds to perfect prediction; we define $\rho_{\text{GS}}$ analogously. 

\begin{figure}[!ht]
  \includegraphics[width=\textwidth]{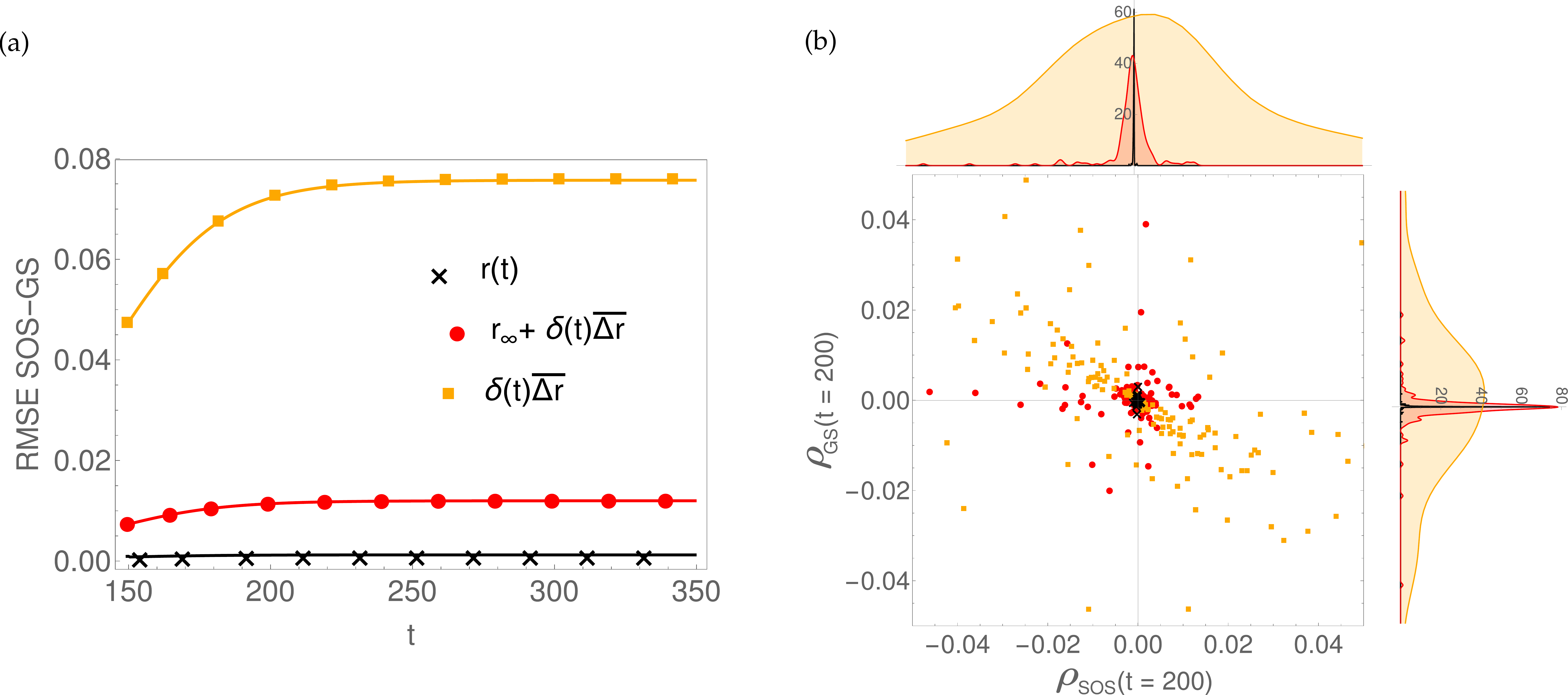}
 \caption[]{\textbf{Accuracy of prediction with conservation laws.} (a) Root Mean Squared Error (RMSE) on subnetwork boundary species GS and SOS (involved in the same conservation law) for the full random force (bold black line with crosses), for the impulse plus persistent approximation (bold red line with filled circles) and for the impulse-only one (bold orange line with filled squares). (b) Scatter plot of the steady state values of $\rho_{\text{SOS}}$ and $\rho_{\text{GS}}$ for 200 bulk initial conditions and for the same approximations of the random force as in (a). The histograms on the respective axes give the dispersion around zero (perfect prediction).}
 \label{fig:cons_law}
\end{figure}

We can numerically verify that approximating the random force \emph{without} the $\bm{r}_{\infty}$-part implies a shift in the overall value of the conservation law, hence leads to a larger error in the prediction of steady states reached at long times compared to an approximation including both $\bm{r}_{\infty}$ and $\overline{\Delta\bm{r}}$, as reflected in the sample-averaged root mean squared values $\rho_{\text{GS}}$ and $\rho_{\text{SOS}}$, see Fig.~\ref{fig:cons_law}(a). In addition, without this constant piece, the dispersion of the prediction errors $\rho_{\text{SOS}}$, $\rho_{\text{GS}}$ is larger [see histograms in Fig.~\ref{fig:cons_law}(b)] and there are correlations in their values as the long time behavior of the two species is correlated via the conservation law [see scatter plot in Fig.~\ref{fig:cons_law}(b)].

\cleardoublepage
\subsection{Stochastic fluctuations}
\begin{figure}[!ht]
 \centering
\includegraphics[width=16cm]{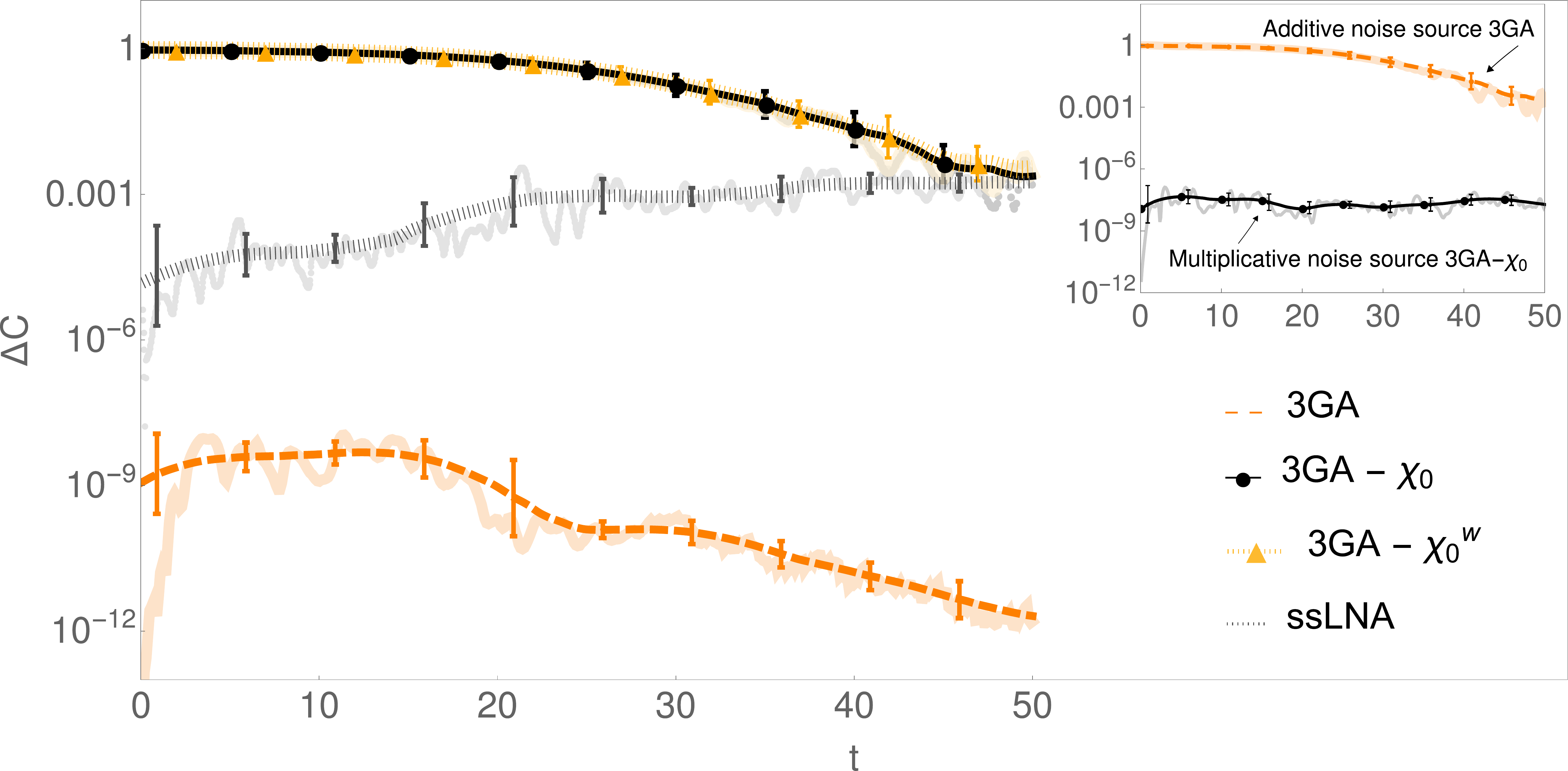}
 \caption[]{\textbf{Accuracy of effective subnetwork noise approximations at $\epsilon = 0.01$}. Equivalent of Fig.\ \ref{fig:egfr_eps}(b) (main text) where the covariance matrix error $\Delta C$ is evaluated {\em with} the normalization by the steady state concentrations, as defined by \eqref{eq:deltaC} in the main text.}
\label{fig:egfr_eps_si}
\end{figure}

\begin{figure}[!ht]
 \centering
  \includegraphics[width=\textwidth]{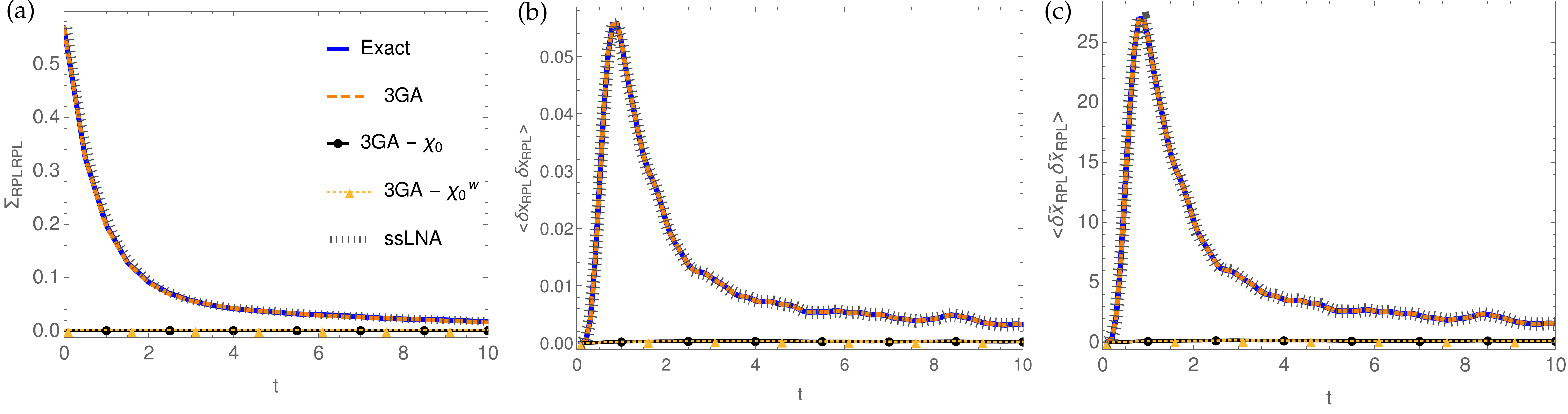}
 \caption[]{\textbf{Fluctuations magnitude for intra-subnetwork species RPL}. (a) RPL is a species with very low steady state concentration ($y_{RPL} = 0.045$ nMol), and here we consider initial conditions determining a large initial deviation from such steady state. Additive noise approximations such as 3GA-$\bm{\chi}_0$ and 3GA$-\bm{\chi}_0^w$ set the noise amplitude, here exemplified by the diagonal noise variance $\bm{\Sigma}_{RPL,RPL}$, at any $t$ to the small value expected at steady state, while multiplicative noise approximations (3GA, ssLNA) keep a dependence of the noise on concentration trajectories as in the exact dynamics. Additive noise approximations hence significantly underestimate the amplitude of noise fluctuations in the transient and hence the variance $\langle \delta x_{RPL} \delta x_{RPL}\rangle$ arising from that source (b). This difference in accuracy is dramatically amplified when considering the normalized covariance $\langle \delta\tilde{x}_{RPL} \delta\tilde{x}_{RPL}\rangle$ due to the small value of $y_{RPL}$ (c).}
\label{fig:egfr_coveps_si}
\end{figure}

\cleardoublepage
\section{Joint sampling of extrinsic and intrinsic noise}

\begin{figure}[!ht]
 \includegraphics[width=\textwidth]{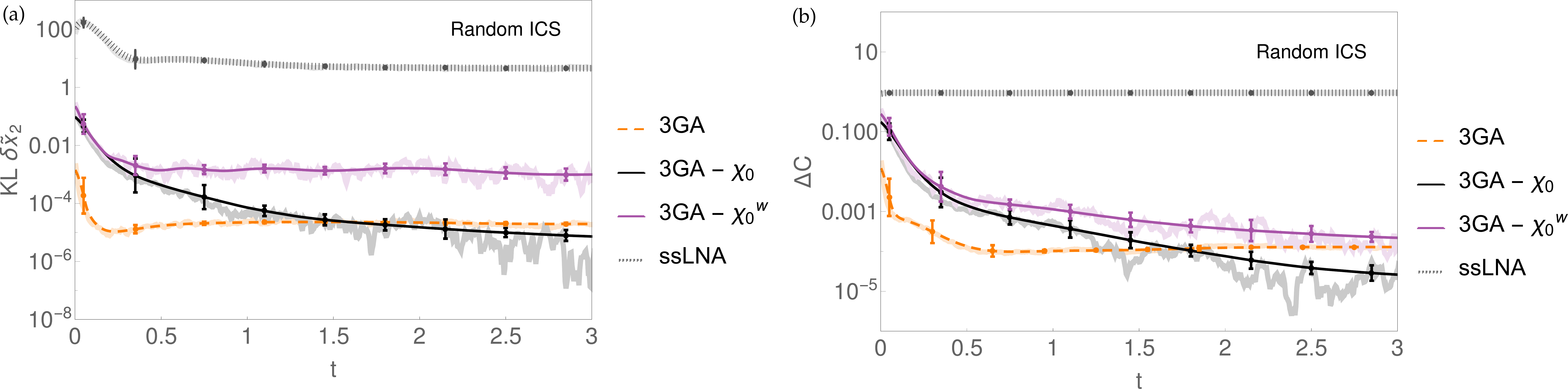}
 \caption{\textbf{Accuracy of approximations of effective subnetwork noise in the simple network model at $\epsilon = \epsilon_0 = 0.1$}. (a) KL of subnetwork boundary species $\delta \tilde{x}_2$ and (b) error $\Delta C$, for 3GA, ssLNA and the intermediate approximations (3GA-$\bm{\chi}_0$ and 3GA-$\bm{\chi}^{w}_0$), estimated from a sample of 1000 realizations of noise ($\epsilon = 0.1$) \emph{and} random bulk initial conditions ($\epsilon_0 = 0.1$, ``Random ICS''). Smoothed curves with corresponding error bars are shown in dark color in addition to the raw numerical data (light color).}
 \label{fig:3p3c3klepseps0}
\end{figure}

\begin{figure}[!ht]
  \includegraphics[width=\textwidth]{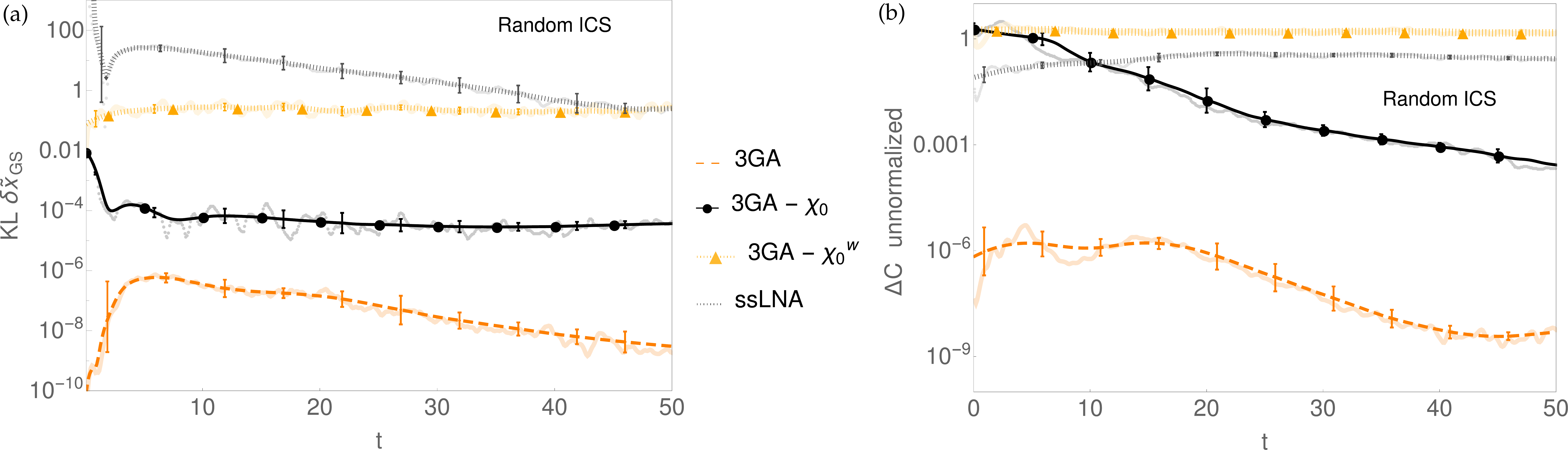}
  \caption[]{\textbf{Accuracy of approximations of effective subnetwork noise in the EGFR network at $\epsilon = \epsilon_0= 0.01$}. (a) KL of boundary species GS in the EGFR network and (b) error $\Delta C$ for 3GA, ssLNA and the intermediate approximations (3GA-$\bm{\chi}_0$ and 3GA-$\bm{\chi}^{w}_0$), estimated from a sample of 200 realizations of noise ($\epsilon = 0.01$) \emph{and} random bulk initial conditions ($\epsilon_0 = 0.01$, ``Random ICS''). Smoothed curves with corresponding error bars are shown in dark color in addition to the raw numerical data (light color).}
  \label{fig:kl_gs_ics}
\end{figure}

In this section we report tests of the accuracy of the various approximations when noise and random initial conditions in the bulk are both present. In this case, the two contributions to the effective noise correspond to the terms proportional to $\bm{\Sigma}$ and $\bm{C}^{\rm bb}(0,0)$ [as visible in Eq.~\eqref{eq:bulk_corr} of main text] and their actual weight depends on the reaction rates in the network that is being considered. For our simple network of Fig.~\ref{fig:3p3c3rf}, the second contribution (initial bulk uncertainty) works out to be the larger one. The ssLNA, by assuming the bulk to be at steady state at each time, sets $\bm{C}^{\rm bb}(0,0)=0$. It thus neglects the dominant fluctuations induced by stochastic initial conditions. This leads to large [close to unity relative errors $\Delta C$ in the time-dependent covariances [Fig.~\ref{fig:3p3c3klepseps0}(b)] and large KL divergences [Fig.~\ref{fig:3p3c3klepseps0}(a)]. The 3GA-$\bm{\chi}_0^{w}$ approximation does include contributions from $\bm{C}^{\rm bb}(0,0)$ so its performance is significantly better than the ssLNA and close to the one of the 3GA-$\bm{\chi}_0$; note that these two approximations account for the $\bm{C}^{\rm bb}(0,0)$ part of the noise in the same way. A difference between 3GA-$\bm{\chi}_0$ and 3GA-$\bm{\chi}_0^{w}$ can still be observed in the KL divergences [Fig.~\ref{fig:3p3c3klepseps0}(a)], where means are also taken into account, especially at long times. In contrast, for the EGFR network, one can verify that the difference w.r.t.\ sampling stochastic noise only is minimal, since the $\bm{C}^{\rm bb}(0,0)$ part does not dominate the subnetwork variation as in the simple network mode, see comparison between Fig.~\ref{fig:egfr_eps} and Fig.~\ref{fig:kl_gs_ics}. 

\cleardoublepage
\begingroup
\def\refname{References}
\def\bibname{References}

\input{main_SI.bbl}
\endgroup